\documentclass[11pt]{article}
\usepackage[T1]{fontenc}
\usepackage{mathtools}
\usepackage{amssymb}
\usepackage{stmaryrd}
\usepackage[mathscr]{euscript}
\usepackage{dsfont}
\usepackage{slashed}
\usepackage{fullpage}
\usepackage{currvita}
\usepackage{tocloft}
\usepackage[numbers,sort,compress]{natbib}
\usepackage[colorlinks=true,linkcolor=blue,citecolor=magenta,linktocpage=true]{hyperref}
\usepackage{braket}
\usepackage{titlesec}
\titleformat*{\section}{\normalsize\bfseries}
\titleformat*{\subsection}{\normalsize\bfseries}
\titleformat*{\subsubsection}{\normalsize\bfseries}
\makeatletter
\renewcommand{\@dotsep}{1000}
\def\be#1\ee{\begin{align}#1\end{align}}
\def\bsub#1\esub{\begin{subequations}#1\end{subequations}}

\def\q{\qquad}
\def\f{\frac}
\def\eps{\varepsilon}
\def\teps{\tilde{\varepsilon}}
\def\lb{\big\lbrace}
\def\rb{\big\rbrace}

\def\ip{\lrcorner\,}
\def\ipp{\ip\!\!\!\ip}

\def\de_\omega{\mathrm{D}}
\def\de{\mathrm{d}}
\def\i{\mathrm{i}}

\def\A{\mathcal{A}}
\def\B{\mathcal{B}}
\def\C{\mathcal{C}}
\def\D{\mathcal{D}}
\def\E{\mathcal{E}}
\def\F{\mathcal{F}}

\def\H{\mathcal{H}}

\def\J{\mathcal{J}}
\def\K{\mathcal{K}}
\def\L{\mathcal{L}}

\def\O{\mathcal{O}}
\def\P{\mathcal{P}}
\def\Q{\mathcal{Q}}

\def\T{\mathcal{T}}

\numberwithin{equation}{section}
\def\ra{\rangle}
\def\la{\langle}
\def\pe{\phantom{\ =}}

\begin{document}

\title{\Large{\textbf{\sffamily Most general theory of 3d gravity:\\ Covariant phase space, dual diffeomorphisms, and more}}}
\author{\sffamily Marc Geiller$^1$, Christophe Goeller$^1$, Nelson Merino$^2$}
\date{\small{\textit{
$^1$Univ Lyon, ENS de Lyon, Univ Claude Bernard Lyon 1,\\ CNRS, Laboratoire de Physique, UMR 5672, F-69342 Lyon, France\\
$^2$Instituto de Ciencias Exactas y Naturales (ICEN), Facultad de Ciencias,\\ Universidad Arturo Prat, Iquique, Chile\\~}}}

\maketitle

\begin{abstract}
We show that the phase space of three-dimensional gravity contains two layers of dualities: between diffeomorphisms and a notion of ``dual diffeomorphisms'' on the one hand, and between first order curvature and torsion on the other hand. This is most elegantly revealed and understood when studying the most general Lorentz-invariant first order theory in connection and triad variables, described by the so-called Mielke--Baekler Lagrangian. By analyzing the quasi-local symmetries of this theory in the covariant phase space formalism, we show that in each sector of the torsion/curvature duality there exists a well-defined notion of dual diffeomorphism, which furthermore follows uniquely from the Sugawara construction. Together with the usual diffeomorphisms, these duals form at finite distance, without any boundary conditions, and for any sign of the cosmological constant, a centreless double Virasoro algebra which in the flat case reduces to the BMS$_3$ algebra. These algebras can then be centrally-extended via the twisted Sugawara construction. This shows that the celebrated results about asymptotic symmetry algebras are actually generic features of three-dimensional gravity at any finite distance. They are however only revealed when working in first order connection and triad variables, and a priori inaccessible from Chern--Simons theory. As a bonus, we study the second order equations of motion of the Mielke--Baekler model, as well as the on-shell Lagrangian. This reveals the duality between Riemannian metric and teleparallel gravity, and a new candidate theory for three-dimensional massive gravity which we call teleparallel topologically massive gravity.
\end{abstract}

\thispagestyle{empty}
\newpage
\setcounter{page}{1}

\tableofcontents


\section{Introduction}

In spite of the apparent simplicity owing to its topological nature, three-dimensional gravity is a very rich theory \cite{MR1637718}. It contains black hole solutions with thermodynamical behavior \cite{Banados:1992wn,Banados:1998ta,Carlip:2005zn}, admits non-perturbative quantizations following different yet complementary techniques \cite{MR974271,Carlip:1993zi,Carlip:1993ze,Carlip:1994ap,Alexandrov:2011ab,Goeller:2019zpz}, allows to describe topology change \cite{Witten:1989sx,Ooguri:1991ni,Carlip:1994tt,Oriti:2006se}, possesses remarkable holographic properties \cite{Brown:1986nw,Coussaert:1995zp,Maldacena:2003nj,Dittrich:2018xuk,Dittrich:2017hnl,Dittrich:2017rvb,Goeller:2019apd,Asante:2019ndj}, gives rise to quantum groups \cite{Witten:1988hf,Witten:1989rw,Bais:2002ye,Noui:2006ku,Meusburger:2008bs,Ballesteros:2010zq,Dupuis:2020ndx}, and can be generalized to include e.g. massive propagating gravitons \cite{Deser:1981wh,Deser:1982vy,Deser:1984kw,Bergshoeff:2009aq,Bergshoeff:2009hq,Bergshoeff:2013xma,Alexandrov:2014oda,Bergshoeff:2015zga,Afshar:2014ffa,Bergshoeff:2014pca,Bergshoeff:2014bia,Merbis:2014vja,Geiller:2018ain,Geiller:2019dpc}.

Gravity in three spacetime dimensions has also proven very useful to understand the symmetry content of general relativity in the presence of boundaries. The seminal work of Brown and Henneaux on AdS$_3$ \cite{Brown:1986nw} is now understood as part of a much bigger picture relating asymptotic symmetries and boundary dynamics, and many recent studies have for example focused on developing an understanding of the BMS$_3$ group associated with flat spacetimes \cite{Ashtekar:1996cd,Barnich:2006av,Barnich:2014kra,Barnich:2015uva,Barnich:2015sca,Barnich:2015mui,Dittrich:2017rvb,Oblak:2016eij,Garbarz:2015lua,Strominger:2017zoo,Bagchi:2019unf}. Studies such as \cite{Grumiller:2016pqb,Grumiller:2017sjh} have also started to classify possible boundary conditions and related asymptotic symmetries.

This wealth of results is in part due to the existence of many (possibly inequivalent) formulations of three-dimensional gravity. The most notorious one is the Chern--Simons formulation (CS hereafter), which itself is, loosely speaking, based on a change of variables from the first order connection-triad formulation of Cartan. At the difference with the metric formulation, the Chern--Simons one puts forward the role of algebraic structures and of internal gauge transformations as opposed to diffeomorphisms.

In order to understand some general and systematic results about three-dimensional gravity, we revisit here the study of the most general diffeomorphism and Lorentz-invariant theory in connection and triad variables. This is described by the Mielke--Baekler (MB) Lagrangian \eqref{MB Lagrangian}, which is given by the sum of a volume (or cosmological constant) term, the usual Hilbert--Palatini term, a Chern--Simons term for the Lorentz connection, and a torsion term \cite{Mielke:1991nn,Baekler:1992ab}. This four-parameter topological theory has first order equations of motion which contain a source of curvature measured by a parameter $p$, and a source of torsion measured by a parameter $q$. When solving for the torsion, the second order equations of motion reduce to the Riemannian Einstein equations $R_{\mu\nu}=2\Lambda g_{\mu\nu}$, where $\Lambda(p,q)$ is the effective cosmological constant. This simple observation has previously led authors to conclude that the MB formulation of three-dimensional gravity is simply a CS formulation whose gauge algebra\footnote{Explicitly, as we work in Lorentzian signature we have $\mathfrak{g}_\Lambda=\mathfrak{iso}(2,1)$ when $\Lambda=0$, $\mathfrak{so}(2,2)$ when $\Lambda<0$, and $\mathfrak{so}(3,1)$ when $\Lambda>0$.} $\mathfrak{g}_\Lambda$ depends on the sign of $\Lambda$ \cite{Cacciatori:2005wz,Giacomini:2006dr}. This CS formulation has in turn been used in many studies of holography, asymptotic symmetries, and black hole entropy in the MB model \cite{Blagojevic:2004hj,Blagojevic:2006jk,Blagojevic:2006hh,Blagojevic:2005pd,Blagojevic:2013bu,Cvetkovic:2018ati,Klemm:2007yu,Ning:2018gfm,Peleteiro:2020ubv}.

Here we take a closer look at the theory by focusing on its covariant phase space (using the eponymous formalism) and its charges. More precisely, we are interested in performing the Sugawara construction and in the study of a certain set of ``dual'' diffeomorphism charges. Together with the ``usual'' diffeomorphisms, these enable to define, at finite distance, for tangent diffeomorphisms, and without any boundary conditions, an algebra of charges consisting in two copies of the Witt algebra (below we will often call this the centreless Virasoro algebra). This algebra is defined even with a positive cosmological constant, and in the flat limit reduces to BMS$_3$. We furthermore explain why this result is actually \textit{not} attainable from CS theory. The advantage of performing this construction on the MB model is that it allows for an exploration of the four-dimensional space of parameters, in which one finds for example the first order Hilbert--Palatini Lagrangian for three-dimensional gravity with cosmological constant, but also Witten's ``exotic'' Lagrangian \cite{MR974271}. We will show in fact that different dual charges can be defined in different sectors of the MB model, in particular the ones with $(p\neq0,q=0)$ and $(p=0,q\neq0)$, thereby even defining a curvature/torsion duality between dual charges. The dual charges agree when $(p=0,q=0)$, in which case their algebra with the usual diffeomorphisms is that of BMS$_3$.

Being diffeomorphism-invariant and topological, the MB theory has the property that its diffeomorphisms can be written on-shell as field-dependent gauge transformations. There are two types of such internal gauge transformations, namely the Lorentz transformations, and the ``translations''\footnote{The name comes from the fact that in the flat case the corresponding generators commute.}. We start this work by actually identifying these translations. Their algebra with Lorentz transformations has three central extensions which depend on the parameters of the theory. Although it is isomorphic to $\mathfrak{g}_\Lambda$, this algebra does actually not have the familiar ``canonical form'' unless one performs a redefinition of the generators. While this is of course always allowed, since it amounts to redefining what we mean by the translations to define ``translations prime'', we explain why it is not natural. Working with the initial translations, we are therefore led to introduce a different CS formulation of MB gravity than the $\mathfrak{g}_\Lambda$ one introduced previously in \cite{Cacciatori:2005wz,Giacomini:2006dr}. Besides this elementary remark and exercise in change of generators, the interest in the translations lies in their ability to provide a simple expression for the Sugawara construction.

Setting aside aspects of conformal field theory, in gravitational terms the Sugawara construction amounts to defining diffeomorphisms as quadratic generators starting from the internal gauge transformations. This reflects the above-mentioned relationship which exists for topological theories and which states that, on-shell, diffeomorphisms are nothing but field-dependent gauge transformations. In algebraic terms, this amounts to starting from the $\mathfrak{g}_\Lambda$ current algebra, and constructing a Virasoro algebra from quadratics in the currents. These quadratics are of course related to the Casimirs, while considering higher order operators leads to $W$-algebras \cite{artamonov2016introduction}. In this work we write down the Sugawara construction leading to the diffeomorphisms of the MB model. Since $\mathfrak{g}_\Lambda$ has two quadratic Casimirs, it is then natural to ask whether there exists another, dual Sugawara construction, and which gauge generator it corresponds to. Here we show that one is naturally led to this discussion when thinking about the relationship between diffeomorphisms and internal gauge transformations. Since this involves \textit{field-dependent} gauge transformations, one should be careful about issues of integrability in the covariant phase space. As we will explain, analyzing the condition of integrability for quadratic generators built from field-dependent Lorentz transformations and translations automatically leads to the identification of the above-mentioned dual diffeomorphism charges.

In the familiar constructions of the BMS$_3$ or double Virasoro asymptotic symmetries, the diffeomorphism charges depend on two directions of the vector field (e.g. $u$ and $\varphi$ in Bondi coordinates). This is, in a sense, the reason for the presence of two factors in the associated asymptotic symmetry algebras. Moreover, it is the non-tangentiality of such diffeomorphisms (i.e. the fact that they probe directions other than $\varphi$) which is responsible for the appearance of a classical central extension in the symmetry algebras. For the sake of completeness we recall here these known facts in the case of the MB model. In light of this, one can then understand better our new result, which is the construction at finite distance and from tangent vector fields of a centreless double Virasoro algebra, which can further be reduced to BMS$_3$ in the flat case. More precisely, the reason for which we obtain centreless algebras is because we focus on tangential vector fields. However, in spite of these vector fields being only along the boundary circle, we still obtain algebras with two factors because of the presence of the dual diffeomorphism charges. Let us stress that these dual charges are well-defined already in the usual first order formulation of three-dimensional gravity. Here we simply study the MB model in order to have, once and for all, an understanding of the most general Lagrangian and of the curvature/torsion duality which it encodes.

The interest in studying these dual charges and their algebra with the usual diffeomorphisms, the Lorentz transformations, and the translations, comes from the following question: What are the boundary symmetries of a given formulation of gravity? If, as argued in \cite{Freidel:2020xyx,Freidel:2020svx,Freidel:2020ayo}, quantum gravity arises from a quantization of the quasi-local symmetries of gravity, it is desirable to unravel and understand the nature of the largest possible symmetry algebra. In a first order theory like the MB model (or even BF theory), this seemingly simple question has no clear answer. Indeed, even though they are equivalent as symmetries of the Lagrangian, it is not clear at the level of the charge algebra whether one should consider the diffeomorphisms as independent from the Lorentz transformations and the translations. One the one hand, we have explained above and will show in detail that the Lorentz transformations and the translations parametrize the independent gauge symmetries. In the presence of boundaries they give rise to a centrally-extended $\mathfrak{g}_\Lambda$ current algebra of charges\footnote{The $\mathfrak{g}_\Lambda$ current algebra is also the asymptotic algebra which appears with the most general AdS$_3$ and flat space boundary conditions of \cite{Grumiller:2016pqb,Grumiller:2017sjh}.}. It is these symmetries which play a physical role e.g. in discussions of surface states and entanglement entropy in condensed matter\footnote{There is no doubt that all these results can also be reinterpreted in terms of diffeomorphisms, although to our knowledge this has not been presented anywhere.} \cite{Chen_2016,Chen_2017,Wen:2016snr,Delcamp:2016eya}. On the other hand, we are also encouraged by intuition from metric gravity to work with the diffeomorphism transformations and their charges, and at infinity they indeed carry important physical information such as the energy and the angular momentum. It is clear however that diffeomorphisms are in a sense not fundamental at finite distance. First, they are not always integrable, while the Lorentz transformations and translations always are. Second, since they are only (particular) quadratics built from the Lorentz and translational charges, there is no reason not to consider other possible quadratics, and even higher order operators. In summary, in this work we take a first step towards extending the quasi-local charge algebra of three-dimensional gravity. We do so by studying carefully the relationship between diffeomorphisms and internal gauge transformations, and show that this naturally leads to the introduction of dual diffeomorphism charges.

After having studied the finite distance covariant phase space of the MB model and introduced the dual charges, we conclude this work with a study of the metric and teleparallel formulations of the theory, both at the level of the equations of motion and of the Lagrangian itself. To start with, one can combine the first order equations of motion of the MB theory into a single second order equation. This is done by decomposing the connection into a reference connection and a contorsion tensor, and then solving for the contorsion. When doing so, one can choose the reference connection to be either torsionless or flat. Choosing the reference connection to be torsionless, the combined first order equations of motion reduce to the usual second order Riemannian Einstein equations $R_{\mu\nu}=2\Lambda g_{\mu\nu}$ mentioned above. Choosing instead the reference connection to be flat, one obtains the equations of motion of teleparallel gravity \cite{Aldrovandi:2013wha}. Interestingly, the metric and teleparallel formulations of gravity can also be understood as a form of duality \cite{Delcamp:2018sef}, which corresponds to choosing to encode the dynamics either in the curvature or in the torsion of the geometry.

Going further, one can then inject the solution to the torsion equation of motion back into the Lagrangian. In the case of usual first order gravity with only the Hilbert--Palatini term, this leads to the second order metric Lagrangian or to the teleparallel Lagrangian, depending on whether the reference connection is chosen to be torsionless or flat \cite{Dupuis:2019unm}. In the case of the MB model however, the presence of a CS term in the Lagrangian has dramatic consequences for the theory when solving and injecting the torsion equation of motion. Indeed, when the reference connection is chosen to be the torsionless Levi--Civita connection, the CS term for this latter becomes third order in derivatives of the metric. This is responsible for the appearance of a massive propagating mode, i.e. a massive graviton, and the corresponding Lagrangian is known as that of topologically massive gravity (TMG) \cite{Deser:1981wh,Deser:1982vy,Baekler:1992ab}. We complete this picture by deriving the analogue result for the teleparallel sector of the theory, and show that when choosing a flat connection the on-shell Lagrangian has a CS term for the Weitzenb\"ock torsion. It is then natural to conjecture that the so-obtained Lagrangian is that of a teleparallel version of topologically massive gravity, which we call TTMG. We postpone its detailed study to future work.

This paper is organized as follows. In section \ref{sec:2} we present the Lagrangian of the MB model and study its symmetries. This enables us to identify the translations as the transformations which, when combined with Lorentz transformations, reproduce on-shell the diffeomorphisms. In section \ref{sec:3} we then study the covariant phase space and the algebra of charges. We also present the Sugawara construction and relate it to diffeomorphisms and their dual. We explain how different dual charges can be identified on subspaces of the MB model, and show how they lead at finite distance to a double Virasoro algebra with BMS$_3$ as the flat limit. Section \ref{sec:4} is then devoted to the study of the CS formulation of the MB model, using the algebra of Lorentz transformations and translations. We then present a general discussion on the integrability of quadratic charges, and a set of dual charges which can be considered for different values of the couplings of the MB model. This explains how the dual charges can be defined for non-tangent diffeomorphisms as well. In section \ref{sec:5} we recall the study of asymptotic symmetries in the MB model using Bondi gauge. Section \ref{sec:6} contains the study of the second order metric and teleparallel sectors of the theory, as well as their massive generalizations obtained when injecting the solution of the torsion equation back into the Lagrangian. This section is independent from the previous ones, and can be read separately for the interested reader. We finally present our conclusions and discuss many directions for future work in section \ref{sec:7}.

In addition to the new results, this article covers very general facts about three-dimensional gravity, and fills-in some gaps about elementary aspects which are often not discussed in the literature. It can therefore be considered also as a pedagogical introduction to the covariant phase space in three-dimensional gravity. In order to make this self-contained, we have recalled details about our notations and conventions in appendix \ref{app:A}, and details about covariant phase space methods in appendix \ref{app:B}. Appendix \ref{app:C} contains detailed proofs of mosts calculations.

\section{Mielke--Baekler Lagrangian}
\label{sec:2}

The Mielke--Baekler (MB) Lagrangian is the most general Lorentz-invariant three-form constructed with a triad one-form $e$ and a connection one-form $\omega$ \cite{Baekler:1992ab,Mielke:1991nn}. It is defined as the sum of the volume, Hilbert--Palatini, Chern--Simons and torsion Lagrangians, each with their own coupling $\sigma_i$, and takes the simple form
\be\label{MB Lagrangian}
L_\text{MB}[e,\omega]
&=\sigma_0L_\text{V}[e]+\sigma_1L_{\text{HP}}[e,\omega]+\sigma_2L_\text{CS}[\omega]+\sigma_3L_\text{T}[e,\omega]\cr
&=\f{\sigma_0}{3}e\wedge[e\wedge e]+2\sigma_1e\wedge F+\sigma_2\omega\wedge\left(\de\omega+\f{1}{3}[\omega\wedge\omega]\right)+\sigma_3e\wedge\de_\omega e.
\ee
Here and throughout this article we work in Lorentzian signature, and use an index-free differential form notation recalled in appendix \ref{app:A}.

\subsection{Equations of motion}

The variation of the MB Lagrangian is
\be\label{MB variation}
\delta L_\text{MB}[e,\omega]
&=\de\Big(2\sigma_1\delta\omega\wedge e+\sigma_2\delta\omega\wedge\omega+\sigma_3\delta e\wedge e\Big)\cr
&\pe+\delta e\wedge\Big(2\sigma_1F+2\sigma_3\de_\omega e+\sigma_0[e\wedge e]\Big)\cr
&\pe+\delta\omega\wedge\Big(2\sigma_2F+2\sigma_1\de_\omega e+\sigma_3[e\wedge e]\Big).
\ee
This reveals the symplectic potential, to which we will come back shortly, as well as the equations of motion. One can see that both equations of motion contain all possible two-forms in a ``symmetrized'' manner, and that likewise the symplectic potential contains all possible polarizations. This evidently results from the presence of all four terms/couplings in the MB Lagrangian.

To understand the physical meaning of the first order equations of motion, it is useful to rewrite them in the form
\be\label{EOMs}
2F+p[e\wedge e]\approx0,
\q\q
2\de_\omega e+q[e\wedge e]\approx0,
\ee
with
\be
p\coloneqq\f{\sigma_0\sigma_1-\sigma_3^2}{\sigma_1^2-\sigma_2\sigma_3},
\q\q
q\coloneqq\f{\sigma_1\sigma_3-\sigma_0\sigma_2}{\sigma_1^2-\sigma_2\sigma_3}.
\ee
This shows that the theory has a source of Lorentzian curvature and torsion measured respectively by the parameters $p$ and $q$. We use here the adjective ``Lorentzian'' because these are properties of the first order connection variable $\omega$. Later on we will talk about Riemannian curvature and torsion when going to second order variable by decomposing $\omega$ into a reference connection and a contorsion tensor\footnote{In particular, as we will see below, the theory can have vanishing Lorentzian curvature but non-vanishing Riemannian curvature. It is this latter which matters when talking about the metric geometry of spacetime.}. Note that we require of course that $\sigma_1^2-\sigma_2\sigma_3\neq0$ otherwise the theory is trivial (or more precisely admits only degenerate triads as solutions). Two useful relations which we will use in many places below are $\sigma_0=p\sigma_1+q\sigma_3$ and $\sigma_3=p\sigma_2+q\sigma_1$.

It is now enlightening to combine the first order equations of motion into a single second order equation. For this, one can decompose the Lorentz connection as $\omega=\Gamma+k$, where $\Gamma$ is the torsionless Levi--Civita connection and $k$ is the contorsion. With this decomposition, the curvature and the torsion become
\be\label{decomposed MB EOM}
F=R+\de_\Gamma k+\f{1}{2}[k\wedge k],
\q\q
\de_\omega e=[k\wedge e],
\ee
where $R$ is the Riemannian curvature of $\Gamma$. Using the torsion equation of motion \eqref{EOMs}, this implies that $k\approx-qe/2$, and the curvature equation of motion then leads to
\be\label{metric Ricci tensor solution}
2R\approx\Lambda[e\wedge e],
\q\q
\Lambda\coloneqq-\left(p+\f{q^2}{4}\right).
\ee
In metric terms this means that $R_{\mu\nu}\approx2\Lambda g_{\mu\nu}$, i.e. that the MB model describes constant curvature spacetimes with cosmological constant $\Lambda$. This is the result which has led authors to conclude that the MB model can be written as a CS theory with gauge algebra $\mathfrak{g}_\Lambda$ \cite{Cacciatori:2005wz,Giacomini:2006dr}. We will make this statement more precise in section \ref{sec:4}, and see that it actually contains subtleties.

Note that here we have obtained a metric second order equation of motion because we have used the torsionless connection $\Gamma$ as the reference connection when decomposing $\omega$. As mentioned in the introduction, one can also consider a flat reference connection instead, and this choice leads to teleparallel equations of motion. We come back to this and analyse in more details the metric and teleparallel formulations of the theory in section \ref{sec:6}.

Finally, let us observe that there are two particular cases for specific values of the couplings. For $\sigma_1\sigma_3=\sigma_0\sigma_2$ we get $q=0$, and the MB Lagrangian reduces to Witten's ``exotic'' Lagrangian \cite{MR974271}. For $\sigma_0\sigma_1>0$ and $\sigma_3=\pm\sqrt{\sigma_0\sigma_1}$, which implies $p=0$, we get a first order Lagrangian with Lorentzian torsion but vanishing Lorentzian curvature. Notice that in this latter case we necessarily have $\Lambda<0$, and therefore an AdS Riemannian geometry.

\subsection{Symmetries}

We now turn to the study of the gauge symmetries of the theory. First, we have the internal Lorentz transformations, which act infinitesimally as
\be\label{infinitesimal Lorentz}
\delta^\text{j}_\alpha e=[e,\alpha],
\q\q
\delta^\text{j}_\alpha\omega=\de_\omega\alpha,
\q\q
\delta^\text{j}_\alpha F=[F,\alpha],
\q\q
\delta^\text{j}_\alpha\de_\omega e=[\de_\omega e,\alpha].
\ee
Then, we have the translations acting on the triad and on the connection as
\be\label{infinitesimal translations}
\delta^\text{t}_\phi e=\de_\omega\phi+q[e,\phi],
\q\q
\delta^\text{t}_\phi\omega=p[e,\phi],
\ee
and on the curvature and torsion as
\be
\delta^\text{t}_\phi F=p\de_\omega[e,\phi],
\q\q
\delta^\text{t}_\phi\de_\omega e=\f{1}{2}\big[2F+p[e\wedge e],\phi\big]+q\de_\omega[e,\phi].
\ee
Using \eqref{MB variation} one can indeed check that these are symmetries as they leave the Lagrangian invariant up to the boundary terms $\delta^\text{j}_\alpha L_\text{MB}=\sigma_2\de(\alpha\de\omega)$ and $\delta^\text{t}_\phi L_\text{MB}=\de\big(2\sigma_1\phi F+\sigma_3\phi\de_\omega e+p[e,\phi]\wedge(\sigma_1e+\sigma_2\omega)\big)$.

While the Lorentz transformations have their usual form, one can observe at this point that there is a priori nothing canonical about the transformations which we have called translations. Indeed, since both parameters $\alpha$ and $\phi$ are Lie algebra-valued scalars, any combination of \eqref{infinitesimal Lorentz} and \eqref{infinitesimal translations} is also a legitimate independent gauge symmetry which we could have identified as the translation. Part of the reason for which \eqref{infinitesimal translations} is actually the natural form for the translations comes from the expression for the diffeomorphisms as field-dependent gauge transformations, which will then be the starting point for the Sugawara construction. Indeed, for diffeomorphisms, which act as the Lie derivative $\delta^\text{d}_\xi=\pounds_\xi$, a simple rewriting reveals that
\bsub\label{diffeos as on-shell gauge}
\be
\delta^\text{d}_\xi e
&=\de(\xi\ip e)+\xi\ip(\de e)\cr
&=\de_\omega(\xi\ip e)+\xi\ip(\de_\omega e)+[e,\xi\ip\omega]\cr
&=\delta^\text{j}_{\xi\ip\omega}e+\delta^\text{t}_{\xi\ip e}e+\f{1}{2}\xi\ip\big(2\de_\omega e+q[e\wedge e]\big),\\
\delta^\text{d}_\xi\omega
&=\de(\xi\ip\omega)+\xi\ip(\de\omega)\cr
&=\de_\omega(\xi\ip\omega)+\xi\ip F\cr
&=\delta^\text{j}_{\xi\ip\omega}\omega+\delta^\text{t}_{\xi\ip e}\omega+\f{1}{2}\xi\ip\big(2F+p[e\wedge e]\big),
\ee
\esub
where one can recognize the equations of motion \eqref{EOMs}. As we will see in the next section, yet another reason for considering \eqref{infinitesimal Lorentz} and \eqref{infinitesimal translations} is that their Hamiltonian generators are, as expected, the spatial pullbacks of the equations of motion enforced respectively by $\omega$ and $e$ in \eqref{MB variation}. Therefore, using the Lorentz transformations and the translations to parametrize the set of gauge transformations is the only choice which exhibits the two symmetrical forms \eqref{MB variation} and \eqref{EOMs} of the equations of motion.

By definition of the Lie derivative, we have that $\delta^\text{d}_\xi L_\text{MB}=\de(\xi\ip L_\text{MB})$. Consistently, one can check that the same result is also obtained when acting with the right-hand side of \eqref{diffeos as on-shell gauge}. This uses the fact that the equations of motion appearing there act as so-called trivial gauge transformations \cite{Banerjee:2012jn}. We note that the diffeomorphisms as field-dependent gauge transformations in the MB model were also discussed in \cite{Banerjee:2009vf,Banerjee:2011cu,Banerjee:2012jn}, although in a much more convoluted way and without explicitly identifying the translations.

Now, by acting twice on the dynamical fields with the translations and Lorentz transformations, one can derive the commutators
\be
\big[\delta^\text{j}_\alpha,\delta^\text{t}_\phi\big]=\delta^\text{t}_{[\alpha,\phi]},
\q\q
\big[\delta^\text{j}_\alpha,\delta^\text{j}_\beta\big]=\delta^\text{j}_{[\alpha,\beta]},
\q\q
\big[\delta^\text{t}_\phi,\delta^\text{t}_\chi\big]=p\delta^\text{j}_{[\phi,\chi]}+q\delta^\text{t}_{[\phi,\chi]}.
\ee
This motivates the introduction of generators $(J_i,T_i)_{i=1,2,3}$ forming the algebra
\be\label{JT algebra}
[J_i,T_j]={\eps_{ij}}^kT_k,
\q\q
[J_i,J_j]={\eps_{ij}}^kJ_k,
\q\q
[T_i,T_j]={\eps_{ij}}^k(pJ_k+qT_k),
\ee
whose quadratic Casimir operators are
\be\label{JT Casimirs}
C_1=pJ^iJ_i+T^iT_i,
\q\q
C_2=J^iT_i+T^iJ_i-qJ^iJ_i.
\ee
We will come back to these Casimirs when studying the Sugawara construction in section \ref{sec:Sugawara}. Defining the new generators
\be\label{global P definition}
P\coloneqq T-\f{q}{2}J
\ee
then maps this algebra to
\be\label{JP algebra}
[J_i,P_j]={\eps_{ij}}^kP_k,
\q\q
[J_i,J_j]={\eps_{ij}}^kJ_k,
\q\q
[P_i,P_j]=-\Lambda{\eps_{ij}}^kJ_k,
\ee
which are the usual $\mathfrak{g}_\Lambda$ commutation relations used in the CS formulation of three-dimensional gravity. The construction of the MB model based on the algebras spanned by $(J,T)$ and $(J,P)$ will be discussed in section \ref{sec:4}, where we will see once again why the $(J,T)$ parametrization of the algebra is more natural.

The commutation relations \eqref{JT algebra} only capture the global part of the algebra of gauge symmetries, whose fine structure is revealed upon inspection of the charges in e.g. the covariant phase space formalism.

\section{Covariant phase space and charge algebra}
\label{sec:3}

The study of the charges and their algebra is best carried out in the covariant phase space formalism, whose ingredients are recalled in appendix \ref{app:B}. The starting point is the symplectic potential, which from \eqref{MB variation} is given by
\be\label{potential}
\theta=2\sigma_1\delta\omega\wedge e+\sigma_2\delta\omega\wedge\omega+\sigma_3\delta e\wedge e.
\ee
This in turn determines the symplectic structure
\be\label{symplectic structure}
\Omega=-\int_\Sigma2\sigma_1\delta\omega\wedge\delta e+\sigma_2\delta\omega\wedge\delta\omega+\sigma_3\delta e\wedge\delta e.
\ee
We are now ready to contract this symplectic structure with the various gauge transformations introduced above. This will identify the Hamiltonian generators as well as the boundary charges. We will then compute the algebra of these charges, and present the Sugawara construction which relates diffeomorphism charges to internal gauge transformations. Along the way we will introduce the dual diffeomorphism charges.

\subsection{Generators and charges}

Let us consider initially that the parameters $(\alpha,\phi,\xi)$ are field-independent. Contracting the gauge transformations with the symplectic structure gives
\be
-\delta^\text{j}_\alpha\ipp\Omega=\delta\J(\alpha),
\q\q
-\delta^\text{t}_\phi\ipp\Omega=\delta\T(\phi),
\q\q
-\delta^\text{d}_\xi\ipp\Omega=\slashed{\delta}\D(\xi),
\ee
where the notation $\slashed{\delta}$ means that, for diffeomorphisms, the field-space 1-form obtained with the contraction is not necessarily integrable, i.e. not necessarily a total variation as we will see below. By construction of the covariant phase space, the generators on the right-hand side of these expressions are the sum of a bulk piece vanishing on-shell and a boundary piece called the charge. Let us now identify these two pieces for the various gauge transformations.

For Lorentz transformations, after computing the contraction with the symplectic structure we find that the Hamiltonian generator is given by
\be\label{Lorentz generator}
\J(\alpha)
&=\int_\Sigma2\de_\omega\alpha\wedge(\sigma_1e+\sigma_2\omega)+\sigma_2\alpha[\omega\wedge\omega]-\sigma_3\alpha[e\wedge e]\cr
&=-\int_\Sigma\alpha\big(2\sigma_2F+2\sigma_1\de_\omega e+\sigma_3[e\wedge e]\big)+2\oint_S\alpha(\sigma_1e+\sigma_2\omega).
\ee
For the translations we find that the Hamiltonian generator is
\be\label{translation generator}
\T(\phi)
&=\int_\Sigma2\de_\omega\phi\wedge(\sigma_1\omega+\sigma_3e)+\sigma_1\phi[\omega\wedge\omega]-\sigma_0\phi[e\wedge e]\cr
&=-\int_\Sigma\phi\big(2\sigma_1F+2\sigma_3\de_\omega e+\sigma_0[e\wedge e]\big)+2\oint_S\phi(\sigma_1\omega+\sigma_3e),
\ee
where we have used the fact that $\sigma_0=p\sigma_1+q\sigma_3$ and $\sigma_3=p\sigma_2+q\sigma_1$. As usual, we have written here the generators in their manifestly differentiable form, which is the first line of each expression. We have then integrated by parts to reveal the constraints (which are the spatial pullbacks of the equations of motion), and therefore the boundary charges which are left when going on-shell.

For the diffeomorphisms, we write the Lie derivative as the second line of \eqref{diffeos as on-shell gauge} for each variable, and find
\be\label{full diffeo charge}
\slashed{\delta}\D(\xi)
&=2\int_\Sigma\Big(\sigma_1\big(\de_\omega(\xi\ip\omega)+\xi\ip F\big)+\sigma_3\big(\de_\omega(\xi\ip e)+\xi\ip(\de_\omega e)+[e,\xi\ip\omega]\big)\Big)\wedge\delta e\cr
&\phantom{=\int}+\Big(\sigma_2\big(\de_\omega(\xi\ip\omega)+\xi\ip F\big)+\sigma_1\big(\de_\omega(\xi\ip e)+\xi\ip(\de_\omega e)+[e,\xi\ip\omega]\big)\Big)\wedge\delta\omega\cr
&=\int_\Sigma\xi\ip\Big(2\delta e\wedge(\sigma_1F+\sigma_3\de_\omega e)+\delta\omega\wedge\big(2\sigma_2F+2\sigma_1\de_\omega e+\sigma_3[e\wedge e]\big)\Big)\cr
&\phantom{=\int}-\delta\Big(2(\xi\ip e)(\sigma_1F+\sigma_3\de_\omega e)+(\xi\ip\omega)\big(2\sigma_2F+2\sigma_1\de_\omega e+\sigma_3[e\wedge e]\big)\Big)\cr
&\phantom{=\int}+\oint_S\delta\big(2\sigma_1(\xi\ip\omega)e+\sigma_2(\xi\ip\omega)\omega+\sigma_3(\xi\ip e)e\big)-\xi\ip\theta.
\ee
We see that half of the equations of motion appear in the bulk (second and fourth terms), while for the other half the term in $\sigma_0$ is missing. However this is consistent because of the identity $\xi\ip\big(\delta e\wedge[e\wedge e]\big)=\delta\big((\xi\ip e)[e\wedge e]\big)$, which means that the corresponding term can indeed be re-introduced. As usual, we see that generic diffeomorphisms are not integrable. The non-integrable piece in the bulk vanishes on-shell, or off-shell for $\xi$ tangent to $\Sigma$, while for the boundary charge we need $\xi$ to be tangent to $S$. If $\xi$ is tangent to both $\Sigma$ and $S$ we have that
\be
\D(\xi)
=\int_\Sigma2\sigma_1\pounds_\xi\omega\wedge e+\sigma_2\pounds_\xi\omega\wedge\omega+\sigma_3\pounds_\xi e\wedge e.
\ee
Let us now go back to the case of a generic vector field $\xi$ and study more closely the non-integrability of the diffeomorphisms. From now on we will focus on the boundary charges. For convenience, we will use a slight abuse of notation and write all the generators on-shell, so that they coincide with their boundary charge and we can discard the bulk once and for all.

We have seen earlier that there is a relationship, given by \eqref{diffeos as on-shell gauge}, between the diffeomorphisms and the internal gauge transformations. Since this involves field-dependent gauge transformations, which are therefore generically non-integrable, we should go back to the notation $\slashed{\delta}$ which keeps track of this fact. With this the Lorentz and translation charges are
\be
\slashed{\delta}\J(\alpha)\approx2\oint_S\alpha\delta(\sigma_1e+\sigma_2\omega),
\q\q
\slashed{\delta}\T(\phi)\approx2\oint_S\phi\delta(\sigma_1\omega+\sigma_3e),
\ee
and the diffeomorphism charge is
\be
\slashed{\delta}\D(\xi)\approx\oint_S\delta\big(2\sigma_1(\xi\ip\omega)e+\sigma_2(\xi\ip\omega)\omega+\sigma_3(\xi\ip e)e\big)-\xi\ip\theta.
\ee
Using the fact that $2(\xi\ip\omega)\delta\omega=\delta\big((\xi\ip\omega)\omega\big)-\xi\ip(\delta\omega\wedge\omega)$, and similarly for $e$, it is then straightforward to see that we have
\be\label{diffeo field-dependent}
\slashed{\delta}\D(\xi)=\slashed{\delta}\J(\xi\ip\omega)+\slashed{\delta}\T(\xi\ip e),
\ee
which reflects \eqref{diffeos as on-shell gauge}. This shows, as expected, how the non-integrability of the diffeomorphism charge comes from the non-integrable contributions of the field field-dependent translations and Lorentz transformations.

A natural question to ask is whether the integrability of $\slashed{\delta}\D(\xi)$ implies separately the integrability of $\slashed{\delta}\J(\xi\ip\omega)$ and $\slashed{\delta}\T(\xi\ip e)$. In the general case this statement is evidently not true, as for example taking $\xi$ tangent to $S$ renders $\slashed{\delta}\D(\xi)$ integrable but neither $\slashed{\delta}\J(\xi\ip\omega)$ nor $\slashed{\delta}\T(\xi\ip e)$. We will come back to this question in section \ref{sec:4} in the case of more generic boundary conditions for non-tangent vector fields at finite distance.

\subsection{Dual diffeomorphisms}
\label{sec:dual diffeo C}

Let us now jump a little bit ahead of ourselves, and introduce a new charge also constructed from field-dependent Lorentz transformations and translations, but with a different field-dependency as that of diffeomorphisms. We will come back to the general idea behind the identification of this charge (and its teleparallel dual) in section \ref{sec:dual charges}. For the time being, we are free to define it, to analyse its integrability, and to consider it along with the other charges introduced so far on the covariant phase space.

More precisely, the object which we want to introduce is the \textit{dual diffeomorphism charge}. For this, in full generality and without any boundary conditions, let us consider the quantity
\be\label{definition C}
\slashed{\delta}\C(\xi)
&\coloneqq p\,\slashed{\delta}\J(\xi\ip e)+\slashed{\delta}\T(\xi\ip\omega)\cr
&\phantom{:}=\oint_S\delta\big(2p\sigma_2(\xi\ip\omega)e+\sigma_1(\xi\ip\omega)\omega+p\sigma_1(\xi\ip e)e\big)-\xi\ip\big(2p\sigma_2\delta\omega\wedge e+\sigma_1\delta\omega\wedge\omega+p\sigma_1\delta e\wedge e\big)\cr
&\phantom{\coloneqq\oint_S}+2(\sigma_3-p\sigma_2)(\xi\ip\omega)\delta e.
\ee
Several important things should be noticed about this definition.
\begin{enumerate}
\item We have introduced this charge by analogy with the diffeomorphism charge \eqref{diffeo field-dependent}, but swapped the fields used in the field-dependent gauge transformations. This is a first justification for the name ``dual diffeomorphism'', which will be further clarified below.
\item It is important to notice that this charge is for the moment non-integrable. However, we see that integrability is achieved without restricting the dynamical fields when $\xi$ is tangent to $S$ and when $\sigma_3=p\sigma_2$ (which means at this stage that either $q=0$ or $\sigma_1=0$). Notice also that for the familiar Hilbert--Palatini Lagrangian, obtained for $\sigma_2=\sigma_3=0$, this new charge does indeed exist.
\item To define the dual charge we have combined the field-dependent gauge transformations with a relative factor of $p$, whose role will be clear when computing the algebra below. However, it is important to point out that, instead of the definition \eqref{definition C}, we could have defined the new charge as $\slashed{\delta}\C(\xi)=\sigma_3\slashed{\delta}\J(\xi\ip e)+\sigma_2\slashed{\delta}\T(\xi\ip\omega)$. It is easy to see that this quantity is \textit{always} integrable, without the need to constrain the couplings $\sigma_i$ of the MB model. This definition however has the disadvantage that it makes not obvious the $\sigma_2=\sigma_3=0$ limit giving the usual Hilbert--Palatini Lagrangian, since it gives the false impression that in this case the dual charge does not exist. Moreover, we will see in a moment and explain in appendix \ref{app: C brackets} that, no mater which coefficients are taken to define $\C$, consistency of its algebra with the gauge charges requires to take $q=0$, which in turn implies $\sigma_3=p\sigma_2$ anyways. This is why it is coherent to simply start with the definition \eqref{definition C}. We should keep this remark in mind because it will also apply to the other dual charge introduced below in section \ref{sec: tele dual charge}.
\end{enumerate}

Now, recall that for diffeomorphisms we had the relations \eqref{diffeos as on-shell gauge} which were reflected in \eqref{diffeo field-dependent}. It is then natural, given the definition of $\C$, to look at a transformation of the fields of the form
\bsub\label{dual diffeos}
\be
\delta^\text{c}_\xi e
&=p\delta^\text{j}_{\xi\ip e}e+\delta^\text{t}_{\xi\ip\omega}e\cr
&=p[e,\xi\ip e]+\de_\omega(\xi\ip\omega)+q[e,\xi\ip\omega]\cr
&=\pounds_\xi\omega-\xi\ip F+p[e,\xi\ip e]+q[e,\xi\ip\omega]\cr
&\approx\pounds_\xi\omega+q[e,\xi\ip\omega],\\
\delta^\text{c}_\xi\omega
&=p\delta^\text{j}_{\xi\ip e}\omega+\delta^\text{t}_{\xi\ip\omega}\omega\cr
&=p\de_\omega(\xi\ip e)+p[e,\xi\ip\omega]\cr
&=p\pounds_\xi e-p\xi\ip(\de_\omega e)\cr
&\approx p\pounds_\xi e-pq[e,\xi\ip e],
\ee
\esub
where in both cases for the last equality we have gone on-shell. We see that, up to terms proportional to $q$, the object $\C$ which we have introduced is the charge of a transformation sending $e$ to the Lie derivative of $\omega$, and vice-versa. This is yet another reason behind the name ``dual diffeomorphism''. To understand what is special about the value $q=0$, for which \eqref{dual diffeos} really reduces to a Lie derivative, we have to study the algebra of the charges.

\subsection{Charge algebra}

Let us now go back to the general case of arbitrary couplings $\sigma_i$. Given the Hamiltonian generator $\delta\H(\alpha)=-\delta_\alpha\ipp\Omega$ of an integrable transformation $\delta_\alpha$, the Poisson bracket between two generators is given by  $\lb\H(\alpha),\H(\beta)\rb=\delta_\alpha\H(\beta)=\delta_\alpha\ipp\big(\delta\H(\beta)\big)=-\delta_\alpha\ipp\delta_\beta\ipp\Omega$. Focusing first on the Lorentz transformations and the translations we find\footnote{Notice that the bracket of translations can be rewritten as
$\lb\T(\phi),\T(\chi)\rb=p\lb\J(\phi),\J(\chi)\rb+q\lb\J(\phi),\T(\chi)\rb$.}

\bsub\label{JT extended algebra}
\be
\lb\J(\alpha),\T(\phi)\rb&=\T([\alpha,\phi])-2\sigma_1\oint_S\alpha\de\phi,\\
\lb\J(\alpha),\J(\beta)\rb&=\J([\alpha,\beta])-2\sigma_2\oint_S\alpha\de\beta,\\
\lb\T(\phi),\T(\chi)\rb&=p\J([\phi,\chi])+q\T([\phi,\chi])-2\sigma_3\oint_S\phi\de\chi.
\ee
\esub
These Poisson brackets form as expected an ultra-local extension of the algebra \eqref{JT algebra}, and contain three classical central extensions. The central charges are simply given by three of the couplings of the MB Lagrangian, which is also a consequence of having chosen \eqref{infinitesimal translations} as the translations. Defining $\K_i\coloneqq\J+k_i\T$ for $i=1,2$ and computing the new brackets, one can easily see that it is always possible to kill any two central extensions with an appropriate choice of $k_i$'s.

Following \eqref{global P definition}, we can also define the new generator
\be
\P(\alpha)\coloneqq\T(\alpha)-\f{q}{2}\J(\alpha),
\ee
which maps the algebra to
\bsub\label{JP extended algebra}
\be
\lb\J(\alpha),\P(\phi)\rb&=\P([\alpha,\phi])+(q\sigma_2-2\sigma_1)\oint_S\alpha\de\phi,\\
\lb\J(\alpha),\J(\beta)\rb&=\J([\alpha,\beta])-2\sigma_2\oint_S\alpha\de\beta,\\
\lb\P(\phi),\P(\chi)\rb&=-\Lambda\left(\J([\phi,\chi])-2\sigma_2\oint_S\phi\de\chi\right).
\ee
\esub
As expected we recognize here the $\mathfrak{g}_\Lambda$ current algebra, whose global part is \eqref{JP algebra}, and find once again three central extensions. In this parametrization of the algebra however, the central charges are more complicated functions of the initial couplings $\sigma_i$ of the Lagrangian. We will see below that these central charges are actually (up to signs) the ones appearing in the asymptotic symmetry algebra \eqref{Bondi symmetry algebra} of AdS spacetimes in Bondi gauge. It is intriguing that these asymptotic central charges, which depend on boundary conditions, can already be read off from the algebra \eqref{JP extended algebra} which is independent of boundary conditions (and is not even an algebra of diffeomorphisms).

We can now complete the algebra of Lorentz transformations and translations by including the diffeomorphisms. However, since generic diffeomorphisms are not integrable without specifying boundary conditions, we focus as explained above on vector fields $\xi$ tangent to $S$. We then find the usual brackets
\bsub\label{DJT algebra}
\be
\lb\D(\xi),\J(\alpha)\rb&=-\J(\pounds_\xi\alpha),\\
\lb\D(\xi),\T(\phi)\rb&=-\T(\pounds_\xi\phi),\\
\lb\D(\xi),\D(\zeta)\rb&=-\D([\xi,\zeta]).
\ee
\esub
The brackets \eqref{JT extended algebra} and \eqref{DJT algebra} describe the algebra of charges of the MB model, with the only assumption being that of tangentiality of the diffeomorphisms.

Finally, in the particular case $\sigma_3=p\sigma_2$, it is natural to continue the logic of extending the symmetry algebra by including as well the dual diffeomorphisms. As we are about to see, their algebra with the charges introduced so far in \eqref{JT extended algebra} and \eqref{DJT algebra} contains an important information. Considering again tangential vector fields, the new brackets are
\bsub\label{C brackets}
\be
\lb\C(\xi),\J(\alpha)\rb&=-\T(\pounds_\xi\alpha)\label{CJ bracket},\\
\lb\C(\xi),\T(\phi)\rb&=-p\J(\pounds_\xi\phi)+q\T([\xi\ip\omega,\phi]),\label{CT bracket}\\
\lb\C(\xi),\D(\zeta)\rb&=-\C([\xi,\zeta])\label{CD bracket},\\
\lb\C(\xi),\C(\zeta)\rb&=-p\D([\xi,\zeta]),\label{CC bracket}
\ee
\esub
for which we give detailed proofs in appendix \ref{app: C brackets}. In these brackets we see the appearance of a new object which requires scrutiny. Indeed, at the difference with all the other charges appearing in \eqref{C brackets} and in the brackets above, the object $\T([\xi\ip\omega,\phi])$ in \eqref{CT bracket} has a field-dependent smearing. This means that it cannot actually be obtained as the charge of a field-dependent translation, since this latter would not be integrable. However, we can reverse the logic and ask under which condition $\slashed{\delta}\T([\xi\ip\omega,\phi])$ does become integrable. Writing this charge explicitly we find
\be\label{integrability of CT}
\slashed{\delta}\T([\xi\ip\omega,\phi])=2\oint_S[\xi\ip\omega,\phi]\delta(\sigma_1\omega+\sigma_3e)=\oint_S\sigma_1\delta([\omega,\xi\ip\omega]\phi)+2\sigma_3[\xi\ip\omega,\phi]\delta e,
\ee
where we have used the tangentiality of $\xi$ to show that the first term is always integrable. We now recall that the charge $\C$, which leads to the above algebra, only exists when $\sigma_3=p\sigma_2$, and that this condition implies either that $q=0$ or $\sigma_1=0$. If $q=0$, we see that the translation term with field-dependent smearing in \eqref{CT bracket} simply drops. On the other hand, if instead we have $\sigma_1=0$, this term survives, and we have to discuss its integrability. Without imposing boundary conditions on the fields, this can only be achieved with $\sigma_3=0$. This in turn trivializes \eqref{integrability of CT}, but most importantly implies once again that $q=0$. In summary, we see that consistency of the brackets \eqref{C brackets} \textit{always} implies that $q=0$. Moreover, we explain at the end of appendix \eqref{app: C brackets} that it is always the condition $q=0$ which is singled out, regardless of which coefficients are used to define $\C$ in \eqref{definition C}. Here we have chosen the definition which leads to the simplest form of the algebra.

It is also enlightening to rewrite the above Poisson brackets in Fourier space. This will be useful in the next section to study the Sugawara construction, and it also reveals the particular role played by the condition $q=0$. With the notations and detailed proofs given in appendix \ref{appendix Fourier}, the Fourier version of the algebra \eqref{JT extended algebra}, \eqref{DJT algebra}, and \eqref{C brackets} is

\bsub\label{JTDC extended algebra Fourier}
\be
\lb\J^i_n,\T^j_m\rb&={\eps^{ij}}_k\T^k_{n+m}-2\i\sigma_1m\eta^{ij}\delta_{n+m,0},\\
\lb\J^i_n,\J^j_m\rb&={\eps^{ij}}_k\J^k_{n+m}-2\i\sigma_2m\eta^{ij}\delta_{n+m,0},\\
\lb\T^i_n,\T^j_m\rb&={\eps^{ij}}_k\big(p\J^k_{n+m}+q\T^k_{n+m}\big)-2\i\sigma_3m\eta^{ij}\delta_{n+m,0},\\
\lb\D_n,\J^i_m\rb&=\i m\J^i_{n+m},\\
\lb\D_n,\T^i_m\rb&=\i m\T^i_{n+m},\\
\lb\D_n,\D_m\rb&=\i(m-n)\D_{n+m},\label{Dn Dm}\\
\lb\C_n,\D_m\rb&=\i(m-n)\C_{n+m},\\
\lb\C_n,\C_m\rb&=\i p(m-n)\D_{n+m},\\
\lb\C_n,\J^i_m\rb&=\i m\T^i_{n+m},\\
\lb\C_n,\T^i_m\rb&=\i pm\J^i_{n+m}-\f{pq\sigma_2}{2(p\sigma_2^2-\sigma_1^2)}{\eps^i}_{jk}\sum_{k\in\mathbb{Z}}\J^j_{n+m+k}\T^k_{-k}.\label{CT Fourier bracket}
\ee
\esub
Note that $\sigma_2\sigma_3-\sigma_1^2\neq0$ is a condition which we have from the beginning, and which here becomes $p\sigma_2^2-\sigma_1^2\neq0$ since we are considering $\C$ in the algebra. Once again, we see that the bracket \eqref{CT Fourier bracket} produces an unusual term, namely a quadratic in the currents which has an open index produced with the epsilon symbol. This is the Fourier space version of the field-dependent translation in \eqref{CT bracket}. From this Fourier expression, it is easier to see that the presence of this term prevents the algebra from closing, as the iterated Poisson bracket of \eqref{CT Fourier bracket} with $\C_n$ produces higher and higher order contractions of the currents. We therefore see once again that the well-definiteness of this algebra singles out the condition $q=0$ (recall that $q=0$ in turn implies $\sigma_3=p\sigma_2$).

The remarkable result now comes from the algebra formed by the dual diffeomorphism $\C_n$ together with the usual diffeomorphism $\D_n$. In the case $q=0$, the cosmological constant becomes $\Lambda=-p$. We can then discuss two particular cases of interest. First, if $p=0$, one can see from the above Poisson brackets that the couple $(\D_n,\C_n)$ forms the algebra
\bsub\label{CD BMS}
\be
\lb\D_n,\D_m\rb&=\i(m-n)\D_{n+m},\\
\lb\C_n,\D_m\rb&=\i(m-n)\C_{n+m},\\
\lb\C_n,\C_m\rb&=0,
\ee
\esub
which we recognize as the BMS$_3$ algebra with vanishing central charge. When $p$ is non-vanishing, we can define
\be\label{Virasoro change of generators}
\D_n^\pm\coloneqq\f{1}{2}\left(\f{1}{\sqrt{p}}\C_{\pm n}\pm\D_{\pm n}\right)_{\!\!p>0},
\q\q
\D_n^\pm\coloneqq\f{1}{2}\left(\f{1}{\i\sqrt{-p}}\C_{\pm n}\pm\D_{\pm n}\right)_{\!\!p<0},
\ee
and in both cases find the algebra
\bsub\label{CD Virasoro}
\be
\lb\D_n^\pm,\D_m^\pm\rb&=\i(m-n)\D_{n+m}^\pm,\\
\lb\D_n^+,\D_m^-\rb&=0.
\ee
\esub
In the AdS case $p>0$ these are two centreless Virasoro algebras. In the dS case, i.e. when $p<0$, this is the algebra of the conformal group of the complex plane, which is the infinite-dimensional enhancement of the $\mathfrak{g}_\Lambda=\mathfrak{so}(3,1)$ isometries.

As announced, the brackets \eqref{CD BMS} and \eqref{CD Virasoro} are one of the main results of this paper, namely the derivation, at finite distance and without boundary conditions, of a centreless BMS or Witt algebra depending on the sign of the cosmological constant. This result is surprising because it is obtained for a generic boundary. It shows that the BMS and Virasoro structures are not solely properties of asymptotic boundaries, but general features of the boundary algebra of quadratic charges. This also reveals an important difference between the two setups. In the asymptotic case, the BMS or Virasoro structure (and their central charge) comes entirely from the diffeomorphism $\D_n$ alone, which has two independent components due to the non-tangentiality of the asymptotic vector field. Here however, the tangentiality of the vector field requires us to use the dual charge $\C_n$ in addition to $\D_n$ to construct the algebras, and is also responsible for the lack of central charge. It is also interesting that this finite distance construction works for any value of the cosmological constant $\Lambda=-p$, which in particular includes the less-studied dS case. Finally, we note that the algebra of $(\C_n,\D_n)$ is automatically obtained in such a way that the flat limit is well-defined. This is to be contrasted with studies of the asymptotic symmetry algebra, where the Fefferman--Graham gauge does not lead to a well-defined flat limit while the Bondi gauge does \cite{Barnich:2012aw,Barnich:2012rz}.

It is important to note that finite distance Virasoro and BMS algebras were introduced before in \cite{Compere:2014cna,Compere:2015knw} using the notion of symplectic symmetries. There it is however the algebra \eqref{Bondi symmetry algebra} formed by the diffeomorphisms $\D$ alone which is brought to any finite distance $r$. We comment below \eqref{Bondi diffeo charge} and in appendix \ref{app:Kosmann} on the definition of the symplectic symmetries in the first order formulation.

In light of these results, a natural question is therefore that of the fate of $\C$ in the asymptotic case and in the case of finite boundaries with non-tangential vector fields. This also evidently raises the question of the geometrical and metric interpretation of the dual diffeomorphism $\C$. We will come back to this in future work, but give brief comments in section \ref{sec:5}. Here we have established a first contact with this dual charge, and presented the simplest and natural setup in which it arises and can be understood. This understanding is deepened by the Sugawara construction, which we now turn to.

\subsection{Classical Sugawara construction}
\label{sec:Sugawara}

We now want to construct objects which are quadratics in the basic currents $\J^i_n$ and $\T^i_n$, and in particular understand their relationship with the generators $\D_n$ and $\C_n$ of diffeomorphisms and their dual. This is the essence of the Sugawara construction, which provides a concrete algebraic implementation of the relations \eqref{diffeo field-dependent} and \eqref{definition C}. In the case of diffeomorphisms this is well-know for usual three-dimensional gravity. Here we extend this result to the case of the MB model. In addition, we give an interpretation of the dual diffeomorphism as a Sugawara quadratic generator. This is best achieved in the Fourier representation \eqref{JTDC extended algebra Fourier}. The various brackets for this section are given in appendix \ref{appendix Sugawara}.

To start with, let us introduce the three quadratics
\be\label{Q quadratics}
\Q_n^1\coloneqq2\sum_{k\in\mathbb{Z}}\J^i_{n+k}\T^i_{-k},
\q\q
\Q_n^2\coloneqq\sum_{k\in\mathbb{Z}}\T^i_{n+k}\T^i_{-k},
\q\q
\Q_n^3\coloneqq\sum_{k\in\mathbb{Z}}\J^i_{n+k}\J^i_{-k}.
\ee
Their brackets with $\J^i_{m}$ and $\T^i_{m}$ are given in \eqref{Qi and Ji Ti brackets}, while the brackets of $\Q_n$'s with themselves are given in \eqref{Qi Qj brackets} and \eqref{Qi Qj brackets cross}. Matching \eqref{Qtilde with J and T} with the action of $\D_n$ on $\J^i_m$ and $\T^i_m$, we find a unique solution which is that
\be
\D_n=\f{-1}{4(\sigma_1^2-\sigma_2\sigma_3)}\big(\sigma_1\Q^1_n-\sigma_2\Q^2_n-\sigma_3\Q^3_n\big).
\ee
Then \eqref{Qtilde with Qtilde} gives \eqref{Dn Dm} as expected. This is the expression of the Sugawara construction for the diffeomorphisms in the MB model.

Now, matching \eqref{Qtilde with J and T} with the action of $\C_n$ on $\J^i_m$ and $\T^i_m$ requires to take $\sigma_3=p\sigma_2$, and we then find a unique solution given by
\be\label{Sugawara construction C}
\C_n=\f{-1}{4(\sigma_1^2-p\sigma_2^2)}\big(\sigma_1\Q^2_n+p\sigma_1\Q^3_n-p\sigma_2\Q^1_n\big).
\ee
This equation is the Sugawara construction for the dual diffeomorphisms in the MB model with the extra condition $\sigma_3=p\sigma_2$. Once again, note that this includes the case of usual first order three-dimensional gravity, which is recovered for $\sigma_2=\sigma_3=0$. The existence of this dual Sugawara construction is yet another argument for why these dual charges are natural to consider.

Finally, we can also connect these Sugawara constructions with the expression \eqref{JT Casimirs} for the Casimirs of the $(J,T)$ algebra, which is the global part of the $(\J,\T)$ one. Indeed, the existence of two Sugawara constructions, and therefore the existence of the dual charges, can in a sense also be understood as a consequence of the existence of two Casimirs. Identifying the terms in \eqref{JT Casimirs} with the quadratics in \eqref{Q quadratics} shows that
\be
\D_n\sim\f{-1}{4(\sigma_1^2-\sigma_2\sigma_3)}\big(\sigma_1C_2-\sigma_2C_1\big),
\q\q
\C_n\sim\f{-1}{4(\sigma_1^2-p\sigma_2^2)}\big(\sigma_1C_1-p\sigma_2C_2-pq\sigma_2\Q^3_n\big).
\ee
Interestingly, we see here resurfacing the condition $q=0$ which has been identified in the previous subsection in order for the action of $\C$ on $\T$ to be stable. While this condition does not appear in \eqref{Sugawara construction C}, it appears here when trying to connect the quadratic generators to the Casimirs. When $q=0$, we get as expected a (formal) linear relationship between the Casimirs of the initial algebra and the generators of diffeomorphisms and their dual.

In summary, we have shown in this section that all the known results about diffeomorphisms, internal Lorentz transformations, and translations, can be extended to the MB model. We have done so by computing the charge algebra, and related the currents to the quadratics by the Sugawara construction. In addition, we have also shown that the $(p\neq0,q=0)$ subspace of the MB model contains a dual diffeomorphism charge. This latter admits a Sugawara interpretation, and forms with the usual diffeomorphisms a finite distance centreless double Virasoro or BMS algebra. In section \ref{sec:4} we will see how these algebras can be centrally-extended when considering non-tangent vector fields. This can also be interpreted in terms of the so-called twisted Sugawara construction.

The twisted Sugawara construction relies on a first order linear twist of the quadratics $\Q_n$ introduced above in \eqref{Q quadratics}. We define these twists in appendix \ref{app:twisted} in terms of fixed internal vectors $\lambda^i$ and $\mu^i$, and also compute all the relevant brackets between the twisted quadratics. To illustrate this construction and the result, we can focus for simplicity on the case $\sigma_2=\sigma_3=0$ (and therefore $q=0$) of usual three-dimensional gravity. In this case, the diffeomorphisms and their dual are given by
\be
\D_n=\f{-1}{4\sigma_1}\check{\Q}^1_n,
\q\q
\C_n=\f{-1}{4\sigma_1}\big(\check{\Q}^2_n+p\check{Q}^3_n\big),
\ee
where the $\check{\Q}_n$'s are the twisted quadratics defined in \eqref{twisted quadratics}. With this twist, the algebra of the diffeomorphisms and their dual picks up central extensions, and takes the form
\bsub
\be
\lb\D_n,\D_m\rb&=\i(m-n)\D_{m+n}+\i\f{\lambda\cdot\mu}{\sigma_1}m^3\delta_{n+m,0},\\
\lb\C_n,\D_m\rb&=\i(m-n)\C_{m+n}+\i\f{1}{2\sigma_1}(\lambda\cdot\lambda+p\mu\cdot\mu)m^3\delta_{n+m,0},\\
\lb\C_n,\C_m\rb&=\i p(m-n)\D_{m+n}+\i p\f{\lambda\cdot\mu}{\sigma_1}m^3\delta_{n+m,0}.
\ee
\esub
With the redefinition of generators \eqref{Virasoro change of generators} this becomes a centrally-extended double Virasoro algebra. In the flat limit $p=0$ this becomes a centrally-extended BMS$_3$ algebra. This shows that the finite distance Virasoro and BMS algebras formed by the diffeomorphisms and their dual can be centrally-extended in the presence of a twist. In section \ref{sec:4} we will interpret this twist as a choice of boundary conditions on the connection and the triad.

\subsection{Teleparallel dual diffeomorphisms}
\label{sec: tele dual charge}

We have now had a first contact with the dual charges, and studied in minute details the example of the dual diffeomorphisms $\C$ defined in the case $(p\neq0,q=0)$. We have in particular discovered the algebra which they form with the usual diffeomorphisms $\D$, and related them to the Sugawara construction.

Given how the MB model elegantly unifies the Lorentzian curvature and torsion sectors of three-dimensional gravity, characterized respectively by $p$ and $q$, it is natural to now ask whether there exists also a dual charge in the case $(p=0,q\neq0)$. With a slight abuse of language we choose to call this the teleparallel sector\footnote{Indeed, recall that $p$ and $q$ are measures of the Lorentzian curvature and torsion, while the Riemannian curvature and torsion sectors of the MB model will be studied in section \ref{sec:6}.}. In this sector of the MB model, the dual charge is given by
\be\label{definition B}
\slashed{\delta}\B(\xi)\coloneqq q\,\slashed{\delta}\T(\xi\ip e)+\slashed{\delta}\T(\xi\ip\omega),
\ee
whose integrability can easily be seen to be achieved with a tangent vector and the condition $\sigma_3=q\sigma_1$. Much like what has been done above for the dual charge $\C$, one can explicitly check that consistency of the bracket of $\B(\xi)$ with $\T(\phi)$ requires that $p=0$. In this sense the charge $\B$ is the torsion/curvature dual to the charge $\C$.

The action of $\B$ on the connection and on the triad can be computed by analogy with \eqref{dual diffeos}. A direct calculation and a little bit of rewriting shows that
\bsub\label{tele dual diffeos}
\be
\delta^\text{b}_\xi e
&=q\delta^\text{t}_{\xi\ip e}e+\delta^\text{t}_{\xi\ip\omega}e\cr
&=q\de_\omega(\xi\ip e)+q^2[e,\xi\ip e]+\de_\omega(\xi\ip\omega)+q[e,\xi\ip\omega]\cr
&=\pounds_\xi\omega-\xi\ip F+q\de_\omega(\xi\ip e)+q^2[e,\xi\ip e]+q[e,\xi\ip\omega]\cr
&=\pounds_\xi\omega+q\pounds_\xi e-\xi\ip F-\f{q}{2}\xi\ip\big(2\de_\omega e+q[e\wedge e]\big)\cr
&\approx\pounds_\xi\omega+q\pounds_\xi e-p[e,\xi\ip e],\\
\delta^\text{b}_\xi\omega
&=q\delta^\text{t}_{\xi\ip e}\omega+\delta^\text{t}_{\xi\ip\omega}\omega\cr
&=pq[e,\xi\ip e]+p[e,\xi\ip\omega].
\ee
\esub
This result shows that when $p=0$ the transformation $\B$ does indeed have a geometrical action on the dynamical fields. At the difference with $\C$ which was completely dualizing the diffeomorphisms by transforming $e$ into the Lie derivative of $\omega$ and vice-versa, we see that $\B$ transforms $e$ but not $\omega$. This therefore also shows that these duality structures on the MB space of theories are very subtle, since e.g. here we could not have intuited the form of \eqref{tele dual diffeos} from the analogue formula for $\D$ and $\C$. This opens the question of the geometrical understanding of the dual transformations $\B$ and $\C$, which we will explore in future work.

Using the results of appendix \ref{app:C}, we can now also perform a Sugawara construction for this new dual charge. Its expression in terms of the quadratics \eqref{Q quadratics} is in fact very simple, and found to be
\be\label{Sugawara construction B}
\B_n=\f{-1}{4\sigma_1}\Q^2_n.
\ee
Again, we can then relate this Sugawara construction to the Casimirs of the $(J,T)$ algebra, and we find that
\be
\B_n\sim\f{-1}{4\sigma_1}\big(C_1-p\Q^3_n\big),
\ee
which bring up again the condition $p=0$ in order to have a correspondance between the Casimirs and the quadratic generators. The other quadratic in the case $(p=0,q\neq0)$ is again given of course by $\D_n$, and the new algebra of diffeomorphisms and their teleparallel dual is found to be
\bsub\label{BD Virasoro}
\be
\lb\D_n,\D_m\rb&=\i(m-n)\D_{m+n},\\
\lb\B_n,\D_m\rb&=\i(m-n)\B_{m+n},\\
\lb\B_n,\B_m\rb&=\i q(m-n)\B_{m+n}.
\ee
\esub
In the case $q=0$ of vanishing (Lorentzian) torsion this is the centreless BMS$_3$ algebra, as we have found in the flat case with the generators $(\D_n,\C_n)$. For $q\neq0$ however, upon rescaling $\B_n$ we recognize two centreless Virasoro algebras in semi-direct sum. One should recall that, although the teleparallel dual diffeomorphism $\B$ is defined when the Lorentzian curvature $p$ is vanishing, the Riemannian curvature is still non-vanishing and given by $\Lambda=-q^2/4$. From this point of view it is perhaps not surprising to find Virasoro algebras, although the novelty here is that they appear with a semi-direct sum. These algebras can also easily be centrally-extended with the twisted Sugawara construction using the results of appendix \ref{app:twisted}.

Consistently, we then find that in the case $(p=0,q=0)$ we get $\B=\C$, so the dual diffeomorphisms agree. This can be seen both in the definitions \eqref{definition C} and \eqref{definition B} and in the Sugawara expressions \eqref{Sugawara construction C} and \eqref{Sugawara construction B}. 

Finally, let us close with a remark about the (im)possibility of extending further the duality. Given how $\B$, $\C$, and $\D$ are defined in terms of field-dependent gauge transformations in equations \eqref{definition B}, \eqref{definition C}, and \eqref{diffeo field-dependent}, and in particular the fact that $\B$ uses the translations, one can wonder if it is possible to use instead the sum of two field-dependent Lorentz transformations. One can indeed consider $\slashed{\delta}\A\coloneqq\alpha\slashed{\delta}\J(\xi\ip e)+\beta\slashed{\delta}\J(\xi\ip\omega)$, which is integrable for tangent vectors if $\beta\sigma_1=\alpha\sigma_2$. However, then a quick calculation along the lines of \eqref{app: CT details} reveals that this charges cannot have a stable action on $\T(\phi)$ unless $\sigma_1^2-\sigma_2\sigma_3=0$, which is not authorized since it corresponds to the degenerate sector of the MB model. This therefore closes the study of the quadratic charges obtained as combinations of field-dependent gauge transformations. We can now step back a little to study the CS formulation of the MB model, and then relax the condition of tangentiality of the vector fields.

\section{Chern--Simons formulation}
\label{sec:4}

We now turn to the Chern--Simons formulation of the MB model. First, we will explain the difference between the $\mathfrak{g}_\Lambda$ basis \eqref{JP algebra} and \eqref{JT algebra}. Then we will focus on the charges and go back to the connection and triad variables in order to identify the dual charges in the case of non-tangent vector fields.

Recall that for any connection decomposed as $A=\omega^iJ_i+e^iK_i$ (we will keep the indices implicit below), for some generators $(J,K)$ whose algebra is irrelevant for the moment, the CS Lagrangian gives
\be
L_\text{CS}[A]
&=A\wedge\left(\de A+\f{1}{3}[A\wedge A]\right)\cr
&=\omega^i\wedge\de\omega^j\la J_i,J_j\ra+\omega^i\wedge\de e^j\la J_i,K_j\ra+e^i\wedge\de\omega^j\la K_i,J_j\ra+e^i\wedge\de e^j\la K_i,K_j\ra\cr
&\pe+\omega^i\wedge\omega^j\wedge e^k\la J_i,[J_j,K_k]\ra+\omega^i\wedge e^j\wedge e^k\la J_i,[K_j,K_k]\ra\cr
&\pe+\f{1}{3}\omega^i\wedge\omega^j\wedge\omega^k\la J_i,[J_j,J_k]+\f{1}{3}e^i\wedge e^j\wedge e^k\la K_i,[K_j,K_k]\ra,
\ee
where $\la\cdot\,,\cdot\ra$ denotes the pairing in the algebra. This formula is always valid, as for the moment we have not used any commutators or pairings in the algebra of $(J,K)$. Let us now look at the cases $K=P$ and $K=T$.

\subsection[Chern--Simons formulation with $(J,P)$]{Chern--Simons formulation with $\boldsymbol{(J,P)}$}

Let us consider the algebra
\be\label{JP algebra 2}
[J_i,P_j]={\eps_{ij}}^kP_k,
\q\q
[J_i,J_j]={\eps_{ij}}^kJ_k,
\q\q
[P_i,P_j]=-\mu_0{\eps_{ij}}^kJ_k,
\ee
and the compatible pairings
\be\label{JP pairings}
\la J_i,P_j\ra=\mu_1\eta_{ij},
\q\q
\la J_i,J_j\ra=\mu_2\eta_{ij},
\q\q
\la P_i,P_j\ra=-\mu_0\mu_2\eta_{ij}.
\ee
With this and $A=\omega J+eP$ we find
\be\label{Witten Lagrangian}
L_\text{CS}[A]
&=L_\text{W}[e,\omega]+\mu_1\de(e\wedge\omega)\cr
&=2\mu_1e\wedge\left(F-\f{\mu_0}{6}[e\wedge e]\right)+\mu_2\left(\omega\wedge\left(\de\omega+\f{1}{3}[\omega\wedge\omega]\right)-\mu_0e\wedge\de_\omega e\right)+\mu_1\de(e\wedge\omega).\q
\ee
This is the sum of the usual Lagrangian with coupling $\mu_1$, and the so-called ``exotic'' Lagrangian with coupling $\mu_2$. We have denoted the sum after Witten since the two terms appeared in \cite{MR974271}. The Lagrangian \eqref{Witten Lagrangian} differs from the MB Lagrangian \eqref{MB Lagrangian} in two ways:
\begin{enumerate}
\item The couplings for the Lagrangians $L_\text{CS}[\omega]$ and $L_\text{T}[e,\omega]$ in \eqref{Witten Lagrangian} are not independent, since they both appear in the exotic Lagrangian with the common factor of $\mu_2$. One could think of relaxing this by introducing a new independent coupling and working with the pairing $\la P_i,P_j\ra=\mu_3\eta_{ij}$, but then we would have $\la J_i,[P_j,P_k]\ra\neq\la P_k,[J_i,P_j]\ra$, so this pairing is not compatible with the algebra.
\item The couplings for the Lagrangians $L_\text{V}[e]$ and $L_\text{HP}[e,\omega]$ in \eqref{Witten Lagrangian} are not independent. One could think of relaxing this by replacing $\mu_0\mapsto\mu_0/\mu_1$, but then the limit $\mu_1\rightarrow0$ would not be well-defined, neither in the algebra nor in the Lagrangian. One way to make the limit well-defined in the Lagrangian however would be to also replace $\mu_2\mapsto\mu_1\mu_2$, but then the Lagrangians $L_\text{HP}[e,\omega]$ and $L_\text{CS}[\omega]$ would not be independent.
\end{enumerate}
All this comes from the fact that we actually need a fourth independent parameter in the game. This is indeed the property which distinguishes the MB model from \eqref{Witten Lagrangian}.

As pointed out in \cite{Cacciatori:2005wz,Giacomini:2006dr}, a fourth coupling can be introduced consistently by considering the connection
\be\label{shifted connection}
A=(\omega+\mu_3e)J+eP.
\ee
As an indication, one can notice that the curvature of this connection is given by
\be
\de A+\f{1}{2}[A\wedge A]=\left(F+\mu_3\de_\omega e+\f{1}{2}(\mu_3^2-\mu_0)[e\wedge e]\right)J+\Big(\de_\omega e+\mu_3[e\wedge e]\Big)P.
\ee
Setting it to zero gives the equations
\be
2F-(\mu_3^2+\mu_0)[e\wedge e]=0,
\q\q
\de_\omega e+\mu_3[e\wedge e]=0,
\ee
which therefore describe a source of Lorentzian curvature and torsion as in the MB model. The Lagrangian obtained from this connection is
\be\label{MB as shift from W}
L_\text{CS}[A]
&=L_\text{W}[e,\omega+\mu_3e]+\mu_1\de(e\wedge\omega)\cr
&=\sigma_0L_\text{V}[e]+\sigma_1L_\text{HP}[e,\omega]+\sigma_2L_\text{CS}[\omega]+\sigma_3L_\text{T}[e,\omega]+\sigma_1\de(e\wedge\omega),
\ee
with the identifications
\be\label{mu sigma identification}
\sigma_0=\mu_1(3\mu_3^2-\mu_0)+\mu_2\mu_3(\mu_3^2-3\mu_0),
\q
\sigma_1=\mu_1+\mu_2\mu_3,
\q
\sigma_2=\mu_2,\q\sigma_3=2\mu_1\mu_3+\mu_2(\mu_3^2-\mu_0).
\ee
Notice how when going from the first to the second line in \eqref{MB as shift from W} the boundary term also picks up a contribution from an integration by parts. One can therefore see that the CS Lagrangian for the shifted connection \eqref{shifted connection} produces indeed the MB Lagrangian. However, the couplings of the MB Lagrangian then have the complicated relationship \eqref{mu sigma identification} with the parameters defining the algebra and its pairings. Moreover, one can also notice that the connection $\omega$ of the MB model is not the connection used to build the composite CS connection \eqref{shifted connection}, since there it is the shift $\omega+\mu_3e$ which appears. This means that, in this particular $(J,P)$ CS formulation, the MB model is not actually determined solely by the algebraic data \eqref{JP algebra 2} and \eqref{JP pairings}, but requites also the shift of the connection.

It turns out that these two points can be improved at once if we use the $(J,T)$ algebra \eqref{JT algebra} instead of $(J,P)$. First, notice that one can invert the relations \eqref{mu sigma identification} to find $\mu_i=f(\sigma_i)$. This yields in particular $\mu_3=q/2$, meaning that in \eqref{shifted connection} instead of shifting the connection one can simply change the generators as $(J,P)\mapsto(J,T)$. With this change of generators the pairings \eqref{JP pairings} lead to the pairings \eqref{JT pairings} given below.

\subsection[Chern--Simons formulation with $(J,T)$]{Chern--Simons formulation with $\boldsymbol{(J,T)}$}

A more transparent CS formulation of the MB model can be obtained by using the algebra $(J,T)$. This puts forward more clearly the role of the couplings and of the curvature and torsion parameters $(p,q)$. Let us consider the algebra
\be
[J_i,T_j]={\eps_{ij}}^kT_k,
\q\q
[J_i,J_j]={\eps_{ij}}^kJ_k,
\q\q
[T_i,T_j]={\eps_{ij}}^k(pJ_k+qT_k),
\ee
and the pairings
\be\label{JT pairings}
\la J_i,T_j\ra=\sigma_1\eta_{ij},
\q\q
\la J_i,J_j\ra=\sigma_2\eta_{ij},
\q\q
\la T_i,T_j\ra=\sigma_3\eta_{ij}.
\ee
For the moment $p$ and $q$ can of course be arbitrary. With this, defining $A=\omega J+eT$ leads to
\be
L_\text{CS}[A]=(p\sigma_1+q\sigma_3)L_\text{V}[e]+\sigma_1L_\text{HP}[e,\omega]+\sigma_2L_\text{CS}[\omega]+\sigma_3L_\text{T}[e,\omega]+\sigma_1\de(e\wedge\omega),
\ee
where in this derivation the only condition which we have to impose on the pairings in order for the torsion Lagrangian to appear, and therefore to get this Lorentz-invariant MB form of the total Lagrangian, is
\be\label{constraint 1}
\la T_i,T_j\ra=p\la J_i,J_j\ra+q\la J_i,T_j\ra\q\Rightarrow\q\sigma_3=p\sigma_2+q\sigma_1.
\ee
This relation is in fact precisely the compatibility condition
\be
\la J_i,[T_j,T_k]\ra=\la T_k,[J_i,T_j]\ra.
\ee
The other condition, i.e. the one involving two $J$'s and one $T$, is already satisfied. We therefore have a priori 5 couplings $(\sigma_1,\sigma_2,\sigma_3,p,q)$ minus this constraint, which gives 4 independent couplings as expected. We can then trade one of these couplings for $\sigma_0$ by further imposing
\be\label{constraint 2}
\sigma_0=p\sigma_1+q\sigma_3.
\ee
One can then check for consistency that the parameters $p$ and $q$ are indeed solutions to \eqref{constraint 1} and \eqref{constraint 2}, so we finally have the very simple result
\be\label{CS MB relation}
L_\text{CS}[A]=L_\text{MB}[e,\omega]+\sigma_1\de(e\wedge\omega).
\ee
This shows, as announced, that the MB Lagrangian is a CS theory for the connection $A=\omega J+eT$ and the $(J,T)$ algebra \eqref{JT algebra}.

\subsection{Charges in Chern--Simons theory}

We now recall basic facts about gauge and diffeomorphism charges in CS theory, before going back to connection and triad variables to analyse the integrability of the diffeomorphisms. The variation of the CS Lagrangian, the potential and the symplectic structure are given respectively by
\be
\delta L_\text{CS}[A]=2\delta A\wedge F+\de(\delta A\wedge A),
\q\q
\theta_\text{CS}=\delta A\wedge A,
\q\q
\Omega_\text{CS}=-\int_\Sigma\delta A\wedge\delta A.
\ee
Consistently with \eqref{CS MB relation}, we have that $\theta_\text{CS}=\theta_\text{MB}+\sigma_1\delta(e\wedge\omega)$. The generator of the CS gauge transformation $\delta_\lambda A=\de_A\lambda$ is computed from $-\delta_\lambda\ipp\Omega=\delta\F(\lambda)$ and given by
\be
\F(\lambda)=-2\int_\Sigma\lambda F+2\oint_S\lambda A,
\q\q
\lb\F(\lambda),\F(\mu)\rb=\F([\lambda,\mu])-2\oint_S\lambda\de\mu.
\ee
We can now decompose $A=\omega J+eT$ and $\lambda=\alpha J+\phi T$. This gives
\be
\F_{JT}(\alpha,\phi)\approx2\oint_S\alpha(\sigma_1e+\sigma_2\omega)+\phi(\sigma_1\omega+\sigma_3e),
\ee
which indeed agrees with the charges \eqref{Lorentz generator} and \eqref{translation generator}. For diffeomorphisms we have as expected that
\be
\slashed{\delta}\D(\xi)
&=2\int_\Sigma\xi\ip(\delta A\wedge F)-\delta\big((\xi\ip A)F\big)+2\oint_S(\xi\ip A)\delta A\cr
&=2\int_\Sigma\xi\ip(\delta A\wedge F)-\delta\big((\xi\ip A)F\big)+\oint_S\delta\big((\xi\ip A)A\big)-\xi\ip\theta,
\ee
which is in turn equal to \eqref{full diffeo charge}. This shows the consistency of our CS formulation.

In CS theory, the relation \eqref{diffeo field-dependent} giving the diffeomorphisms as field-dependent gauge transformations becomes simply $\slashed{\delta}\D(\xi)=\slashed{\delta}\F(\xi\ip A)$. Just as in the connection/triad formulation, generic diffeomorphisms are not integrable, and integrability can be achieved with either tangent vector fields or with specific boundary conditions on the dynamical fields. We also note very interesting recent work on the possibility of obtaining integrable diffeomorphism charges by considering \textit{field-dependent vector fields} \cite{Adami:2020ugu}, which we think could be extended to the MB model as well. So far we have focused here on tangent diffeomorphisms in order to define the charges $\D$ and their dual $\C$. Here we would like to relax this condition and instead use boundary conditions on the dynamical fields themselves. In the following subsection we therefore recall how this is traditionally done in CS theory, and then what happens when going back to connection/triad variables.


\subsection{Dual diffeomorphisms again}
\label{sec:dual charges}

Let us now look at generic diffeomorphism charges, and consider an arbitrary field-independent vector field of the form $\xi=(\xi^t,\xi^r,\xi^\varphi)$. With this the (a priori non-integrable) CS diffeomorphism charge is
\be
\slashed{\delta}\D(\xi)=2\oint_S(\xi^tA_t+\xi^rA_r+\xi^\varphi A_\varphi)\delta A_\varphi.
\ee
The angular $\varphi$ piece is always integrable. The $(t,r)$ pieces are integrable if we choose either $(A_t,A_r)\big|_S\propto A_\varphi$, or $(A_t,A_r)\big|_S=\text{constant}$, or some mixture of these conditions. Usually this choice is constrained by the form of the boundary conditions chosen for the variational principle. Typical boundary conditions are the chiral ones $A_t=\tau A_\varphi$, which indeed set $\theta_\text{CS}\big|_\Delta=0$ on the time-like boundary. With a gauge choice it is also customary to fix $A_r$ to be a constant $a$. With these conditions the diffeomorphism charge becomes\footnote{Recall that here $\tau$ is a number while $a$ is a Lie algebra element.}
\be\label{integrable diffeo}
\D(\xi)=\oint_S\xi^\tau(A_\varphi)^2+2\xi^raA_\varphi,
\ee
where we have introduced $\xi^\tau\coloneqq\tau\xi^t+\xi^\varphi$. To compute the algebra of these charges, we need two further assumptions. The first one is that $\xi$ depends only on the angular variable $\varphi$. This ensures that the derivative of vector fields which appear when computing the bracket can be rewritten as the Lie bracket. The second one is that $\xi^\tau=\kappa\xi^\varphi$. This is required in order for the bracket to close. With these assumptions, we find that the bracket is
\be\label{general diffeo bracket}
\lb\D(\xi),\D(\zeta)\rb=-\kappa\D([\xi,\zeta])-2a^2\oint_S\xi^r\partial_\varphi\zeta^r.
\ee
This is the centrally-extended $\text{diff}(S)\oplus\mathfrak{u}(1)$ algebra, where $r$ is the Abelian direction (which is the one carrying the central extension). When the radial direction is then related to the angular one with $\xi=(\xi^t,\xi^r,\xi^\varphi)=(0,\gamma\partial_\varphi\bar{\xi},\bar{\xi})$, this becomes the centrally-extended Virasoro algebra
\be\label{CS Virasoro}
\lb\D(\xi),\D(\zeta)\rb=-\kappa\D([\xi,\zeta])+2a^2\gamma^2\oint_S\bar{\xi}\partial^3_\varphi\bar{\zeta}.
\ee
This simple and well-known result shows how the central extension of the diffeomorphism algebra arises from the non-tangentiality of the vector fields\footnote{Note that in appendix \ref{app:B}, when presenting the covariant phase space, we have reproduced the proof of \cite{Speranza:2017gxd} that the non-integrable piece $\xi\ip\theta$ of the diffeomorphism charge (which is the piece vanishing when $\xi$ is tangent) is the one responsible for the appearance of a central extension. In fact in this appendix we have extended this proof to the presence of a corner term along the lines of \cite{Harlow:2019yfa,Geiller:2019bti,Freidel:2020xyx}. This is equation \eqref{app: diffeo central charge}. Note however that this computation does \textit{not} allow to recover the Virasoro or BMS central extension, because it assumes that the boundary term $b$ is covariant. A more general formula for the central extension was given in \cite{Chandrasekaran:2020wwn}, and it would be interesting to explicitly show how this formula reproduces the Virasoro and BMS central extension.}. In this case we have to pick boundary conditions in order for the charge to be integrable, and we have chosen here the conditions leading to \eqref{integrable diffeo} as an example. In section \ref{sec:Sugawara} and in appendix \ref{app:twisted} we have also explained how the non-tangentiality of the diffeomorphisms and the appearance of a central charge can be understood in terms of the so-called twisted Sugawara construction. Note that in \eqref{CS Virasoro} and in the twisted Sugawara construction, although the vector fields are not tangent, they only depend on a single parameter since the radial and angular components are related.

It is now enlightening to go back to the MB model in connection/triad variables. Denoting $a=A_r=(e_r,\omega_r)=(b,c)$, the diffeomorphism charge \eqref{integrable diffeo} for the MB model is
\be\label{integrable diffeo e omega}
\D(\xi)=\oint_S\xi^\tau\big(2\sigma_1\omega_\varphi e_\varphi+\sigma_2(\omega_\varphi)^2+\sigma_3(e_\varphi)^2\big)+2\xi^r\big(\sigma_1ce_\varphi+\sigma_1b\omega_\varphi+\sigma_2c\omega_\varphi+\sigma_3be_\varphi\big).
\ee
For tangent diffeomorphisms, we have seen previously that the field-dependent Lorentz transformations and translations in \eqref{diffeo field-dependent} are \textit{not} separately integrable. With the vector fields and boundary conditions chosen here, we find that these field-dependent charges are also non-integrable and given by
\bsub\label{separate integrability condition on J T}
\be
\slashed{\delta}\J(\xi\ip\omega)&=\oint_S\delta\Big(\xi^\tau\sigma_2(\omega_\varphi)^2+2\xi^r\big(\sigma_1ce_\varphi+\sigma_2c\omega_\varphi\big)\Big)+2\xi^\tau\sigma_1\omega_\varphi\delta e_\varphi,\\
\slashed{\delta}\T(\xi\ip e)&=\oint_S\delta\Big(\xi^\tau\sigma_3(e_\varphi)^2+2\xi^r\big(\sigma_1b\omega_\varphi+\sigma_3be_\varphi\big)\Big)+2\xi^\tau\sigma_1e_\varphi\delta\omega_\varphi.
\ee
\esub
The first piece in each expression is integrable, but because of the cross term in $\sigma_1$, it is only the combination of these charges into diffeomorphisms which is integrable. The charges are therefore generically not integrable on their own.

Interestingly, the obstruction to separate integrability of the charges in \eqref{separate integrability condition on J T} comes from $\sigma_1\neq0$, and we can therefore make these charges integrable separately by working in the sector $\sigma_1=0$ of the theory. This possibility exists because we are considering the MB model, which is non-trivial even when $\sigma_1=0$, while the charges of usual three-dimensional Hilbert--Palatini gravity would trivialize in this case. In order to find charges which are integrable in the case $\sigma_1\neq0$, this then suggests to look at a natural alternative, which is to consider the ``dual'' type of field-dependency for the gauge charges. Instead of \eqref{separate integrability condition on J T} we are then led to consider the dual charges
\bsub\label{dual quadratic J T}
\be
\slashed{\delta}\J(\xi\ip e)&=\oint_S\delta\Big(\xi^\tau\sigma_1(e_\varphi)^2+2\xi^r\big(\sigma_2b\omega_\varphi+\sigma_1be_\varphi\big)\Big)+2\xi^\tau\sigma_2e_\varphi\delta\omega_\varphi,\\
\slashed{\delta}\T(\xi\ip\omega)&=\oint_S\delta\Big(\xi^\tau\sigma_1(\omega_\varphi)^2+2\xi^r\big(\sigma_3ce_\varphi+\sigma_1c\omega_\varphi\big)\Big)+2\xi^\tau\sigma_3\omega_\varphi\delta e_\varphi.
\ee
\esub
Looking at \eqref{separate integrability condition on J T} and \eqref{dual quadratic J T}, it is now evident to recognize the integrable combinations which we have introduced in the previous section. Indeed, if we put a condition on the couplings we can get new integrable charges defined as follows:
\bsub
\be
\B(\xi)&\coloneqq q\T(\xi\ip e)+\T(\xi\ip\omega)\q\text{when}\q p=0,\\
\C(\xi)&\coloneqq p\J(\xi\ip e)+\T(\xi\ip\omega)\q\text{when}\q q=0,\\
\D(\xi)&\coloneqq\phantom{p}\J(\xi\ip\omega)+\T(\xi\ip e)\q\text{whenever}.
\ee
\esub
The generator $\D$ of diffeomorphisms is of course always integrable, while the dual charges $\B$ and $\C$ each require an extra condition. As explained in the previous section we need $p=0$ and $q=0$ respectively for $\B$ and $\C$ to have a consistent algebra with the rest of the charges, and furthermore we then get $p=0\Rightarrow\sigma_3=q\sigma_1$ and $q=0\Rightarrow\sigma_3=q\sigma_2$.

In the case of the boundary conditions used in this section, where the vector fields are not tangent, we have the explicit expressions
\bsub
\be
\B(\xi)&=\oint_S\xi^\tau\big(2\sigma_3\omega_\varphi e_\varphi+\sigma_1(\omega_\varphi)^2+q\sigma_3(e_\varphi)^2\big)+2\xi^r\big(\sigma_3ce_\varphi+\sigma_3b\omega_\varphi+\sigma_1c\omega_\varphi+q\sigma_3be_\varphi\big),\\
\C(\xi)&=\oint_S\xi^\tau\big(2\sigma_3\omega_\varphi e_\varphi+\sigma_1(\omega_\varphi)^2+p\sigma_1(e_\varphi)^2\big)+2\xi^r\big(\sigma_3ce_\varphi+\sigma_3b\omega_\varphi+\sigma_1c\omega_\varphi+p\sigma_1be_\varphi\big).
\ee
\esub
One can compare this to \eqref{integrable diffeo e omega} to see that these new generators can schematically be obtained from $\D$ upon changing the coupling constants according to the pattern
\be\label{swapping BCD}
\begin{tabular}{ccccc}
$\C$ & & $\D$ & & $\B$\vspace{-0.3cm}\\
\vspace{-0.3cm}
& $\longleftarrow$ & & $\longrightarrow$ &\\
$(\sigma_3,\sigma_1,p\sigma_1)$ & & $(\sigma_1,\sigma_2,\sigma_3)$ & & $(\sigma_3,\sigma_1,q\sigma_3)$
\end{tabular}
\ee
where in addition we should recall that $\sigma_3=q\sigma_1$ for $\B$ and $\sigma_3=p\sigma_2$ for $\C$. Just like the algebra of the diffeomorphisms gets centrally-extended as in \eqref{general diffeo bracket} because of the non-tangentiality of the vector field, the brackets involving $\B$ and $\C$ will also receive central extensions. This shows how in the case of non-tangent vector fields it is still possible to define the dual charges, and moreover to obtain central extensions for the finite distance algebras $(\D,\C)$ or $(\D,\B)$. Moreover this can be understood in terms of a twisted Sugawara construction, as we explain in appendix \ref{app:twisted}. We keep the detailed investigation of the central charges for future work.

Finally, an important observation following from the construction of the dual charges is that it is a priori not possible to obtain them directly from CS theory. Indeed, as we have recalled in the previous subsection, the diffeomorphisms of the MB model are obtained from its CS formulation by considering the field dependent gauge charges $\F(\xi\ip A)$. It is however clear that it is not possible to consider any other field dependency, such as the ones defining $\B$ and $\C$. CS theory does simply not allow to write down such objects. It could perhaps be possible to define dualities directly at the level of CS theory, using e.g. a connection $\tilde{A}=\omega T+eJ$ or defining an operation swapping the couplings as in \eqref{swapping BCD}, but this is for the moment speculative and should be investigated further.


\section{Asymptotic charges and symmetries}
\label{sec:5}

Now that we have unraveled the finite distance covariant phase space structure of the MB model, introduced the dual charges, and studied the charge algebra, we would like to briefly recall results about the asymptotic structure of the MB model. This paves the road for future work on the study of the dual charges in the asymptotic context.

\subsection{BTZ black hole}

We start by recalling some basic facts about the BTZ black hole in the MB model. In Schwarzschild coordinates, the Lorentzian line element is given by
\be
\de s^2=-N(r)^2\de t^2+\f{1}{N(r)^2}\de r^2+r^2\big(\de\varphi+N_\varphi(r)\de t\big)^2,
\ee
where
\be
N(r)^2=-8Gm+\f{r^2}{\ell^2}+\f{16G^2j^2}{r^2},
\q\q
N_\varphi(r)=-\f{4Gj}{r^2},
\ee
with $m$ and $j$ the mass and angular momentum. The horizons are located at
\be
r_\pm^2=4Gm\ell^2\left(1\pm\sqrt{1-\left(\f{j}{m\ell}\right)^2}\right).
\ee
The cosmological constant here is related to that in the solution \eqref{metric Ricci tensor solution} given above by
\be
\f{1}{\ell^2}=-\Lambda=p+\f{q^2}{4}.
\ee
The coordinates are the time $t\in\mathbb{R}$, the radius $0<r<\infty$, and the angle $0\leq\varphi\leq2\pi$. With the internal metric $\eta_{ij}=\text{diag}(-1,1,1)$, an adapted set of triad components such that $g_{\mu\nu}=e^i_\mu e^j_\nu\eta_{ij}$, and the corresponding connection components
\be
\omega^i_\mu=\Gamma^i_\mu-\f{q}{2}e^i_\mu,
\q\q
\Gamma^i_\mu=\f{1}{2}{\eps^i}_{jk}g^{\alpha\beta}e^j_\alpha\nabla_\mu e^k_\beta,
\ee
are given by\footnote{The columns and rows correspond respectively to the $i$ and $\mu$ indices.}
\be
\everymath={\displaystyle}
e^i_\mu=
\begin{pmatrix}
N&0&rN_\varphi\\
0&1/N&0\\
0&0&r
\end{pmatrix},
\q\q
\everymath={\displaystyle}
\omega^i_\mu=
\begin{pmatrix}
0&0&r/\ell^2\\
0&-N_\varphi/N&0\\
N&0&rN_\varphi
\end{pmatrix}
-\f{q}{2}e^i_\mu.
\ee
Notice how the connection has a contribution from $q$, which comes from the fact that we have decomposed it as $\omega=\Gamma+k$, where the contorsion is fixed to $k\approx-qe/2$ by the torsion equation of motion in \eqref{decomposed MB EOM}.

As usual, the energy and angular momentum are given by the temporal and angular components of the diffeomorphism charge for the vector field $\xi^\mu=(\xi^t,0,\xi^\varphi)$. In order to write down the charge for such a non-tangent vector field, recall that the boundary condition used on the time-like boundary is $(\theta+\delta b)\big|_\Delta=0$, where $b=\sigma_1(e\wedge\omega)$ is the first order Gibbons--Hawking--York (GHY) term\footnote{Note that, as always, this boundary term is automatically provided by the CS formulation, since from \eqref{CS MB relation} we get $\theta_\text{CS}=\theta_\text{MB}+\sigma_1\delta(e\wedge\omega)$.}. With this boundary condition, the non-integrable piece $-\xi\ip\theta$ for a generic diffeomorphism charge \eqref{full diffeo charge} becomes integrable. Indeed, as explained in the appendix leading to formula \eqref{D with b and c}, where $c=0$ because the first order GHY has no derivative and therefore no corner term, the diffeomorphism charge is then
\be\label{diffeo charge with GHY}
\D(\xi)=\oint_SQ+\sigma_1\xi\ip(e\wedge\omega),
\ee
where the Noether charge and the extra contribution are
\be
Q=2\sigma_1(\xi\ip\omega)e+\sigma_2(\xi\ip\omega)\omega+\sigma_3(\xi\ip e)e,
\q\q
\xi\ip(e\wedge\omega)\big|_S=-\xi^t\left(N^2-r^2N_\varphi^2+\f{r^2}{\ell^2}\right).
\ee
We then find that the total diffeomorphism charge is finite for any value of $r$ and given by
\be\label{BTZ diffeo charge}
\D(\xi)=\xi^tE+\xi^\varphi L,
\ee
with
\be\label{BTZ E and L}
\f{E}{16\pi G}\coloneqq(2\sigma_1-q\sigma_2)\f{m}{2}-\f{\sigma_2}{\ell^2}j,
\q\q
\f{L}{16\pi G}\coloneqq-(2\sigma_1-q\sigma_2)\f{j}{2}+\sigma_2m.
\ee
These conserved charges for the MB model were derived previously in \cite{Blagojevic:2006jk} using the canonical analysis (i.e. as the boundary terms in the Hamiltonian formalism) instead of the covariant phase space. We can see here the efficiency of the latter over the former. It is interesting to note that the energy and angular momentum are in fact a linear combination of the mass and spin of the black hole, even in the case of vanishing Lorentzian torsion $q$.

We can complete this analysis with a brief discussion of the thermodynamical properties of the BTZ black hole in the MB model. In terms of the radii, the temperature, mass, and spin are given by
\be
T=\f{r_+^2-r_-^2}{2\pi\ell^2r_+},
\q\q
m=\f{r_+^2+r_-^2}{8G\ell^2},
\q\q
j=\f{r_+r_-}{4G\ell}.
\ee
With the angular velocity of the outer horizon given by $\Omega=N_\varphi(r_+)$, the first law $\de E=T\de S+\Omega\de L$ gives an entropy \cite{Ning:2018gfm}
\be
S=8\sigma_1\pi^2r_+-4\sigma_2\pi^2\left(qr_++\f{2r_-}{\ell}\right).
\ee
Carefully keeping track of the conventions to match the first order action with the Einstein--Hilbert one, as in \eqref{EH and HP Lagrangians}, shows that we need $\sigma_1=(16\pi G)^{-1}$. The first term then indeed agrees with the usual result $S=\text{perimeter}/4G$, but is modified by $\sigma_2$ even in the absence of Lorentzian torsion $q$.

\subsection{AdS spacetimes and flat limit}

Having studied the fixed BTZ black hole solution, we can now study asymptotic symmetries for a family of spacetimes. We would like to do so in a way which enables to study both the AdS and flat cases by taking a flat limit. For this, we follow \cite{Barnich:2012aw} and choose Bondi null coordinates $(u,r,\varphi)$ to write the line element 
\be
\de s^2=\left(\mathscr{M}(u,\varphi)-\f{r^2}{\ell^2}\right)\de u^2-2\de u\,\de r+\mathscr{N}(u,\varphi)\de u\,\de\varphi+r^2\de\varphi^2,
\ee
where $\ell^2\partial_u\mathscr{M}=\partial_\varphi\mathscr{N}$ and $\partial_u\mathscr{N}=\partial_\varphi\mathscr{M}$. This is the most general solution to the Einstein field equations $R_{\mu\nu}=-\ell^{-2}g_{\mu\nu}$. With the internal metric
\be\label{internal Bondi metric}
\eta_{ij}=
\begin{pmatrix}
0&1&0\\
1&0&0\\
0&0&1
\end{pmatrix},
\ee
a compatible set of triad and connection components can be chosen in the form
\be\label{Bondi e omega}
\everymath={\displaystyle}
e^i_\mu=
\begin{pmatrix}
\big(\mathscr{M}-r^2/\ell^2\big)/2&1&0\\
-1&0&0\\
\mathscr{N}/2&0&r
\end{pmatrix},
\q\q
\everymath={\displaystyle}
\omega^i_\mu=
\begin{pmatrix}
\mathscr{N}/(2\ell^2)&0&r/\ell^2\\
0&0&0\\
\big(\mathscr{M}-r^2/\ell^2\big)/2&1&0
\end{pmatrix}
-\f{q}{2}e^i_\mu.
\ee
The on-shell asymptotic Killing vectors are $\xi^\mu=(\xi^u,\xi^r,\xi^\varphi)$ with
\be\label{Bondi diffeo}
\xi^u=f,
\q\q
\xi^r=\partial^2_\varphi f-r\partial_\varphi g-\mathscr{N}\f{\partial_\varphi f}{2r},
\q\q
\xi^\varphi=g-\f{\partial_\varphi f}{r},
\ee
and the conditions $\partial_rf=0=\partial_rg$, $\ell^2\partial_ug=\partial_\varphi f$, and $\partial_uf=\partial_\varphi g$. Note that this expansion of the vector field is exact, and stops at next-to-next-to-leading order.

With the fields given above, the only non-vanishing component of the symplectic potential \eqref{potential} is $\theta_{\varphi u}=\sigma_1\delta\mathscr{M}$. This is the component along the time-like boundary $\Delta$, so we need a boundary condition. This is again given by $(\theta+\delta b)\big|_\Delta=0$ with $b=\sigma_1(e\wedge\omega)$. With this we then have
\be
\xi\ip(e\wedge\omega)\big|_S=\xi^u\left(\mathscr{M}-\f{2r^2}{\ell^2}\right)-\xi^r,
\ee
and the full diffeomorphism charge \eqref{diffeo charge with GHY} is found to be
\be\label{Bondi diffeo charge}
\D(\xi)=\oint_Sf\left((2\sigma_1-q\sigma_2)\f{\mathscr{M}}{2}+\f{\sigma_2}{\ell^2}\mathscr{N}\right)+g\left((2\sigma_1-q\sigma_2)\f{\mathscr{N}}{2}+\sigma_2\mathscr{M}\right)+\O(r^{-1}).
\ee
The subleading term in this charge is exact and of order $r^{-1}$. This is to be contrasted with the corresponding charges computed when $\sigma_2=\sigma_3=0$ in the metric formulation, in which case the $r$ dependency drops completely and the charges can then be defined at any finite distance $r$ (see e.g. equation (2.13) of \cite{Compere:2014cna}). This latter case is an example of so-called symplectic symmetries \cite{Compere:2014cna,Compere:2015knw,Grumiller:2019ygj}. This slight discrepancy between \eqref{Bondi diffeo charge} and the corresponding metric result (in the case $\sigma_2=\sigma_3=0$), which prevents \eqref{Bondi diffeo} from being interpreted as symplectic symmetries (although they remain of course asymptotic ones),  can be corrected if we replace the Lie derivative on the first order variables by the Kosmann derivative. The use of this derivative in first order tetrad gravity was advocated in \cite{Jacobson:2015uqa,Prabhu:2015vua} and studied in more details in \cite{DePaoli:2018erh,Oliveri:2019gvm,Oliveri:2020xls}. It amounts to correcting the ordinary Lie derivative by a field-dependent Lorentz transformation. We explain how this is realized in appendix \ref{app:Kosmann}, where we also illustrate the same result in Fefferman--Graham gauge instead of the Bondi gauge used here.

We now come back to asymptotic symmetries, and focus on the charges \eqref{Bondi diffeo charge} in the limit $r\to\infty$. We are interested in their Poisson bracket. For this, we need to know how the functions $\mathscr{M}$ and $\mathscr{N}$ transform. This information can be found by looking at $\delta^\text{d}_\xi g_{\mu\nu}=\pounds_\xi g_{\mu\nu}$. Denoting the angular derivative $\partial_\varphi$ by a prime, we find
\bsub
\be
\delta^\text{d}_\xi\mathscr{M}&=g\mathscr{M}'+2\mathscr{M}g'-2g'''+\f{1}{\ell^2}\big(2\mathscr{N}f'+f\mathscr{N}'\big),\\
\delta^\text{d}_\xi\mathscr{N}&=f\mathscr{M}'+2\mathscr{M}f'-2f'''+2\mathscr{N}g'+g\mathscr{N}'.
\ee
\esub
To compute the algebra, let us denote the null and angular components of the diffeomorphism charge by
\be
\E(f)\coloneqq\oint_Sf\left((2\sigma_1-q\sigma_2)\f{\mathscr{M}}{2}+\f{\sigma_2}{\ell^2}\mathscr{N}\right),
\q\q
\L(g)\coloneqq\oint_Sg\left((2\sigma_1-q\sigma_2)\f{\mathscr{N}}{2}+\sigma_2\mathscr{M}\right).
\ee
Note that with the identifications $\mathscr{M}=8Gm$ and $\mathscr{N}=-8Gj$ this agrees with the BTZ charges $E$ and $L$ which we have computed in \eqref{BTZ E and L}, with the factor of $2\pi$ coming from the angular integral. With these charges we then find the algebra
\bsub\label{Bondi symmetry algebra}
\be
\lb\L(g),\E(f)\rb&=-\E([g,f])+(2\sigma_1-q\sigma_2)\oint_Sgf''',\\
\lb\L(g_1),\L(g_2)\rb&=-\L([g_1,g_2])+2\sigma_2\oint_Sg_1g_2''',\\
\lb\E(f_1),\E(f_2)\rb&=-\f{1}{\ell^2}\left(\L([f_1,f_2])-2\sigma_2\oint_Sf_1f_2'''\right),
\ee
\esub
where $[f,g]=fg'-f'g$. The flat limit $\ell\to\infty$ of this algebra is well-defined. It reproduces as expected a centrally-extended BMS$_3$ algebra, where the central charges are functions of the couplings of the MB model. Note that, up to signs, they are the same as in the algebra \eqref{JP extended algebra}. For a finite cosmological constant, the algebra is most easily understood in Fourier modes, in terms of which it becomes
\bsub
\be
\lb\L_n,\E_m\rb&=\i(m-n)\E_{n+m}+(2\sigma_1-q\sigma_2)\i m^3\delta_{n+m,0},\\
\lb\L_n,\L_m\rb&=\i(m-n)\L_{n+m}+2\sigma_2\i m^3\delta_{n+m,0},\\
\lb\E_n,\E_m\rb&=\f{1}{\ell^2}\big(\i(m-n)\L_{n+m}+2\sigma_2\i m^3\delta_{n+m,0}\big).
\ee
\esub
Introducing
\be\label{Virasoro D}
\D^\pm_n\coloneqq\f{1}{2}\big(\ell\E_{\pm n}\pm\L_{\pm n}\big),
\ee
we obtain
\bsub
\be
\lb\D^\pm_n,\D^\pm_m\rb&=\i(m-n)\D^\pm_{n+m}+\f{c^\pm}{12}\i m^3\delta_{m+n,0},\\
\lb\D^\pm_n,\D^\mp_m\rb&=0,
\ee
\esub
which is a double Virasoro algebra with central charges $c^\pm=6\big(\ell(2\sigma_1-q\sigma_2)\pm2\sigma_2\big)$ depending once again non-trivial on the couplings of the MB model.

These central charges can be used to compute the entropy using the Cardy formula \cite{Blagojevic:2006hh,Cvetkovic:2018ati}. For this, let us first notice that in appendix \ref{appendix Fourier} we have defined the Fourier representation with angular integrals on $\varphi\in[0,1]$. We therefore first restore the factor of $2\pi$ in the central charges and introduce
\be
c^\pm=(2\pi)6\big(\ell(2\sigma_1-q\sigma_2)\pm2\sigma_2\big),
\q\q
D^\pm=\f{1}{2}(\ell E\pm L),
\ee
where $D^\pm$ is the zero-mode of \eqref{Virasoro D} defined in terms of \eqref{BTZ E and L}. With this the Cardy formula gives
\be
S=2\pi\left(\sqrt{\f{c^+D^+}{6}}+\sqrt{\f{c^-D^-}{6}}\right)=8\sigma_1\pi^2r_+-4\sigma_2\pi^2\left(qr_++\f{2r_-}{\ell}\right),
\ee
which matches the result derived from the first law.

In this section we have studied the asymptotic diffeomorphism charges for AdS spacetimes in Bondi gauge, and shown how the couplings of the MB model affect the central charges. The asymptotic algebras have two factors (i.e. the two Virasoros or the rotation/translation parts of BMS$_3$) because the asymptotic Killing vector fields are along $u$ and $\varphi$, and are centrally-extended because these vectors fields are not tangent. At finite distance and for tangent vector fields along $\varphi$ only, we have seen in \eqref{CD BMS} and \eqref{CD Virasoro} that the BMS and double Virasoro algebras exist because of the availability of the dual charges. We have also seen in section \ref{sec:dual charges} and with the twisted Sugawara construction that the algebra of dual charges also becomes centrally-extended in the case of non-tangent vector fields (although we have not studied the central extensions in details). Together, these results and observations point towards an important step for future work, namely the study of the \textit{asymptotic dual charges}. Were these charges to be non-trivial, they could extend the asymptotic symmetry algebra \eqref{Bondi symmetry algebra} by potentially doubling it and bringing new central extensions. This requires to first analyse whether there exist spacetimes and vector fields for which $\B(\xi)$ and/or $\C(\xi)$ can be defined to act non-trivially as asymptotic symmetries. We keep the detailed investigation of this question for future work.

\section{Metric, teleparallel, and massive sectors}
\label{sec:6}

We finally come to the last part of this work\footnote{This part is completely independent from the previous ones.}, which concerns the second order metric and teleparallel formulations of the MB model. The MB Lagrangian \eqref{MB Lagrangian} is first order in the connection and triad variables. As such, and as we have seen, it leads to two independent first order equations of motion. They can be read on \eqref{MB variation}, or rewritten in the elegant form \eqref{EOMs} displaying the sources of Lorentzian curvature and torsion. In such a theory, two natural questions then arise:
\begin{enumerate}
\item How to combine the two first order equations of motion into a single second order equation? What is the physical interpretation of this latter?
\item Can we inject part or all of the equations of motion back into the initial Lagrangian? What is the physical interpretation of the newly obtained Lagrangian?
\end{enumerate}
The goal of this section is to study these questions. Similar work was done in \cite{Dupuis:2019unm} in the case of the three-dimensional Hilbert--Palatini Lagrangian, and here we extend the results to the MB model with arbitrary couplings. As we are about to see, this is the source of new surprises, and in particular of a new description of three-dimensional massive gravity.

\subsection{Second order equations of motion}

We start by decomposing the connection as $\omega=\varpi+k$, where $\varpi$ is an arbitrary reference connection whose curvature will be denoted by $R$, and $k$ an arbitrary tensor\footnote{\label{quantum group paper}As a remark, we note that picking a non-dynamical internal vector $n$ (i.e. such that $\delta n=0$) and choosing $k=[e,n]$ leads to the Iwasawa decomposition which was used in \cite{Dupuis:2020ndx} to explain the classical origin of quantum group symmetry in three-dimensional gravity. This study can potentially be extended to the MB model, and it would be interesting to see how its symmetries are deformed.}
In terms of this decomposed connection, the MB Lagrangian becomes
\be\label{decomposed MB Lagrangian}
L_\text{MB}[e,\varpi+k]
&=L_\text{MB}[e,\varpi]+2\sigma_2k\wedge R+(2\sigma_1e+\sigma_2k)\wedge\de_\varpi k\cr
&\pe+\sigma_1e\wedge[k\wedge k]+\sigma_3k\wedge[e\wedge e]+\f{\sigma_2}{3}k\wedge[k\wedge k]+\sigma_2\de(k\wedge\varpi).
\ee
Notice that choosing $k=\kappa e$ gives again $L_\text{MB}[e,\varpi]$, but with new coupling constants. This is the ``stability'' property of the MB Lagrangian. The equations of motion obtained by varying the above decomposed Lagrangian with respect to $e$ and $k$ respectively are
\bsub
\be
&2\sigma_1R+2\sigma_3\de_\varpi e+2\sigma_1\de_\varpi k+\sigma_0[e\wedge e]+\sigma_1[k\wedge k]+2\sigma_3[k\wedge e]\approx0,\\
&2\sigma_2R+2\sigma_1\de_\varpi e+2\sigma_2\de_\varpi k+\sigma_3[e\wedge e]+\sigma_2[k\wedge k]+2\sigma_1[k\wedge e]\approx0.
\ee
\esub
Notice that varying with respect to $\varpi$ leads of course to the same equation of motion as varying along $k$. The equations of motion can be rewritten in the form
\be\label{general torsion EOM}
2R+2\de_\varpi k+[k\wedge k]+p[e\wedge e]\approx0,
\q\q
2\de_\varpi e+2[k\wedge e]+q[e\wedge e]\approx0,
\ee
which is evidently the same as plugging the decomposition $\omega=\varpi+k$ into \eqref{EOMs}.

Although the initial equations of motion were quite ``symmetrical'', we see that in \eqref{general torsion EOM} they are not exactly on the same footing anymore, as it is the second one, i.e. the torsion equation, which can easily be solved for $k$. In order to do so, let us consider the relation $[k\wedge e]=M$, where $M$ is an arbitrary Lie algebra-valued 2-form. In components, this is
\be
{\eps^i}_{jk}\big(k^j_\mu e^k_\nu-k^j_\nu e^k_\mu\big)=M^i_{\mu\nu}.
\ee
When the triad is invertible, this relation can be inverted to find
\be
k^i_\mu=\left(\delta^\alpha_\mu\delta^i_j-\f{1}{4}e^i_\mu\hat{e}^\alpha_j\right){\eps^j}_{kl}M^k_{\alpha\beta}\hat{e}^{\beta l}=\f{1}{2}{\eps^i}_{jl}\left(\delta^\alpha_\mu\delta^j_k+\f{1}{2}e_{\mu k}\hat{e}^{\alpha j}\right)M^k_{\alpha\beta}\hat{e}^{\beta l},
\ee
where $\hat{e}$ denotes the inverse triad. It will be convenient to write this in index-free notation with a linear operator such that $k=\mathbb{E}(M)$. With this compact notation at hand, solving the second equation of motion in \eqref{general torsion EOM} tells us that
\be\label{general solution for k}
k\approx-\f{q}{2}e-\mathbb{T},
\q\q
\mathbb{T}\coloneqq\mathbb{E}(\de_\varpi e),
\q\q
[\mathbb{T}\wedge e]=\de_\varpi e,
\ee
where we have defined the tensor $\mathbb{T}$ to be the image under the above map of the torsion of the reference connection. We can then inject this in the first equation of motion in \eqref{general torsion EOM} to find
\be\label{general on-shell EOM}
2R-2\de_\varpi\mathbb{T}+[\mathbb{T}\wedge\mathbb{T}]\approx\Lambda[e\wedge e].
\ee
This is the second order form of the equations of motion of the MB model. It encodes a duality, which is that between the metric and teleparallel formulations of gravity. When the reference connection is the torsionless Levi--Civita connection $\varpi=\Gamma$, the tensor $k$ is the contorsion, we have $\de_\varpi e=\de_\Gamma e=0=\mathbb{T}$, and we recover the metric equations of motion $R_{\mu\nu}\approx2\Lambda g_{\mu\nu}$. When the reference connection is a flat connection, $\varpi=\Upsilon=\vartheta\de\vartheta^{-1}$, we have $R=0$ and the quantity $\de_\varpi e=\de_\Upsilon e\neq0$ is the Weitzenb\"ock torsion. We recover in this case the teleparallel equations of motion with a cosmological constant \cite{Aldrovandi:2013wha}. We can see that in both cases the only parameter is $\Lambda$.

\subsection{Second and higher order Lagrangians}

Now that we have described the second order equations of motion of the MB model, we can ask which second order Lagrangian (if any) they follow from. This can be investigated by injecting the solution \eqref{general solution for k} into the original Lagrangian. In the usual case of Hilbert--Palatini gravity (i.e. for $\sigma_2=\sigma_3=0$), choosing the torsionless Levi--Civita connection as the reference connection leads to the second order Einstein--Hilbert action in triad variables. Choosing instead a flat reference connection leads to the teleparallel Lagrangian as in \cite{Dupuis:2019unm}. As we are about to see, such a manipulation in the MB model leads to a Lagrangian with drastically different properties from the initial first order one.

Let us consider again an arbitrary decomposition of the connection as $\omega=\varpi+k$, and inject the solution \eqref{general solution for k} of the torsion equation of motion into the decomposed Lagrangian \eqref{decomposed MB Lagrangian}. A lengthy rewriting making repeated use of $[\mathbb{T}\wedge e]=\de_\varpi e$ shows that this leads to the ``half on-shell'' Lagrangian
\be\label{half on-shell Lagrangian}
L_\text{h}[e,\varpi]
&\approx\f{1}{2}(2\sigma_1-q\sigma_2)\Big(L_\text{HP}[e,\varpi]-\Lambda L_\text{V}[e]-e\wedge\de_\varpi\mathbb{T}\Big)+\sigma_2L_\text{CS}[\varpi]\cr
&\pe-2\sigma_2\mathbb{T}\wedge R+\sigma_2\mathbb{T}\wedge\left(\de_\varpi\mathbb{T}-\f{1}{3}[\mathbb{T}\wedge\mathbb{T}]\right)+\de(\sigma_1e\wedge\mathbb{T}+\sigma_2k\wedge\varpi),
\ee
where for simplicity we have kept $k$ in the boundary term. This result is the general form of the MB Lagrangian on-shell of the torsion equation, for an arbitrary reference connection $\varpi$. Several important things can be observed.
\begin{enumerate}
\item Let us first consider the Riemannian metric geometry where $\varpi=\Gamma$ and $\mathbb{T}=0$. In this case, up to a boundary term we obtain the Lagrangian
\be\label{TMG}
L_\text{TMG}[e]=\f{1}{2}(2\sigma_1-q\sigma_2)\Big(L_\text{HP}[e]-\Lambda L_\text{V}[e]\Big)+\sigma_2L_\text{CS}[\Gamma],
\ee
which depends only on $e$ since the reference connection is the torsionless connection. Because of the presence of the CS term for $\Gamma(e)$, this Lagrangian is actually third order in derivatives of $e$. This is responsible for the appearance of a massive propagating mode. The equations of motion contain the Cotton tensor, and describe the propagation of a massive graviton. As already pointed out in \cite{Baekler:1992ab,Mielke:1991nn}, this result shows that the Riemannian metric sector of the MB model is a theory of massive gravity, known as topologically massive gravity (TMG) \cite{Deser:1981wh,Deser:1982vy}. The massless sector is obtained only for $\sigma_2=0$. This shows that the Riemannian metric equation of motion $2R=\Lambda[e\wedge e]$ in \eqref{general on-shell EOM} with an arbitrary $\Lambda$ \textit{cannot} be obtained from the Lagrangian \eqref{TMG}. It can only be obtained when $\sigma_2=0$.
\item In the teleparallel sector of the theory, i.e. when the reference connection is flat (let us choose the representative $\varpi=0$), up to a boundary term we obtain the Lagrangian
\be\label{TTMG}
L_\text{TTMG}[e]=\f{1}{2}(2\sigma_1-q\sigma_2)\Big(e\wedge\de\tilde{\mathbb{T}}-\Lambda L_\text{V}[e]\Big)+\sigma_2L_\text{CS}[\tilde{\mathbb{T}}],
\ee
where we have denoted $\tilde{\mathbb{T}}=-\mathbb{T}$. When $\sigma_2=0$, this is the Lagrangian for teleparallel gravity with a cosmological constant. When $\sigma_2\neq0$ however, we see the appearance of a Chern--Simons term for $\tilde{\mathbb{T}}$, and the Lagrangian \eqref{TTMG} is exactly the teleparallel dual of \eqref{TMG}. In particular, because of the CS term it is also of third order in derivatives of $e$. It is therefore natural to conjecture that this describes a teleparallel version of topologically massive gravity, which is why we have chosen to call this Lagrangian TTMG. The detailed study of this Lagrangian is deferred to future work.
\item When $\sigma_2=0$, up to a boundary term and an integration by parts the teleparallel Lagrangian in \eqref{half on-shell Lagrangian} can be written as $\de_\varpi e\wedge\mathbb{T}$. Since $\mathbb{T}=\mathbb{E}(\de_\varpi e)$, this Lagrangian can in turn be written in the suggestive form $\de_\varpi e\wedge\star\de_\varpi e$, where the non-standard $\star$ operation is defined by the action of the linear operator $\mathbb{E}$. This is the Yang--Mills form of the teleparallel Lagrangian \cite{Aldrovandi:2013wha}.
\item Finally, it is interesting to note that the couplings of \eqref{half on-shell Lagrangian} actually correspond to the central charges of \eqref{JP extended algebra} and \eqref{Bondi symmetry algebra}.
\end{enumerate}
We therefore see that inserting half of the equations of motion of the MB model back into the Lagrangian has a drastic effect: it leads to a third order Lagrangian describing a theory of three-dimensional massive gravity. This result was known for the Riemannian metric sector of the theory, described by \eqref{TMG}, and we have extended it to the dual teleparallel sector \eqref{TTMG}, which describes the TTMG dual of TMG.

There is a vast literature on three-dimensional massive gravity \cite{Deser:1981wh,Deser:1982vy,Deser:1984kw,Bergshoeff:2009aq,Bergshoeff:2009hq,Bergshoeff:2013xma,Alexandrov:2014oda,Bergshoeff:2015zga,Afshar:2014ffa,Bergshoeff:2014pca,Bergshoeff:2014bia,Merbis:2014vja,Geiller:2018ain,Geiller:2019dpc}. Among all the available models, that of \cite{Geiller:2018ain} resembles a bit the MB model. Indeed, the simple theory of massive gravity introduced in \cite{Geiller:2018ain} is a first order Lagrangian built with the connection and the triad, and contains all possible terms which can be written in terms of these two fields, much like in the MB model. The only difference with the MB model is that the Lagrangian of \cite{Geiller:2018ain} breaks internal Lorentz invariance, and this is the mechanism responsible for the appearance of a massive propagating mode. At the level of the MB Lagrangian \eqref{MB Lagrangian}, a two-field first order Lagrangian for massive gravity in the family of \cite{Geiller:2018ain} can be obtained by detuning any covariant derivative and replacing e.g. $\de_\omega e\mapsto\de e+\mu[\omega\wedge e]$.

As a last manipulation, we can now insert the second order equation of motion \eqref{general on-shell EOM} into the half on-shell Lagrangian \eqref{half on-shell Lagrangian}. To do so, notice that taking a wedge product of \eqref{general on-shell EOM} with $e$ implies that
\be
L_\text{HP}[e,\varpi]
&\approx3\Lambda L_\text{V}[e]+e\wedge[\mathbb{T}\wedge\mathbb{T}]-2\de(e\wedge\mathbb{T})\cr
&=3\Lambda L_\text{V}[e]+e\wedge\de_\varpi\mathbb{T}-\de(e\wedge\mathbb{T}),
\ee
while a wedge product with $\mathbb{T}$ gives
\be
2\mathbb{T}\wedge R\approx2\mathbb{T}\wedge\de_\varpi\mathbb{T}-\mathbb{T}\wedge[\mathbb{T}\wedge\mathbb{T}]+\Lambda\mathbb{T}\wedge[e\wedge e].
\ee
Using these two relations we can go once again on-shell in \eqref{half on-shell Lagrangian} to obtain the ``full on-shell'' Lagrangian
\be
L_\text{f}[e,\varpi]
&\approx(2\sigma_1-q\sigma_2)\Lambda L_\text{V}[e]+\sigma_2L_\text{CS}[\varpi]-\sigma_2\mathbb{T}\wedge\left(\de_\varpi\mathbb{T}-\f{2}{3}[\mathbb{T}\wedge\mathbb{T}]+\Lambda[e\wedge e]\right)+\sigma_2\de(k\wedge\varpi).
\ee
Once again, for $\sigma_2=0$ we recover the expected result, namely that the full on-shell Lagrangian is non-dynamical and just proportional to the volume term. For $\sigma_2\neq0$ however we obtain a dynamical Lagrangian with an independent Chern--Simons term for both the metric and teleparallel sectors.

\section{Perspectives}
\label{sec:7}

We have shown that the phase space of three-dimensional gravity contains two layers of rich dualities, which are elegantly revealed when considering the most general first order Lagrangian depending on the connection and triad variables. This Lagrangian is that of the Mielke--Baekler model \eqref{MB Lagrangian}, and it describes a theory with Lorentzian curvature and torsion measured by the parameters $p$ and $q$ respectively. This model therefore has two natural subspaces, corresponding to $(p\neq0,q=0)$ and $(p=0,q\neq0)$, which represents a first layer of duality between Lorentzian curvature and torsion. In each of these sectors, we have exhibited a second layer of duality encoded in the existence of new integrable charges, called the dual diffeomorphisms, which arise as particular field-dependent gauge transformations, or equivalently as quadratics in the Sugawara construction. We have first studied the algebra of these dual diffeomorphisms together with the usual diffeomorphisms in the generic case of finite boundaries with no boundary conditions, in which case integrability requires tangent vector fields. In the case $(p\neq0,q=0)$, the resulting centreless algebra consists in two commuting copies of Virasoro, and is defined even for a positive cosmological constant. In the flat limit it reduces to BMS$_3$. In the case $(p=0,q\neq0)$ the two Virasoro algebras are found to be in semi-direct sum, and also reduce to BMS$_3$ in the flat limit. To the best of our knowledge, it is the first time that such algebras, which are typically thought to be properties of asymptotic boundaries, are seen to appear at finite distance for arbitrary subregions and without boundary conditions (nor the use of symplectic symmetries \cite{Compere:2014cna,Compere:2015knw}).

This result is particularly interesting in light of recent work on asymptotic dual charges in four-dimensional gravity \cite{Godazgar:2018qpq,Godazgar:2018dvh,Godazgar:2019dkh,Godazgar:2020kqd,Godazgar:2020gqd,Oliveri:2020xls} (see also \cite{Kol:2019nkc,Kol:2020vet} for a soft theorem interpretation of these dual charges). It begs the question of whether the dual diffeomorphisms $\B$ and $\C$ which we have identified act non-trivially at infinity, and whether they can be expressed in terms of second order metric variables. Important future work will now be to understand these dual charges in the case of asymptotic boundaries, where they could further enlarge the asymptotic symmetry algebra. In particular, it would be interesting to study the dual charges using the most general flat and AdS$_3$ boundary conditions presented in \cite{Grumiller:2016pqb,Grumiller:2017sjh}.

Coming back to finite distance and arbitrary subregions, this enlargement of the boundary symmetry algebra could have important consequences for the quantization of gravity. The program initiated in \cite{Freidel:2020xyx,Freidel:2020ayo,Freidel:2020svx} aims at constructing quantum gravity from the quantization of the quasi-local algebra of charges, and from this perspective it is important to understand the nature and the physical significance of the largest available symmetry algebra. In terms of representation theory, a related question is that of the relationship between representations of the $\mathfrak{g}_\Lambda$ current algebras \eqref{JT extended algebra} or \eqref{JP extended algebra} and the algebra of diffeomorphisms. In principle, the current algebra seen as a universal enveloping algebra contains all possible polynomials in the currents, and therefore includes the diffeomorphisms and their dual. Understanding this at the level of representation theory is however still an open issue, especially given the fact that dual diffeomorphisms do not seem to have been discussed before.

In summary, by a detailed analysis of the covariant phase space and the boundary symmetries of the MB model, we have shown that if we consider the stable quadratics in the enveloping algebra \eqref{JT extended algebra}, at finite distance we are automatically lead to the double Virasoro algebras \eqref{CD Virasoro} and \eqref{BD Virasoro} (and their flat BMS$_3$ limit) depending on which sector of the duality between $(p\neq0,q=0)$ and $(p=0,q\neq0)$ we consider. It is noteworthy that \eqref{CD Virasoro} is a direct sum, while \eqref{BD Virasoro} is a semi-direct one, and that it is therefore \eqref{CD Virasoro} which is the finite distance counterpart of the asymptotic result \eqref{Bondi symmetry algebra} (minus the central extensions).

After having presented a detailed study of the dual charges at finite distance, we have discussed in section \ref{sec:6} the second order equations of motion arising from the MB model as well as the on-shell Lagrangians. This represents the second order version of the duality between Lorentzian curvature and torsion, which then becomes a duality between the Riemannian metric and teleparallel formulations of gravity. In the metric formulation the dynamics of general relativity is encoded in the curvature and the geometry is torsionless, while in the teleparallel formulation the geometry is flat and the dynamics is encoded in the torsion. The MB Lagrangian has the particularity that, when rewritten on-shell, it actually becomes third order and describes a propagating massive graviton. While this was known in the metric case \cite{Baekler:1992ab,Mielke:1991nn}, we have extended the result to the teleparallel dual. This has revealed a new candidate theory of massive gravity, described by the Lagrangian \eqref{TTMG} and called TTMG for teleparallel topologically massive gravity.

We close this article with a list of possible directions for future work. It would be particularly interesting to pursue these directions in the case of the MB model, because as we have seen here this latter reveals dual structures which are hard (if not impossible) to guess when working with the usual Hilbert--Palatini or Chern--Simons Lagrangians for three-dimensional gravity.

\begin{enumerate}
\item \textbf{Asymptotic dual charges.} As already explained above, it is now important to study whether the dual charges presented in this work have a realization at infinity and potential analogues in second order metric variables.
\item \textbf{Holography.} Boundary symmetries in three-dimensional gravity take a particularly important physical meaning when studying holography and the boundary CFT \cite{Balasubramanian:1999re,Fareghbal:2013ifa}. Three-dimensional gravity being exactly soluble, one can expect it to always be dual to a CFT or an integrable model, and to admit a quasi-local holographic description as constructed in \cite{Dittrich:2018xuk,Dittrich:2017hnl,Dittrich:2017rvb,Goeller:2019apd,Asante:2019ndj}. Some aspects of holography in the MB model were studied previously in \cite{Blagojevic:2013bu}. It would be interesting to push this study further, and to investigate whether the dual gravitational charges play a role in this holographic description. In this respect, it is particularly interesting to note that we have obtained, thanks to the availability of the dual charge $\C$ in the case $(p\neq0,q=0)$, a double Virasoro algebra defined for all values of the cosmological constant $\Lambda=-p\neq0$. This could play an important role in the understanding of dS holography.
\item \textbf{Boundary dynamics.} An interesting property of topological three-dimensional gravity is the possibility of writing down a boundary dynamics invariant under the boundary symmetries. Following the seminal work in AdS$_3$ spacetimes \cite{Coussaert:1995zp,Banados:1998ta}, this has been generalized to dS and the flat case \cite{Barnich:2012rz,Barnich:2013yka,Compere:2014cna} (see also \cite{Carlip:2016lnw}). This boundary dynamics is typically obtained via a reduction from the WZNW boundary theory. It would be enlightening to perform this construction in the case of the MB model.
\item \textbf{Relation between first order and metric charges.} One intriguing property of the MB model is that a given metric geometry obeying the second order equation of motion \hbox{$2R=\Lambda[e\wedge e]$} can actually be obtained from several inequivalent first order equations of motion. Indeed, since $\Lambda=-(p+q^2/4)$, there are actually infinitely many choices of $(p,q)$ leading to a given $\Lambda$. While this in itself might seem innocent, it becomes more intriguing in light of the fact that the sectors $(p\neq0,q=0)$ and $(p=0,q\neq0)$ have different dual charges with a slightly different algebraic structure (i.e. a direct vs semi-direct sum). The flat case is even more puzzling, since it can be obtained with $p=q=0$, where the dual charges $\B$ and $\C$ exist and agree, but also for $4p=-q^2$, where a priori there is no well-defined dual charge (see however the next point below). If the dual charges turn out to exist in the metric theory, this therefore raises the question of which version of the first order dual charges they follow from.
\item \textbf{Dual charges for all couplings.} As we have explained in section \ref{sec:3}, the dual charges $\C$ and $\B$ (and even a version of this latter with $\T$ replaced by $\J$, which we could call $\A$) can actually \textit{always} be made integrable on the phase space. It is the additional requirement of stability of the action of the dual charges on $\T(\phi)$ which has singled out the consistency conditions $q=0$ and $p=0$ for $\C$ and $\B$ respectively, and prevented $\A$ from existing. One can therefore wonder if it is possible to define dual charges in the case $(p,q)\neq0$, i.e. for arbitrary couplings of the MB model. This would however evidently require a deeper understanding of the enveloping algebra and of the ``open index quadratics'' generated e.g. in \eqref{CT Fourier bracket}.
\item \textbf{Coupling to point particles.} Three-dimensional gravity can easily be coupled to massive spinning point particles  \cite{Carlip:1989nz,MR1080700,Hooft:1993gz,Hooft:1993nj,Matschull:1997du,Matschull:2001ec}. In the first order formulation, the mass acts as a source of curvature, and is therefore a distributional contribution to $p$, while the spin is a source of torsion and contributes to $q$. The MB model, which treats curvature and torsion on an equal footing, could therefore provide a new understanding of the coupling of three-dimensional gravity to point particles. One question is whether the dualities studied in this paper extend in the presence of point particles, and if so which physical meaning they acquire. Moreover, particles can also be understood in terms of boundary defects and excitations \cite{Buffenoir:2003zu,Dittrich:2016typ}, and carry representations of some deformation of the symmetry algebra \cite{Bais:2002ye,Noui:2004iz,Noui:2006ku,Schroers:2007ey} (typically a quantum group or quantum double). Now that we have unraveled the structure of the bulk and boundary symmetries of the MB model, one can wonder what type of deformation plays a role in the quantum theory and the coupling to particles. This is in turn related to the next two points below.
\item \textbf{Quantum groups and deformations.} It is known since the seminal work of Witten that quantum groups appear upon quantization of three-dimensional gravity \cite{MR974271,Witten:1988hf,Witten:1989rw}, and in a related manner in the presence of particles \cite{Bais:2002ye,Noui:2006ku,Schroers:2007ey,Meusburger:2008bs}. It was recently proposed that quantum groups have a classical underpinning in terms of classical deformed symmetries \cite{Dupuis:2020ndx}. A first step towards understanding quantum groups in the MB model could therefore be to reproduce the classical analysis of \cite{Dupuis:2020ndx} (see footnote \ref{quantum group paper}). The understanding of quantum groups is related to the issue of discretizing and regularizing the current algebra of boundary charges. First steps in this direction were taken in \cite{FGL-toappear} in the flat case, and should be extended to the full MB model in order to understand the fate of the dualities in the discrete setting.
\item \textbf{Quantization.} Three-dimensional gravity admits many complementary quantization schemes \cite{Carlip:1993zi,Carlip:1993ze,Carlip:1994ap}, but it is not clear which one could be most adapted to quantizing the MB model. In first order connection and triad variables, it is particularly useful to use the techniques of canonical loop quantum gravity and spin foam models \cite{Noui:2004ja,Alexandrov:2011ab}. However, this has so far only been studied in the case of the usual Lagrangian obtained from \eqref{MB Lagrangian} with $\sigma_2=\sigma_3=0$. Moreover, these constructions depend crucially on the signature and the sign of the cosmological constant. In the flat case the partition function is described by the Ponzano--Regge model \cite{Freidel:2000uq,Freidel:2004vi,Barrett:2008wh}, while in Euclidean signature with positive cosmological constant it is given by the Turaev--Viro model at root of unity \cite{1992PhRvL..68.1795M,Turaev:1992hq}. The latter being a quantum deformation of the former related to the passage from $(p,q)=0$ to $(p\neq0,q=0)$, it is natural to ask if there is a quantum deformation which could describe the dual case $(p=0,q\neq0)$. A dual loop quantization of three-dimensional gravity was proposed in \cite{Delcamp:2018sef}, and shown to lead to Dijkgraaf--Witten models, but there the duality was proposed in the case $(p,q)=0$ only. In full generality, the question is that of the role of the couplings of the MB model as deformation parameters in the quantum theory, and how classical dualities are lifted to the quantum theory.
\item \textbf{Extension to four- and two-dimensional gravity.} A natural question is that of the extension of the present work to four-dimensional gravity. There are several aspects to this. First, one could try to construct dual diffeomorphisms by mimicking what has been done in section \ref{sec:dual diffeo C}. This seems subtle because, since four-dimensional gravity is not topological, it has no analogue of relation \eqref{diffeos as on-shell gauge}. Even in four-dimensional topological BF theory, the difference of form degree between the connection $\omega$ and the $B$ field prevents from considering charges with field-dependent smearings where the fields $\omega$ and $B$ are swapped. Second, we could instead consider in the Lagrangian of four-dimensional first order gravity all possible terms which are allowed by Lorentz and diffeomorphism invariance. It has already been suggested that such terms lead to dual asymptotic charges \cite{Godazgar:2018qpq,Godazgar:2018dvh,Godazgar:2019dkh,Godazgar:2020kqd,Godazgar:2020gqd,Oliveri:2020xls}, and the systematic investigation of general results at finite distance has been initiated in \cite{Freidel:2020xyx,Freidel:2020svx,Freidel:2020ayo}. In these last references it has been demonstrated that working in first order variables and adding topological terms to the Lagrangian changes drastically the physical content of the finite distance charge algebra. Finally, one could also contemplate going towards lower dimensions, and investigating the existence and the role of dual diffeomorphisms in two-dimensional gravity (which also admits a first order formulation as a BF theory \cite{Grumiller:2002nm}). A first step in this direction could be to perform a dimensional reduction from the MB model to two-dimensional spacetimes, along the lines of \cite{Mertens:2018fds,Gaikwad:2018dfc}.
\item \textbf{Teleparallel topologically massive gravity.} We have conjectured that \eqref{TTMG} is the teleparallel dual to the third order Lagrangian \eqref{TMG} of topologically massive gravity (TMG), and proposed the name teleparallel topologically massive gravity (TTMG). This is very natural given the construction of section \ref{sec:6} and how these two Lagrangians were obtained. For \eqref{TTMG} to earn its name, it remains however to rigorously prove that it is indeed a theory of massive gravity. For this, we should first analyse the linearized theory around a flat background and derive the equation for the propagating graviton. Then, we should perform a full non-linear analysis by counting the degrees of freedom in the Hamiltonian formalism. To our knowledge there is so far no teleparallel description of massive gravity. This could be interesting in its own right, but also for potential extensions to four-dimensional massive gravity, where theoretical model building is much more constrained. Finally, we mention that aspects of holography in TMG were studied in \cite{Skenderis:2009nt}, and could also be extended to TTMG, as well as the study of the asymptotic symmetries and charges.
\item \textbf{Extensions of the MB model.} Finally, we note that the MB model has been generalized to include e.g. higher spin fields \cite{Peleteiro:2020ubv} or Maxwell algebras \cite{Concha:2018zeb,Adami:2020xkm}. The study of the covariant phase space and of the dual diffeomorphisms could be extended to these models, as well as that of the teleparallel formulation along the lines of section \ref{sec:6}.
\end{enumerate}

\section*{Acknowledgements}

We would like to thank Glenn Barnich, Laurent Freidel, Etera Livine, and Antony Speranza for discussions and comments, Geoffrey Comp\`ere for pointing out references and bringing symplectic symmetries to our attention, and Omar Vadivia for pointing out a useful reference. N. M. was funded by the Fondecyt grant 11180894.

\appendix

\section{Notations and conventions}
\label{app:A}

\paragraph{Indices}~\\
Spacetime indices are denoted by $\mu,\nu,\ldots\in\{0,1,2\}$ and spatial indices by $a,b,\ldots\in\{1,2\}$. Internal indices are denoted by $i,j,\ldots\in\{0,1,2\}$, and are lowered and raised with the $\mathfrak{so}(1,2)$ metric $\eta_{ij}$. Anti-symmetrization of $n$ indices is defined with a factor $1/n!$.

\paragraph{Levi--Civita symbol and tensor}~\\
We denote by $\teps_{\mu\nu\rho}$ and $\teps^{\mu\nu\rho}=\teps_{\mu\nu\rho}$ the densities of weight $-1$ and $+1$ respectively, which we define to be such that $\teps_{012}=1$ in every coordinate system. We then define the true tensor (i.e. the density of weight 0) $\eps_{\mu\nu\rho}\coloneqq\sqrt{|g|}\teps_{\mu\nu\rho}$, so that with our convention the tensor with upper indices is
\be
\eps^{\mu\nu\rho}=g^{\mu\alpha}g^{\nu\beta}g^{\rho\gamma}\eps_{\alpha\beta\gamma}=g^{\mu\alpha}g^{\nu\beta}g^{\rho\gamma}\sqrt{|g|}\teps_{\alpha\beta\gamma}=g^{-1}\sqrt{|g|}\teps_{\mu\nu\rho}=\f{\text{sgn}(g)}{\sqrt{|g|}}\teps_{\mu\nu\rho}=\f{\text{sgn}(g)}{\sqrt{|g|}}\teps^{\mu\nu\rho}.
\ee
For internal indices we also use the tensor $\eps_{ijk}$ defined by $\eps_{012}=1$, which satisfies
\bsub
\be
\eps_{ijk}\eps^{lmn}&=-\big(\delta^l_i\delta^m_j\delta^n_k-\delta^l_i\delta^m_k\delta^n_j+\delta^m_i\delta^n_j\delta^l_k-\delta^m_i\delta^n_k\delta^l_j+\delta^n_i\delta^l_j\delta^m_k-\delta^n_i\delta^l_k\delta^m_j\big),\label{epsilon1}\\
\eps_{ijk}\eps^{lmk}&=-\big(\delta^l_i\delta^m_j-\delta^l_j\delta^m_i\big),\label{epsilon2}\\
\eps_{ijk}\eps^{ljk}&=-2\delta^l_i,\\
\eps_{ijk}\eps^{ijk}&=-3!.
\ee
\esub
With these we have that
\be\label{determinant e}
|e|\coloneqq\det(e^i_\mu)=\f{1}{6}\teps^{\mu\nu\rho}\eps_{ijk}e^i_\mu e^j_\nu e^k_\rho,
\q\q
|e|\eps^{ijk}=-\teps^{\mu\nu\rho}e^i_\mu e^j_\nu e^k_\rho.
\ee

\paragraph{Differential forms}~\\
When written in terms of components, a $p$-form is given by
\be
P=\f{1}{p!}P_{\mu_1\dots\mu_p}\de x^{\mu_1}\wedge\dots\wedge\de x^{\mu_p}.
\ee
The wedge product between a $p$-form $P$ and a $q$-form $Q$ has components given by
\be
(P\wedge Q)_{\mu_1\dots\mu_{p+q}}=\f{(p+q)!}{p!q!}P_{[\mu_1\dots\mu_p}Q_{\mu_{p+1}\dots\mu_q]}.
\ee
Introducing the notation $\de^dx\coloneqq\de x^1\wedge\dots\wedge\de x^d$, we have
\be
\de x^{\mu_1}\wedge\dots\wedge\de x^{\mu_d}=\teps_{\mu_1\dots\mu_d}\de^dx=\teps^{\mu_1\dots\mu_d}\de^dx=\eps^{\mu_1\dots\mu_d}\text{sgn}(g)\sqrt{|g|}\de^dx.
\ee
In particular, for $d=3$ we therefore have that
\be
e\wedge F
&=\f{1}{3!}(e\wedge F)_{\mu\nu\rho}\de x^\mu\wedge\de x^\nu\wedge\de x^\rho\cr
&=\f{1}{3!}(e\wedge F)_{\mu\nu\rho}\teps^{\mu\nu\rho}\de^3x\cr
&=\f{1}{3!}\f{(1+2)!}{1!2!}e_{[\mu}F_{\nu\rho]}\teps^{\mu\nu\rho}\de^3x\cr
&=\f{1}{2}e_\mu F_{\nu\rho}\teps^{\mu\nu\rho}\de^3x,
\ee
and
\be
e\wedge[e\wedge e]=e_\mu[e_\nu,e_\rho]\teps^{\mu\nu\rho}\de^3x.
\ee

\paragraph{Index-free notation}~\\
In our notation there is always an implicit pairing of Lie algebra indices, and $[\cdot\,,\cdot]$ denotes the Lie algebra commutator. Explicitly, the corresponding internal indices are
\be
e\wedge F=\eta_{ij}e^i\wedge F^j,
\q\q
e\wedge[e\wedge e]=\eta_{ij}e^i\wedge[e\wedge e]^j=\eps_{ijk}e^i\wedge e^j\wedge e^k.
\ee
The curvature and torsion are given in index-free notation by
\be
F=\de\omega+\f{1}{2}[\omega\wedge\omega],
\q\q
\de_\omega e=\de\omega+[\omega\wedge e],
\ee
which in components translates to
\be
F^i_{\mu\nu}=\partial_\mu\omega^i_\nu-\partial_\nu\omega^i_\mu+{\eps^i}_{jk}\omega^j_\mu\omega^k_\nu,
\q\q
(\de_\omega e^i)_{\mu\nu}=\partial_\mu e^i_\nu-\partial_\nu e^i_\mu+{\eps^i}_{jk}(\omega^j_\mu e^k_\nu-\omega^j_\nu e^k_\mu).
\ee
When switching from the fundamental representation of the Lie algebra to the adjoint representation, we have introduced $\omega^i_\mu$ such that $\omega^{ij}_\mu\eqqcolon-{\eps^{ij}}_k\omega^k_\mu$, and therefore
\be
F^{ij}_{\mu\nu}=\partial_\mu\omega^{ij}_\nu-\partial_\nu\omega^{ij}_\mu+\omega^i_{\mu k}\omega^{kj}_\nu-\omega^i_{\nu k}\omega^{kj}_\mu=-{\eps^{ij}}_k\big(\partial_\mu\omega^k_\nu-\partial_\nu\omega^k_\mu+{\eps^k}_{mn}\omega^m_{\mu}\omega^n_\nu\big)=-{\eps^{ij}}_kF^k_{\mu\nu}.
\ee
For $(p,q,r)$-forms $(P,Q,R)$ we have
\be
[P\wedge Q]\wedge R=(-1)^{(p+q)r}[R\wedge P]\wedge Q,
\q\q
[P\wedge Q]=(-1)^{pq+1}[Q\wedge P],
\ee
the Leibniz rule
\be
\de_\omega[P\wedge Q]=[\de_\omega P\wedge Q]+(-1)^p[P\wedge\de_\omega Q],
\ee
the integration by parts
\be
\de_\omega P\wedge Q=\de(P\wedge Q)+(-1)^{p+1}P\wedge\de_\omega Q,
\ee
and the squared action of the covariant derivative
\be
\de_\omega\de_\omega P=[F\wedge P].
\ee
For two 1-forms $e$ and $\omega$, and two 0-forms $\alpha$ and $\beta$, we have the useful formulas
\be
[\alpha,e]\wedge[\beta,\omega]+[\alpha,\omega]\wedge[\beta,e]=[\alpha,\beta][e\wedge\omega],
\q\q
\de_\omega\alpha\wedge\de_\omega\beta=[\alpha,\beta]F+\de(\alpha\de_\omega\beta).
\ee

\paragraph{Einstein--Hilbert Lagrangian}~\\
With these notations and conventions, using $\sqrt{|g|}=|e|$ and $2|e|\hat{e}^{[\mu}_i\hat{e}^{\nu]}_j=\teps^{\mu\nu\rho}\eps_{ijk}e^k_\rho$, where $\hat{e}$ denotes the inverse triad, we get that
\be\label{EH and HP Lagrangians}
\sqrt{|g|}\,R
&=|e|g^{\mu\rho}g^{\nu\sigma}R_{\mu\nu\rho\sigma}\cr
&=|e|\hat{e}^\mu_i\hat{e}^\rho_k\hat{e}^\nu_j\hat{e}^\sigma_lR_{\mu\nu\rho\sigma}\eta^{ik}\eta^{jl}\cr
&=|e|\hat{e}^{[\mu}_i\hat{e}^{\nu]}_j\hat{e}^\rho_k\hat{e}^\sigma_lR_{\mu\nu\rho\sigma}\eta^{ik}\eta^{jl}\cr
&=|e|\hat{e}^{[\mu}_i\hat{e}^{\nu]}_jF^{ij}_{\mu\nu}\cr
&=\f{1}{2}\teps^{\mu\nu\rho}\eps_{ijk}e^k_\rho F^{ij}_{\mu\nu}\cr
&=\teps^{\mu\nu\rho}e^i_\mu F^j_{\nu\rho}\eta_{ij}\cr
&=2e\wedge F,
\ee
where we have omitted the $\de^3x$ for simplicity. Note that in the main text we reserve $R$ for the curvature of the torsionless connection.

\section{Covariant phase space}
\label{app:B}

Here we review basic facts about the covariant phase space, which we use repeatedly in the main text. We first present the notations and conventions, and then the general formula for the Hamiltonian charges. We follow the logic of \cite{Harlow:2019yfa,Geiller:2019bti,Freidel:2020xyx}, which is to pay attention to possible corner terms in the boundary conditions.

\subsection{Notations and conventions}

Consider a Lagrangian, which may contain a boundary term, and which we will write as $L+\de\ell$. The variation of this Lagrangian is
\be
\delta(L+\de\ell)=\text{E}\wedge\delta\Phi+\de(\theta+\delta\ell),
\ee
where $\text{E}$ are the equations of motion, $\Phi$ is the set of fields, and $\theta=\theta[\delta]$ is the potential $(d-1,1)$-form coming from the bulk Lagrangian. Here it appears shifted by the variation of the boundary term. The symplectic current is the field-space variation of the total potential, namely
\be
\delta(\theta+\delta b)=\delta\theta=\delta\theta[\delta]=\delta_1\theta[\delta_2]-\delta_2\theta[\delta_1],
\ee
where we have made explicit the fact that variations in field-space are anti-symmetrized, and used the fact that $\delta^2=0$. From the variation of the Lagrangian, one can see that $\de\delta\theta\approx0$. Integrating this identity on a spacetime manifold $M$ with boundary $\partial M=\Sigma_1\cup\Sigma_2\cup\Delta$, where $\Delta$ is the time-like boundary, we get that
\be
\int_M\de\delta\theta=\left(\oint_{\Sigma_1}-\oint_{\Sigma_2}+\oint_\Delta\right)\delta\theta\approx0.
\ee
For the symplectic structure
\be
\Omega\coloneqq\int_\Sigma\delta\theta
\ee
to be independent of the slice $\Sigma$, we see that we have to require the vanishing of symplectic flux through $\Delta$. In order to understand how this can be achieved in general, let us go back to the variational principle. For this latter to be well-defined in the variation of the action, the most general boundary condition which we can consider is
\be
(\theta+\delta b)\big|_\Delta=\de c,
\ee
for some $(d-2,1)$-form $c$ on the corner. Note that this is a condition on the components (say) $\theta_{t\varphi}$ along $\Delta$. This is the most general boundary condition but also, consistently, the most general condition for the vanishing of symplectic flux, provided we take into account the corner term arising from $\partial\Delta=S_1\cup S_2$ in the case $c\neq0$. Indeed, in this case the conserved symplectic structure acquires a boundary term and is
\be
\Omega=\int_\Sigma\delta\theta-\oint_S\delta c.
\ee
This is simply saying that $\theta-\de c$ is the potential which leads to a conserved symplectic structure in the case $c\neq0$.

Once we have defined the symplectic structure $(0,2)$-form, we can contract it with variations in field-space such as gauge transformations $\delta_\epsilon$. Heuristically, viewing these gauge transformations as vectors and the field-space variations as forms, the contraction is
\be
\delta_\epsilon\ipp\Omega=\delta_\epsilon\ipp\big(\Omega[\delta_1,\delta_2]\big)=\Omega[\delta_\epsilon,\delta]-\Omega[\delta,\delta_\epsilon].
\ee
Analogously to what happens when contracting vectors with differential forms, which lowers the spacetime form degree, one can see that the contraction of the transformation $\delta_\epsilon$ with the $(0,2)$-form $\Omega$ lowers the field-space degree, and produces a $(0,1)$-form.

\subsection{Noether and Hamiltonian charges}

Consider $\delta_\epsilon L$ with $\epsilon\in(\xi,\alpha)$ either a diffeomorphism $\xi$ or an internal gauge transformation with parameter $\alpha$. Suppose that we have $\delta_\epsilon L=\de m$. The variation of the Lagrangian is
\be
\delta L=\text{E}\wedge\delta\Phi+\de\theta.
\ee
Taking the variation to be a gauge transformation, we get
\be
\delta_\epsilon L=\text{E}\wedge\delta_\epsilon\Phi+\de(\delta_\epsilon\ipp\theta)=\de m,
\ee
which leads to
\be
\text{E}\wedge\delta_\epsilon\Phi+\de(\delta_\epsilon\ipp\theta-m)=0.
\ee
The Noether current $(d-1,0)$-form is defined as
\be
J\coloneqq\delta_\epsilon\ipp\theta-m,
\ee
so on-shell we have $\de J\approx0$, which implies that $J\approx\de Q$ where the $(d-2,0)$-form $Q$ is the Noether charge. We also have
\be
\text{E}\wedge\delta_\epsilon\Phi=\epsilon N-\de P=-\de P
\ee
by Noether's second theorem, so we can write
\be
J=P+\de Q.
\ee
The quantity $P$ is therefore the off-shell part of the Noether current. We then get
\be
\slashed{\delta}\H(\epsilon)
&=-\delta_\epsilon\ipp\Omega\cr
&=\int_\Sigma\delta(\delta_\epsilon\ipp\theta)-\delta_\epsilon\theta-\oint_S\delta(\delta_\epsilon\ipp c)-\delta_\epsilon c\cr
&=\int_\Sigma\delta P+\delta m-\delta_\epsilon\theta+\oint_S\delta(Q-\delta_\epsilon\ipp c)+\delta_\epsilon c.
\ee
Now, consider the equality
\be
0=\delta\delta_\epsilon L-\delta_\epsilon\delta L=\de(\delta m-\delta_\epsilon\theta),
\ee
which holds because $\delta_\epsilon(\text{E}\wedge\delta\Phi)=0$. This implies that there exists a $(d-2,1)$-form $M$ such that
\be
\delta m-\delta_\epsilon\theta=\de M.
\ee
In particular, for a diffeomorphism we have $\delta_\epsilon=\pounds_\xi$ and $m=\xi\ip L$, which implies that
\be
\delta m-\delta_\epsilon\theta=\xi\ip(\text{E}\wedge\delta\Phi)+\xi\ip\de\theta-\pounds_\xi\theta=\xi\ip(\text{E}\wedge\delta\Phi)-\de(\xi\ip\theta),
\ee
and we get
\be
\slashed{\delta}\D(\xi)
&=\int_\Sigma\delta P+\xi\ip(\text{E}\wedge\delta\Phi)+\oint_S\delta(Q-\pounds_\xi\ipp c)+\pounds_\xi c-\xi\ip\theta\cr
&=\int_\Sigma\delta P+\xi\ip(\text{E}\wedge\delta\Phi)+\oint_S\delta(Q-\pounds_\xi\ipp c)+\xi\ip(\de c-\theta)\cr
&=\int_\Sigma\delta P+\xi\ip(\text{E}\wedge\delta\Phi)+\oint_S\delta(Q+\xi\ip b-\pounds_\xi\ipp c).
\ee
This gives an integrable charge
\be\label{D with b and c}
\D(\xi)=\oint_SQ(\xi)+\xi\ip b-\pounds_\xi\ipp c.
\ee
In the case where $b=0=c$, we have instead the usual formula
\be
\slashed{\delta}\D(\xi)=\int_\Sigma\delta P+\xi\ip(\text{E}\wedge\delta\Phi)+\oint_S\delta Q-\xi\ip\theta,
\ee
which we use in the main text when studying arbitrary diffeomorphisms without boundary conditions.

Under the assumption that $b$ is covariant, we can compute the bracket of the charges \eqref{D with b and c} along the lines of \cite{Speranza:2017gxd}, we first note that
\be
\pounds_\xi\ipp\big(\delta Q(\zeta)\big)
&=\pounds_\xi\ipp(\zeta\ip\delta\omega e+\zeta\ip\omega\delta e)\cr
&=\zeta\ip\pounds_\xi\omega e+\zeta\ip\omega\pounds_\xi e\cr
&=\pounds_\xi(\zeta\ip\omega e)-[\pounds_\xi,\zeta\ip]\omega e\cr
&=\pounds_\xi(\zeta\ip\omega e)-[\xi,\zeta]\ip\omega e\cr
&=\pounds_\xi Q(\zeta)-Q([\xi,\zeta]),
\ee
which we have here illustrated in intermediate steps with the example of the MB charge for $2\sigma_1=1$ and $\sigma_2=\sigma_3=0$. Furthermore, the Noether charge is defined by
\be
\pounds_\xi\ipp\theta-\xi\ip L=\pounds_\xi\ipp(\de c-\delta b)-\xi\ip L=\pounds_\xi\ipp\de c-\pounds_\xi b-\xi\ip L\approx\de Q(\xi).
\ee
With this we find the Poisson bracket
\be\label{app: diffeo central charge}
\lb\D(\xi),\D(\zeta)\rb
&=\pounds_\xi\ipp\big(\delta\D(\zeta)\big)\cr
&=\oint_S\pounds_\xi\ipp\big(\delta Q(\zeta)+\zeta\ip\delta b-\delta(\pounds_\zeta\ipp c)\big)\cr
&=-\oint_SQ([\xi,\zeta])-\pounds_\xi Q(\zeta)-\zeta\ip\pounds_\xi b+\pounds_\xi(\pounds_\zeta\ipp c)\cr
&=-\oint_SQ([\xi,\zeta])-\xi\ip\de Q(\zeta)-\zeta\ip\pounds_\xi b+\xi\ip\de(\pounds_\zeta\ipp c)\cr
&\approx-\oint_SQ([\xi,\zeta])+\xi\ip\pounds_\zeta b+\xi\ip\zeta\ip L-\zeta\ip\pounds_\xi b+\xi\ip\big(\de(\pounds_\zeta\ipp c)-\pounds_\zeta\ipp\de c\big)\cr
&=-\oint_SQ([\xi,\zeta])+[\xi,\zeta]\ip b+\pounds_\zeta(\xi\ip b)+\xi\ip\zeta\ip L-\zeta\ip\pounds_\xi b\cr
&=-\D([\xi,\zeta])-\oint_S\xi\ip\zeta\ip(L+\de b).
\ee
This is the result for the bracket of the integrable diffeomorphism charges \eqref{D with b and c} under the assumption that $b$ is covariant. It shows how the contributions in addition to the Noether charge in the Hamiltonian charge are responsible for the appearance of a central term in the algebra. Note however that this central extension is \textit{not} the one of e.g. the BMS or Virasoro algebra derived in section \ref{sec:5}, as can be checked by an explicit calculation. This is because covariance of $b$ was assumed in order to derive \eqref{app: diffeo central charge}. Instead, a more general formula is given in \cite{Chandrasekaran:2020wwn}. It would be interesting to apply it on an explicit example, and show that it indeed reproduces e.g. the BMS or Virasoro central charge.

\section{Brackets and algebra}
\label{app:C}

\subsection[Poisson brackets involving $\C(\xi)$]{Poisson brackets involving $\boldsymbol{\C(\xi)}$}
\label{app: C brackets}

To compute the brackets appearing in \eqref{C brackets} we use $\sigma_3=p\sigma_2$ and a tangential vector field $\xi$. With the definition $\C(\xi)=p\J(\xi\ip e)+\T(\xi\ip\omega)$, for the bracket \eqref{CJ bracket} we find
\be
\lb\C(\xi),\J(\alpha)\rb
&=p\lb\J(\xi\ip e),\J(\alpha)\rb+\lb\T(\xi\ip\omega),\J(\alpha)\rb\cr
&=p\J([\xi\ip e,\alpha])+\T([\xi\ip\omega,\alpha])-2\oint_S\big(\sigma_1(\xi\ip\omega)+p\sigma_2(\xi\ip e)\big)\de\alpha\cr
&=-2\oint_S\big(\sigma_1(\xi\ip\omega)+\sigma_3(\xi\ip e)\big)\de\alpha\cr
&=-2\oint_S(\sigma_1\omega+\sigma_3e)(\xi\ip\de\alpha)\cr
&=-\T(\pounds_\xi\alpha),
\ee
where we have used the tangentiality of $\xi$ to write
\be
\T([\xi\ip\omega,\alpha])
&=2\oint_S[\xi\ip\omega,\alpha](\sigma_1\omega+\sigma_3e)\cr
&=2\oint_S[\xi\ip\omega,\alpha]\sigma_3e\cr
&=2p\oint_S[\xi\ip\omega,\alpha]\sigma_2e\cr
&=-2p\oint_S[\xi\ip e,\alpha]\sigma_2\omega\cr
&=-2p\oint_S[\xi\ip e,\alpha](\sigma_1e+\sigma_2\omega)\cr
&=-p\J([\xi\ip e,\alpha]).
\ee
Similarly, for the bracket \eqref{CT bracket} we find
\be
\lb\C(\xi),\T(\phi)\rb
&=p\lb\J(\xi\ip e),\T(\phi)\rb+\lb\T(\xi\ip\omega),\T(\phi)\rb\cr
&=p\T([\xi\ip e,\phi])+p\J([\xi\ip\omega,\phi])+q\T([\xi\ip\omega,\phi])-2\oint_S\big(p\sigma_1(\xi\ip e)+\sigma_3(\xi\ip\omega)\big)\de\phi\cr
&=q\T([\xi\ip\omega,\phi])-2p\oint_S\big(\sigma_1(\xi\ip e)+\sigma_2(\xi\ip\omega)\big)\de\phi\cr
&=-p\J(\pounds_\xi\phi)+q\T([\xi\ip\omega,\phi]),
\ee
where for the third equality we have used the tangentiality of $\xi$ to write
\be
\T([\xi\ip e,\phi])
&=2\oint_S[\xi\ip e,\phi](\sigma_1\omega+\sigma_3e)\cr
&=2\oint_S[\xi\ip e,\phi]\sigma_1\omega\cr
&=-2\oint_S[\xi\ip\omega,\phi]\sigma_1e\cr
&=-2\oint_S[\xi\ip\omega,\phi](\sigma_1e+\sigma_2\omega)\cr
&=-\J([\xi\ip\omega,\phi]).
\ee
In order to compute the bracket \eqref{CD bracket}, we write
\be
\lb\C(\xi),\D(\zeta)\rb
&=-\lb\D(\zeta),\C(\xi)\rb\cr
&=-p\lb\D(\zeta),\J(\xi\ip e)\rb-\lb\D(\zeta),\T(\xi\ip\omega)\rb\cr
&=p\J\big(\pounds_\zeta(\xi\ip e)\big)-p\J\big(\xi\ip(\pounds_\zeta e)\big)+\T\big(\pounds_\zeta(\xi\ip\omega)\big)-\T\big(\xi\ip(\pounds_\zeta\omega)\big)\cr
&=-p\J([\xi,\zeta]\ip e)-\T([\xi,\zeta]\ip\omega)\cr
&=-\C([\xi,\zeta]).
\ee
Finally, for the bracket \eqref{CC bracket} we have
\be
\lb\C(\xi),\C(\zeta)\rb
&=p\lb\C(\xi),\J(\zeta\ip e)\rb+\lb\C(\xi),\T(\zeta\ip\omega)\rb\cr
&=-p\T\big(\pounds_\xi(\zeta\ip e)\big)-p\J\big(\pounds_\xi(\zeta\ip\omega)\big)+p\J\big(\zeta\ip(\delta^\text{c}_\xi e)\big)+\T\big(\zeta\ip(\delta^\text{c}_\xi\omega)\big)\cr
&\approx-p\T\big(\pounds_\xi(\zeta\ip e)\big)-p\J\big(\pounds_\xi(\zeta\ip\omega)\big)+p\J\big(\zeta\ip(\pounds_\xi\omega)\big)+p\T\big(\zeta\ip(\pounds_\xi e)\big)\cr
&=-p\D([\xi,\zeta]).
\ee
In the second equality, the first two terms come from the field-independent brackets \eqref{CJ bracket} and \eqref{CT bracket}, and the last two terms come from the action of $\C(\xi)$ on the field dependency of the argument. The third equality uses the on-shell expression \eqref{dual diffeos} for the dual diffeomorphisms. We have used everywhere the fact that the cubic charges vanish identically in the case of tangent vector fields $\xi$ and $\zeta$, i.e. $\J([\xi\ip\Phi_1,\zeta\ip\Phi_2])=0=\T([\xi\ip\Phi_1,\zeta\ip\Phi_2])$ for any two fields $\Phi_1,\Phi_2\in(e,\omega)$.

Finally, let us point out that, if we define $\C(\xi)$ as $\C(\xi)=\alpha\J(\xi\ip e)+\beta\T(\xi\ip\omega)$, the integrability condition obtained in \eqref{definition C} becomes $\alpha\sigma_2=\beta\sigma_3$. The bracket \eqref{CT bracket}, on the other hand, becomes
\be\label{app: CT details}
\lb\C(\xi),\T(\phi)\rb
&=-\alpha\J(\pounds_\xi\phi)+(\beta p-\alpha)\J([\xi\ip\omega,\phi])+\beta q\T([\xi\ip\omega,\phi])\cr
&=-\alpha\J(\pounds_\xi\phi)+2\big(\beta(p\sigma_1+q\sigma_3)-\alpha\sigma_1\big)\oint_S[\xi\ip\omega,\phi]e\cr
&=-\alpha\J(\pounds_\xi\phi)+2(\beta\sigma_0-\alpha\sigma_1)\oint_S[\xi\ip\omega,\phi]e.
\ee
One can therefore see that the last term can be cancelled by choosing $\alpha\sigma_1=\beta\sigma_0$. Together with the integrability condition $\alpha\sigma_2=\beta\sigma_3$, this in turn implies that $\sigma_1\sigma_3=\sigma_0\sigma_2$, i.e. that $q=0$. We therefore recover consistently the condition given in the main text.

\subsection{Fourier representation of the algebra}
\label{appendix Fourier}

Let us first introduce the distributional currents $\J^i$ and $\T^i$, which are functions on $S$ defined by
\be
\J(\alpha)=\oint_S\de\varphi\,\J^i(\varphi)\alpha_i(\varphi),
\q\q
\T(\phi)=\oint_S\de\varphi\,\T^i(\varphi)\phi_i(\varphi).
\ee
In terms of these distributional currents, the Poisson brackets in \eqref{JT extended algebra} become
\bsub\label{JT extended algebra distrib}
\be
\lb\J^i(\varphi),\T^j(\varphi')\rb&={\eps^{ij}}_k\T^k(\varphi)\delta(\varphi-\varphi')+2\sigma_1\eta^{ij}\partial_\varphi\delta(\varphi-\varphi'),\\
\lb\J^i(\varphi),\J^j(\varphi')\rb&={\eps^{ij}}_k\J^k(\varphi)\delta(\varphi-\varphi')+2\sigma_2\eta^{ij}\partial_\varphi\delta(\varphi-\varphi'),\\
\lb\T^i(\varphi),\T^j(\varphi')\rb&={\eps^{ij}}_k\big(p\J^k(\varphi)+q\T^k(\varphi)\big)\delta(\varphi-\varphi')+2\sigma_3\eta^{ij}\partial_\varphi\delta(\varphi-\varphi').
\ee
\esub
Note the change of sign in the central terms with respect to \eqref{JT extended algebra}. Let us now define the associated Fourier modes
\be
\J^i_n\coloneqq\oint_S\de\varphi\,e^{-\i n\varphi}\J^i(\varphi),
\q\q
\T^i_n\coloneqq\oint_S\de\varphi\,e^{-\i n\varphi}\T^i(\varphi).
\ee
The Fourier brackets are easily computed. For example, we have
\be
\lb\J^i_n,\J^j_m\rb
&=\oint_S\de\varphi\,\de\varphi'\,e^{-\i n\varphi}e^{-\i m\varphi'}\lb\J^i(\varphi),\J^j(\varphi')\rb\cr
&={\eps^{ij}}_k\oint_S\de\varphi\,e^{-\i(n+m)\varphi}\J^k(\varphi)+2\sigma_2\eta^{ij}\oint_S\de\varphi\,\de\varphi'\,e^{-\i n\varphi-\i m\varphi'}\partial_\varphi\delta(\varphi-\varphi')\cr
&={\eps^{ij}}_k\oint_S\de\varphi\,e^{-\i(n+m)\varphi}\J^k(\varphi)-2\sigma_2\eta^{ij}\oint_S\de\varphi\,\de\varphi'\,\partial_\varphi e^{-\i n\varphi-\i m\varphi'}\delta(\varphi-\varphi')\cr
&={\eps^{ij}}_k\oint_S\de\varphi\,e^{-\i(n+m)\varphi}\J^k(\varphi)+2\i\sigma_2n\eta^{ij}\oint_S\de\varphi\,e^{-\i(n+m)\varphi}\cr
&={\eps^{ij}}_k\J^k_{n+m}+2\i\sigma_2n\eta^{ij}\delta_{n+m,0}\cr
&={\eps^{ij}}_k\J^k_{n+m}-2\i\sigma_2m\eta^{ij}\delta_{n+m,0},
\ee
and similarly for the other brackets. Let us now introduce the Fourier modes
\be
\D_n\coloneqq\D(\xi=e^{-\i n\varphi}),
\q\q
\C_n\coloneqq\C(\xi=e^{-\i n\varphi}),
\ee
where we have to recall that integrability of $\C$ requires tangentiality and $\sigma_3=p\sigma_2$. For the brackets of $\D_n$ with the previous Fourier modes, we first compute
\be
\lb\D_n,\J(\alpha)\rb 
&=-\oint_S\de\varphi\,\J^i(\varphi)e^{-\i n\varphi}\partial_\varphi\alpha_i(\varphi)\cr
&=\oint_S\de\varphi\,\partial_\varphi\big(\J^i(\varphi)e^{-\i n\varphi}\big)\alpha_i(\varphi)\cr
&=\oint_S\de\varphi\,\lb\D_{n},\J^i(\varphi)\rb\alpha_i(\varphi),
\ee
which gives
\be
\lb\D_n,\J^i_m\rb
&=\oint_S\de\varphi\,e^{-\i m\varphi}\lb\D_n,\J^i(\varphi)\rb\cr
&=\oint_S\de\varphi\,e^{-\i m\varphi}\partial_\varphi\big(\J^i(\varphi)e^{-\i n\varphi}\big)\cr
&=-\oint_S\de\varphi\,\partial_\varphi e^{-\i m\varphi}\J^i(\varphi)e^{-\i n\varphi}\cr
&=\i m\J^i_{n+m},
\ee
and similarly for $\lb\D_n,\T^i_m\rb$. Finally, computing $\lb\D_n,\D_m\rb$ is immediate using the commutator $[\xi,\zeta]=\xi\zeta'-\zeta\xi'$, and we find
\be
\lb\D_n,\D_m\rb=-\i(n-m)\D_{n+m}.
\ee
We proceed in the same way to obtain the brackets involving $\C$. The only subtlety is for the bracket \eqref{CT bracket}, which has a field-dependent term on the right-hand side. To evaluate this in Fourier space, we first compute
\be
\lb\C_n,\T^i_m\rb
&=\oint_S\de\varphi\,e^{-\i m\varphi}\lb\C_n,\T^i(\varphi)\rb\cr
&=p\oint_S\de\varphi\,e^{-\i m\varphi}\partial_\varphi\big(\J^i(\varphi)e^{-\i n\varphi}\big)-q{\eps^i}_{jk}\oint_S\de\varphi\,e^{-\i(n+m)\varphi}\omega^j_\varphi\T^k(\varphi)\cr
&=\i pm\J^i_{n+m}-q{\eps^i}_{jk}\oint_S\de\varphi\,e^{-\i(n+m)\varphi}\omega^j_\varphi\T^k(\varphi).
\ee
The first term is part of the what we expect. For the second term, we use
\be
\J^i=2(\sigma_1e_\varphi^i+\sigma_2\omega_\varphi^i),
\q\q
\T^i=2(\sigma_1\omega_\varphi^i+\sigma_3e_\varphi^i)=2(\sigma_1\omega_\varphi^i+p\sigma_2e_\varphi^i),
\ee
to write
\be
\omega_\varphi^i=\f{p\sigma_2\J^i-\sigma_1\T^i}{2(p\sigma_2^2-\sigma_1^2)},
\ee
which gives
\be
\lb\C_n,\T^i_m\rb=\i pm\J^i_{n+m}-\f{pq\sigma_2}{2(p\sigma_2^2-\sigma_1^2)}{\eps^i}_{jk}\oint_S\de\varphi\,e^{-\i(n+m)\varphi}\J^j(\varphi)\T^k(\varphi).
\ee
The last step of the computation is then to insert a Dirac delta in the remaining integral to obtain the Fourier expression
\be
\lb\C_n,\T^i_m\rb
&=\i pm\J^i_{n+m}-\f{pq\sigma_2}{2(p\sigma_2^2-\sigma_1^2)}{\eps^i}_{jk}\oint_S\de\varphi\,e^{-\i(n+m)\varphi}\delta(\varphi-\varphi')\J^j(\varphi) \T^k(\varphi') \cr
&=\i pm\J^i_{n+m}-\f{pq\sigma_2}{2(p\sigma_2^2-\sigma_1^2)}{\eps^i}_{jk}\sum_{k\in\mathbb{Z}}\oint_S\de\varphi\,e^{-\i(n+m+k)\varphi}\J^j(\varphi)\oint_S\de\varphi'\,e^{-\i\varphi'(-k)}\T^k(\varphi')\cr
&=\i pm\J^i_{n+m}-\f{pq\sigma_2}{2(p\sigma_2^2-\sigma_1^2)}{\eps^i}_{jk}\sum_{k\in\mathbb{Z}}\J^j_{n+m+k}\T^k_{-k}.
\ee

%

\subsection{Details on the Sugawara construction}
\label{appendix Sugawara}

The brackets of the quadratics $\Q_n$ with $\J^i_{m}$ and $\T^i_{m}$ are given by
\bsub\label{Qi and Ji Ti brackets}
\be
\lb\Q_n^1,\J^i_m\rb&=-4\i m\big(\sigma_1\J^i_{n+m}+\sigma_2\T^i_{n+m}\big),\\
\lb\Q_n^1,\T^i_m\rb&=-4\i m\big(\sigma_3\J^i_{n+m}+\sigma_1\T^i_{n+m}\big)-2q{\eps^i}_{jk}\sum_{k\in\mathbb{Z}}\J^j_{n+m+k}\T^k_{-k},\\
\lb\Q_n^2,\J^i_m\rb&=-4\i\sigma_1m\T^i_{n+m},\\
\lb\Q_n^2,\T^i_m\rb&=-4\i\sigma_3m\T^i_{n+m}+2p{\eps^i}_{jk}\sum_{k\in\mathbb{Z}}\J^j_{n+m+k}\T^k_{-k},\\
\lb\Q_n^3,\J^i_m\rb&=-4\i\sigma_2m\J^i_{n+m},\\
\lb\Q_n^3,\T^i_m\rb&=-4\i\sigma_1m\J^i_{n+m}-2{\eps^i}_{jk}\sum_{k\in\mathbb{Z}}\J^j_{n+m+k}\T^k_{-k}.
\ee
\esub
For the brackets of the $\Q_n$'s with themselves we find
\bsub\label{Qi Qj brackets}
\be
\lb\Q_n^1,\Q_m^1\rb&=-4\i\sigma_1(m-n)\Q^1_{n+m}-4\i\sigma_2(m-n)\Q_{n+m}^2-4\i\sigma_3(m-n)\Q_{n+m}^3,\label{Q1 Q1 bracket}\\
\lb\Q_n^2,\Q_m^2\rb&=-4\i\sigma_3(m-n)\Q^2_{n+m},\\
\lb\Q_n^3,\Q_m^3\rb&=-4\i\sigma_2(m-n)\Q^3_{n+m},
\ee
\esub
and
\bsub\label{Qi Qj brackets cross}
\be
\lb\Q_n^1,\Q_m^2\rb
&=-4\i\sigma_3m\Q^1_{n+m}-4\i\sigma_1(m-n)\Q^2_{n+m}-8\i\sigma_3\sum_{k\in\mathbb{Z}}k\J^i_{n+m+k}\T^i_{-k},\label{Q1 Q2 bracket}\\
\lb\Q_n^1,\Q_m^3\rb
&=-4\i\sigma_2m\Q^1_{n+m}-4\i\sigma_1(m-n)\Q^3_{n+m}-8\i\sigma_2\sum_{k\in\mathbb{Z}}k\J^i_{-k}\T^i_{m+n+k}\cr
&=4\i\sigma_2n\Q^1_{n+m}-4\i\sigma_1(m-n)\Q^3_{n+m}+8\i\sigma_2\sum_{k\in\mathbb{Z}}k\J^i_{n+m+k}\T^i_{-k},\label{Q1 Q3 bracket}\\
\lb\Q_n^3,\Q_m^2\rb&=-4\i\sigma_1m\Q^1_{n+m}-8\i\sigma_1\sum_{k\in\mathbb{Z}}k\J^i_{n+m+k}\T^i_{-k}.
\ee
\esub
The bracket \eqref{Q1 Q1 bracket} follows from the fact that
\be
\lb\Q_n^1,\Q_m^1\rb
&=2\sum_{k\in\mathbb{Z}}\lb\Q^1_n,\J^i_{m+k}\T^i_{-k}\rb\cr
&=2\sum_{k\in\mathbb{Z}}\big(\J^i_{m+k}\lb\Q^1_n,\T^i_{-k}\rb+\lb\Q^1_n,\J^i_{m+k}\rb\T^i_{-k}\big)\cr
&=-4\i\sum_{k\in\mathbb{Z}}\Big((-k)\big(\sigma_3\J^i_{n-k}\J^i_{m+k}+\sigma_1\T^i_{n-k}\J^i_{m+k}\big)+(m+k)\big(\sigma_1\J^i_{m+n+k}\T^i_{-k}+\sigma_2\T^i_{m+n+k}\T^i_{-k}\big)\Big)\cr
&=-4\i\sum_{k\in\mathbb{Z}}\Big(-(n+k)\big(\sigma_3\J^i_{-k}\J^i_{n+m+k}+\sigma_1\T^i_{-k}\J^i_{n+m+k}\big)+(m+k)\big(\sigma_1\J^i_{m+n+k}\T^i_{-k}+\sigma_2\T^i_{m+n+k}\T^i_{-k}\big)\Big)\cr
&=-4\i\sigma_1(m-n)\Q^1_{n+m}-4\i\sigma_2\sum_{k\in\mathbb{Z}}(m+k)\T^i_{n+m+k}\T^i_{-k}+4\i\sigma_3\sum_{k\in\mathbb{Z}}(n+k)\J^i_{n+m+k}\J^i_{-k}\cr
&=-4\i\sigma_1(m-n)\Q^1_{n+m}-4\i\sigma_2m\Q_{n+m}^2+4\i\sigma_3n\Q_{n+m}^3-4\i\sum_{k\in\mathbb{Z}}k\big(\sigma_2\T^i_{n+m+k}\T^i_{-k}+\sigma_3\J^i_{n+m+k}\J^i_{-k}\big)\cr
&=-4\i\sigma_1(m-n)\Q^1_{n+m}+4\i\sigma_2n\Q_{n+m}^2-4\i\sigma_3m\Q_{n+m}^3+4\i\sum_{k\in\mathbb{Z}}k\big(\sigma_2\T^i_{n+m+k}\T^i_{-k}+\sigma_3\J^i_{n+m+k}\J^i_{-k}\big),\cr
\ee
where for the last equality we have changed the summation variable as $k\to-(n+m+k)$. This allow to switch the sign of the sum over $k$ without changing the argument, and we can then add the last two lines to find the result. The same trick has been used for the brackets \eqref{Q1 Q3 bracket} and \eqref{Q1 Q2 bracket}.

In order to find the combination of quadratics which is well-behaved and reproduces $\D_n$ and $\C_n$, we define $\tilde{\Q}_n\coloneqq\alpha\Q^1_n+\beta\Q^2_n+\gamma\Q^3_n$. With this we find
\bsub\label{Qtilde with J and T}
\be
\lb\tilde{\Q}_n,\J^i_m\rb&=-4\i m(\alpha\sigma_1+\gamma\sigma_2)\J^i_{m+n}-4\i m(\beta\sigma_1+\alpha\sigma_2)\T^i_{m+n},\\
\lb\tilde{\Q}_n,\T^i_m\rb&=-4\i m(\gamma\sigma_1+\alpha\sigma_3)\J^i_{m+n}-4\i m(\beta\sigma_3+\alpha\sigma_1)\T^i_{m+n}-2{\eps^i}_{jk}(\alpha q-\beta p+\gamma)\sum_{k\in\mathbb{Z}}\J^j_{n+m+k}\T^k_{-k},
\ee
\esub
and
\be\label{Qtilde with Qtilde}
\lb\tilde{\Q}_{n},\tilde{\Q}_m\rb=
-4\i(m-n)\Big(&(\alpha^2\sigma_1+\beta\gamma\sigma_1+\alpha\gamma\sigma_2+\alpha\beta\sigma_3)\Q^1_{n+m}\cr
+\,&(2\alpha\beta\sigma_1+\alpha^2\sigma_2+\beta^2\sigma_3)\Q^2_{n+m}\cr
+\,&(2\alpha\gamma\sigma_1+\gamma^2\sigma_2+\alpha^2\sigma_3)\Q^3_{n+m}\Big).
\ee
These brackets can be compared with the ones involving $\D_n$ and $\C_n$ in \eqref{JTDC extended algebra Fourier} in order to identify which triple $(\alpha,\beta,\gamma)$ give the Sugawara expression $\tilde{\Q}_n$ for $\D_n$ and $\C_n$. A similar calculation can be done for the teleparallel dual charge $\B_n$.


\subsection{Twisted Sugawara construction and central extensions}
\label{app:twisted}

The Sugawara construction can be generalized to obtain centrally-extended algebras of quadratics, such as the diffeomorphisms or their duals introduced here. This is done via the so-called twisted Sugawara construction. This corresponds in fact to a particular choice of boundary conditions such that the quadratic charges (e.g. of diffeomorphisms) are also defined for non-tangent vector fields, as was the case in e.g. \eqref{integrable diffeo}.

The twisted construction is based on a first order linear twist of the quadratics $\Q_n$ introduced in \eqref{Q quadratics}. More specifically, we define
\bsub\label{twisted quadratics}
\be
\check{\Q}_n^1&\coloneqq2\sum_{k\in\mathbb{Z}}\J^i_{n+k}\T^i_{-k}+2n\lambda^i\J_n^i+2n\mu^i\T_n^i,\\ 
\check{\Q}_n^2&\coloneqq\sum_{k\in\mathbb{Z}}\T^i_{n+k}\T^i_{-k}+2n\lambda^i\T_n^i,\\
\check{\Q}_n^3&\coloneqq\sum_{k\in\mathbb{Z}}\J^i_{n+k}\J^i_{-k}+2n\mu^i\J_n^i,
\ee
\esub
where $\lambda^i$ and $\mu^i$ are given vectors. The brackets between these twisted quadratics and the initial currents become
\bsub
\be
\lb\check{\Q}_n^1,\J^i_m\rb&=-4\i m\Big(\sigma_1\big(\J^i_{n+m}+n\mu^i\delta_{n+m,0}\big)+\sigma_2\big(\T^i_{n+m}+n\lambda^i\delta_{n+m,0}\big)\Big)\\
&\pe-2{\eps^i}_{jk}\big(n\lambda^j\J_{n+m}^k+n\mu^j\T^k_{n+m}\big),\cr
\lb\check{\Q}_n^1,\T^i_m\rb&=-4\i m\Big(\sigma_3\big(\J^i_{n+m}+n\mu^i\delta_{n+m,0}\big)+\sigma_1\big(\T^i_{n+m}+n \lambda^i\delta_{n+m,0}\big)\Big)\\
&\pe-2{\eps^i}_{jk}\big(pn\mu^j\J^k_{n+m}+n(\lambda^j+q\mu^j)\T^k_{n+m}\big)-2q{\eps^i}_{jk}\sum_{s\in\mathbb{Z}}\J^j_{n+m+s}\T^k_{-s},\cr
\lb\check{\Q}_n^1,\check{\Q}_m^1\rb&=-4\i(m-n)\sigma_1\check{\Q}^1_{m+n}-4\i(m-n)\sigma_2\check{\Q}^2_{m+n}-4\i(m-n)\sigma_3\check{\Q}^3_{m+n}\\
&\pe+8\i m^3\big(2\sigma_1(\lambda\cdot\mu)+\sigma_2(\lambda\cdot\lambda)+\sigma_3(\mu\cdot\mu)\big)-4(m-n)q\eps_{ijk}\sum_{s\in\mathbb{Z}}\mu^i\J^j_{m+n+s}\T^{k}_{-s},\nonumber
\ee
\esub
\bsub
\be
\lb\check{\Q}_n^2,\J^i_m\rb&=-4\i m\sigma_1\big(\T^i_{n+m}+n\lambda^i\delta_{n+m,0}\big)-2n{\eps^i}_{jk}\lambda^j\T^k_{n+m},\\
\lb\check{\Q}_n^2,\T^i_m\rb&=-4\i m\sigma_3\big(\T^i_{n+m}+n\lambda^i\delta_{n+m,0}\big)-2n{\eps^i}_{jk}\lambda^j\big(p\J^k_{n+m}+q\T^k_{n+m}\big)\cr
&\pe+2p{\eps^i}_{jk}\sum_{s\in\mathbb{Z}}\J^j_{n+m+s}\T^k_{-s},\\
\lb\check{\Q}_n^2,\check{\Q}_m^2\rb&=-4\i(m-n)\sigma_3\check{\Q}^2_{m+n}+8\i m^3\sigma_3(\lambda\cdot\lambda)+4(m-n)p\eps_{ijk}\sum_{s\in\mathbb{Z}}\lambda^i\J^j_{m+n+s}\T^{k}_{-s},\q\label{twisted B algebra}
\ee
\esub
\bsub
\be
\lb\check{\Q}_n^3,\J^i_m\rb&=-4\i m\sigma_2\big(\J^i_{n+m}+n\mu^i\delta_{n+m,0}\big)-2n{\eps^i}_{jk}\mu^j\J^k_{n+m},\\
\lb\check{\Q}_n^3,\T^i_m\rb&=-4\i m\sigma_1\big(\J^i_{n+m}+n\mu^i\delta_{n+m,0}\big)-2n{\eps^i}_{jk}\mu^j\T^k_{n+m}-2{\eps^i}_{jk}\sum_{s\in\mathbb{Z}}\J^j_{n+m+s}\T^k_{-s},\\
\lb\check{\Q}_n^3,\check{\Q}_m^3\rb&=-4\i (m-n)\sigma_2\check{\Q}^3_{m+n}+8\i m^3\sigma_2(\mu\cdot\mu).
\ee
\esub
We can now compute the cross brackets between $\check{\Q}_n^i$, to find
\bsub
\be
\lb\check{\Q}^1_n,\check{\Q}^2_m\rb 
=&-4\i\sigma_1\big((m-n)\Q^2_{m+n}-2m^3(\lambda\cdot\lambda)\delta_{n+m,0}\big)\cr
&-4\i\sigma_3\big(m\Q^1_{m+n}-2m^3(\lambda\cdot\mu)\delta_{n+m,0}\big)\cr
&-8\i\sigma_3\left(\sum_ss\J^i_{n+m+s}\T^i_{-s}-mn\lambda^i\J^i_{n+m}-(m^2+n^2+mn)\mu^i\T^i_{n+m}\right)\cr
&-4np\eps_{ijk}\mu^i\left(\sum_s\J^j_{n+m+s}\T^k_{-s}-m\lambda^j\J^k_{n+m}\right)\cr
&-4mq\eps_{ijk}\lambda^i\left(\sum_s\J^j_{n+m+s}\T^k_{-s}+n\mu^j\T^k_{n+m}\right),\\
\lb\check{\Q}^1_n,\check{\Q}^3_m\rb
=&-4\i\sigma_1\big((m-n)\Q^3_{m+n}-2m^3(\mu\cdot\mu)\delta_{n+m,0}\big)\cr
&+4\i\sigma_2\big(n\Q^1_{m+n}+2m^3(\lambda\cdot\mu)\delta_{n+m,0}\big)\cr
&+8\i\sigma_2\left(\sum_ss\J^i_{n+m+s}\T^i_{-s}-mn\lambda^i\J^i_{n+m}-(m^2+n^2+mn)\mu^i\T^i_{n+m}\right)\cr
&+4n\eps_{ijk}\mu^i\left(\sum_s\J^j_{n+m+s}\T^k_{-s}-m\lambda^j\J^k_{n+m}\right),\\
\lb\check{\Q}^3_n,\check{\Q}^2_m\rb
=&-4\i\sigma_1\big(m\Q^1_{m+n}-2m^3(\mu\cdot\lambda)\delta_{n+m,0}\big)\cr
&-8\i\sigma_1\left(\sum_ss\J^i_{n+m+s}\T^i_{-s}-mn\lambda^i\J^i_{n+m}-(m^2+n^2+mn)\mu^i\T^i_{n+m}\right)\cr
&-4n\eps_{ijk}\lambda^i\left(\sum_s\J^j_{n+m+s}\T^k_{-s}+m\mu^j\J^k_{n+m}\right).
\ee
\esub
This coincides with the untwisted results when $\lambda=\mu=0$. We can then combine these elementary twisted quadratics to form the diffeomorphisms and their duals, whose algebra becomes centrally extended as well. This is easily seen for example in the case $p=0$ on the teleparallel dual diffeomorphism $\B_n$, whose Sugawara expression is proportional to $\Q^2_n$. Twisting this generator leads to the centrally-extended bracket \eqref{twisted B algebra}.

\section{Symplectic symmetries, Kosmann derivative, and Fefferman--Graham gauge}
\label{app:Kosmann}

We have seen in \eqref{Bondi diffeo charge} that the first order diffeomorphism charge for the vector field \eqref{Bondi diffeo} contains an $r^{-1}$ dependency. This is fine as long as we focus on asymptotic symmetries. Without the asymptotic limit, the full diffeomorphism charge \eqref{Bondi diffeo charge} is given by
\be\label{full Bondi diffeo charge}
\D(\xi)=\oint_S\text{finite}-\f{\partial_\varphi f}{r}\left(\sigma_2\mathscr{M}+(2\sigma_1-q\sigma_2)\f{\mathscr{N}}{2}\right).
\ee
We therefore see that even in the case $q\sigma_2=0$, where the charge can also be computed in metric variables, this charge does not a priori match the corresponding metric expression. Indeed, one can see e.g. in equation (2.13) of \cite{Compere:2014cna} that the $r^{-1}$ dependency drops there, and that the asymptotic symmetries then become so-called symplectic symmetries which can be defined at any finite distance $r$.

This discrepancy is due to the fact that we are here considering a first order formulation in terms of connection and triad variables. Clearly, we have to do so because it is the only formulation in which the general Lagrangian \eqref{MB Lagrangian} can be written. We therefore cannot go back to a metric formulation for the general diffeomorphism charges with all independent couplings present. However, a way to resolve the tension is to use the Kosmann derivative. In three spacetime dimensions it is defined as
\be
\mathscr{K}_\xi\coloneqq\pounds_\xi+\delta^\text{j}_{\lambda(\xi,e)},
\ee
which is the composition of the ordinary Lie derivative with a field-dependent Lorentz transformation whose parameter is
\be
\lambda(\xi,e)=-\f{1}{2}[\hat{e}\ip\pounds_\xi e],
\q\q
\lambda^i(\xi,e)=-\f{1}{2}{\eps^i}_{jk}\hat{e}^{\mu j}\pounds_\xi e^k_\mu,
\ee
where $\hat{e}^{\mu i}=g^{\mu\nu}e^i_\nu$ is the inverse triad (see \cite{Jacobson:2015uqa,Prabhu:2015vua,DePaoli:2018erh,Oliveri:2019gvm,Oliveri:2020xls} for the four-dimensional definition and other properties). This derivative has the property that $\mathscr{K}_\xi e=0$ when $\xi$ is Killing. In order to see this, we can compute explicitly the field-dependent Lorentz transformation
\be
\delta^\text{j}_{\lambda}e^i_\alpha
&=[e_\alpha,\lambda]^i\cr
&={\eps^i}_{jk}e^j_\alpha\lambda^k\cr
&=-\f{1}{2}{\eps^i}_{jk}{\eps^k}_{lm}e^j_\alpha\hat{e}^{\mu l}\pounds_\xi e^m_\mu\cr
&=-\f{1}{2}(\pounds_\xi e^i_\alpha-\eta_{jm}e^j_\alpha\hat{e}^{\mu i}\pounds_\xi e^m_\mu)\cr
&=-\pounds_\xi e^i_\alpha,
\ee
where we have used \eqref{epsilon2}, and in the last step the fact that $\xi$ is an exact Killing vector to write $\pounds_\xi g_{\mu\nu}=0=\eta_{jm}(\pounds_\xi e^j_\mu e^m_\nu+e^j_\mu\pounds_\xi e^m_\nu)$.

In order to compute the charge associated with the Kosmann derivative, we can simply add to \eqref{Bondi diffeo charge} the charge $\slashed{\delta}\J(\lambda)$ of the field-dependent Lorentz transformation. This latter is found to be
\be
\slashed{\delta}\J(\lambda)=\f{1}{r}\oint_S\partial_\varphi f\left(\sigma_2\delta\mathscr{M}+(2\sigma_1-q\sigma_2)\f{\delta\mathscr{N}}{2}\right),
\ee
and is therefore integrable. We see that this charge cancels exactly the $r^{-1}$ contribution in \eqref{full Bondi diffeo charge}, thereby leading to an $r$-independent charge associated with the Kosmann derivative. The Bondi gauge asymptotic symmetries \eqref{Bondi diffeo} can then be promoted to symplectic symmetries and one finds that the algebra of charges is valid at any place in the bulk. This mechanism can also be studied in Fefferman--Graham gauge.

Let us consider the solution to Einstein equations with Brown--Henneaux boundary conditions in Fefferman--Graham gauge. With coordinates $x^\mu=(x^+,x^-,r)$, this is given by
\be
\de s^2=\f{\ell^2}{r^2}\de r^2-\left(r\de x^+-\f{\ell^2}{r}L^-(x^-)\de x^-\right)\left(r\de x^--\f{\ell^2}{r}L^+(x^+)\de x^+\right).
\ee
Working with the internal metric \eqref{internal Bondi metric}, a compatible set of triad and connection components are given by
\be\label{Bondi e omega}
\everymath={\displaystyle}
e^i_\mu=
\begin{pmatrix}
\ell^2L_+/(2r)&r&0\\
-r/2&-\ell^2L_-/r&0\\
0&0&\ell/r
\end{pmatrix},
\q\q
\everymath={\displaystyle}
\omega^i_\mu=
\begin{pmatrix}
\ell L_+/(2r)&r/\ell&0\\
r/(2\ell)&\ell L_-/r&0\\
0&0&0
\end{pmatrix}
-\f{q}{2}e^i_\mu.
\ee
The asymptotic Killing vector field for this geometry is $\xi=(\xi^+,\xi^-,\xi^r)$ with
\be
\xi^\pm=f^\pm(x^\pm)+\f{\ell^2}{2r^2}\partial_\mp^2f^\mp(x^\mp)+\O(r^{-4}),
\q\q
\xi^r=-\f{r}{2}\big(\partial_+f^+(x^+)+\partial_-f^-(x^-)\big).
\ee
With this, the diffeomorphism charge with the GHY boundary term is found to be
\be
\D(\xi)
&=\oint_SQ+\sigma_1\xi\ip(e\wedge\omega)\\
&=\f{1}{2}\oint_S\xi^+L^+\big(\ell(2\sigma_1-q\sigma_2)+2\sigma_2\big)+\xi^-L^-\big(\ell(2\sigma_1-q\sigma_2)-2\sigma_2\big)\cr
&=\oint_S\text{finite}+\f{\ell^2}{4r^2}\Big(\partial^2_-f^-L^+\big(\ell(2\sigma_1-q\sigma_2)+2\sigma_2\big)+\partial^2_+f^+L^-\big(\ell(2\sigma_1-q\sigma_2)-2\sigma_2\big)\Big)+\O(r^{-4}),\nonumber
\ee
where in these coordinates the integral is along $\varphi=(x^+-x^-)/2$. We can now compute the charge of the field-dependent Lorentz transformation entering the definition of the Kosmann derivative. It is found to be
\be
\slashed{\delta}\J(\lambda)=-\f{\ell^2}{4r^2}\oint_S\partial^2_-f^-\delta L^+\big(\ell(2\sigma_1-q\sigma_2)+2\sigma_2\big)+\partial^2_+f^+\delta L^-\big(\ell(2\sigma_1-q\sigma_2)-2\sigma_2\big)+\O(r^{-4}),
\ee
and we see at order $r^{-2}$ that this cancels the opposite contribution in the first order diffeomorphism charge. The same thing presumably happens at subleading order.

\bibliography{Biblio.bib}

\providecommand{\href}[2]{#2}\begingroup\raggedright\begin{thebibliography}{100}

\bibitem{MR1637718}
S.~Carlip, \emph{Quantum gravity in {$2+1$} dimensions}.
\newblock Cambridge Monographs on Mathematical Physics. Cambridge University
  Press, Cambridge, 1998,
  \href{http://dx.doi.org/10.1017/CBO9780511564192}{10.1017/CBO9780511564192}.

\bibitem{Banados:1992wn}
M.~Banados, C.~Teitelboim and J.~Zanelli, \emph{{The Black hole in
  three-dimensional space-time}},
  \href{http://dx.doi.org/10.1103/PhysRevLett.69.1849}{\emph{Phys. Rev. Lett.}
  {\bfseries 69} (1992) 1849--1851},
  [\href{https://arxiv.org/abs/hep-th/9204099}{{\ttfamily hep-th/9204099}}].

\bibitem{Banados:1998ta}
M.~Banados, T.~Brotz and M.~E. Ortiz, \emph{{Boundary dynamics and the
  statistical mechanics of the (2+1)-dimensional black hole}},
  \href{http://dx.doi.org/10.1016/S0550-3213(99)00069-3}{\emph{Nucl. Phys.}
  {\bfseries B545} (1999) 340--370},
  [\href{https://arxiv.org/abs/hep-th/9802076}{{\ttfamily hep-th/9802076}}].

\bibitem{Carlip:2005zn}
S.~Carlip, \emph{{Conformal field theory, (2+1)-dimensional gravity, and the
  BTZ black hole}},
  \href{http://dx.doi.org/10.1088/0264-9381/22/12/R01}{\emph{Class. Quant.
  Grav.} {\bfseries 22} (2005) R85--R124},
  [\href{https://arxiv.org/abs/gr-qc/0503022}{{\ttfamily gr-qc/0503022}}].

\bibitem{MR974271}
E.~Witten, \emph{{$2+1$}-dimensional gravity as an exactly soluble system},
  \href{http://dx.doi.org/10.1016/0550-3213(88)90143-5}{\emph{Nuclear Phys. B}
  {\bfseries 311} (1988/89) 46--78}.

\bibitem{Carlip:1993zi}
S.~Carlip, \emph{{Six ways to quantize (2+1)-dimensional gravity}},  in
  \emph{{5th Canadian Conference on General Relativity and Relativistic
  Astrophysics (5CCGRRA)}}, pp.~0215--234, 5, 1993.
\newblock \href{https://arxiv.org/abs/gr-qc/9305020}{{\ttfamily
  gr-qc/9305020}}.

\bibitem{Carlip:1993ze}
S.~Carlip and J.~Nelson, \emph{{Equivalent quantizations of (2+1)-dimensional
  gravity}}, \href{http://dx.doi.org/10.1016/0370-2693(94)90197-X}{\emph{Phys.
  Lett. B} {\bfseries 324} (1994) 299--302},
  [\href{https://arxiv.org/abs/gr-qc/9311007}{{\ttfamily gr-qc/9311007}}].

\bibitem{Carlip:1994ap}
S.~Carlip and J.~Nelson, \emph{{Comparative quantizations of (2+1)-dimensional
  gravity}}, \href{http://dx.doi.org/10.1103/PhysRevD.51.5643}{\emph{Phys. Rev.
  D} {\bfseries 51} (1995) 5643--5653},
  [\href{https://arxiv.org/abs/gr-qc/9411031}{{\ttfamily gr-qc/9411031}}].

\bibitem{Alexandrov:2011ab}
S.~Alexandrov, M.~Geiller and K.~Noui, \emph{{Spin Foams and Canonical
  Quantization}}, \href{http://dx.doi.org/10.3842/SIGMA.2012.055}{\emph{SIGMA}
  {\bfseries 8} (2012) 055}, [\href{https://arxiv.org/abs/1112.1961}{{\ttfamily
  1112.1961}}].

\bibitem{Goeller:2019zpz}
C.~Goeller, E.~R. Livine and A.~Riello, \emph{{Non-Perturbative 3D Quantum
  Gravity: Quantum Boundary States and Exact Partition Function}},
  [\href{https://arxiv.org/abs/1912.01968}{{\ttfamily 1912.01968}}].

\bibitem{Witten:1989sx}
E.~Witten, \emph{{Topology Changing Amplitudes in (2+1)-Dimensional Gravity}},
  \href{http://dx.doi.org/10.1016/0550-3213(89)90591-9}{\emph{Nucl. Phys. B}
  {\bfseries 323} (1989) 113--140}.

\bibitem{Ooguri:1991ni}
H.~Ooguri, \emph{{Partition functions and topology changing amplitudes in the
  3-D lattice gravity of Ponzano and Regge}},
  \href{http://dx.doi.org/10.1016/0550-3213(92)90188-H}{\emph{Nucl. Phys. B}
  {\bfseries 382} (1992) 276--304},
  [\href{https://arxiv.org/abs/hep-th/9112072}{{\ttfamily hep-th/9112072}}].

\bibitem{Carlip:1994tt}
S.~Carlip and R.~Cosgrove, \emph{{Topology change in (2+1)-dimensional
  gravity}}, \href{http://dx.doi.org/10.1063/1.530760}{\emph{J. Math. Phys.}
  {\bfseries 35} (1994) 5477--5493},
  [\href{https://arxiv.org/abs/gr-qc/9406006}{{\ttfamily gr-qc/9406006}}].

\bibitem{Oriti:2006se}
D.~Oriti, \emph{{The group field theory approach to quantum gravity}},
  [\href{https://arxiv.org/abs/gr-qc/0607032}{{\ttfamily gr-qc/0607032}}].

\bibitem{Brown:1986nw}
J.~D. Brown and M.~Henneaux, \emph{{Central Charges in the Canonical
  Realization of Asymptotic Symmetries: An Example from Three-Dimensional
  Gravity}}, \href{http://dx.doi.org/10.1007/BF01211590}{\emph{Commun. Math.
  Phys.} {\bfseries 104} (1986) 207--226}.

\bibitem{Coussaert:1995zp}
O.~Coussaert, M.~Henneaux and P.~van Driel, \emph{{The Asymptotic dynamics of
  three-dimensional Einstein gravity with a negative cosmological constant}},
  \href{http://dx.doi.org/10.1088/0264-9381/12/12/012}{\emph{Class. Quant.
  Grav.} {\bfseries 12} (1995) 2961--2966},
  [\href{https://arxiv.org/abs/gr-qc/9506019}{{\ttfamily gr-qc/9506019}}].

\bibitem{Maldacena:2003nj}
J.~M. Maldacena, \emph{{TASI 2003 lectures on AdS / CFT}},  in \emph{{Progress
  in string theory. Proceedings, Summer School, TASI 2003, Boulder, USA, June
  2-27, 2003}}, pp.~155--203, 2003.
\newblock \href{https://arxiv.org/abs/hep-th/0309246}{{\ttfamily
  hep-th/0309246}}.

\bibitem{Dittrich:2018xuk}
B.~Dittrich, C.~Goeller, E.~R. Livine and A.~Riello, \emph{{Quasi-local
  holographic dualities in non-perturbative 3d quantum gravity}},
  \href{http://dx.doi.org/10.1088/1361-6382/aac606}{\emph{Class. Quant. Grav.}
  {\bfseries 35} (2018) 13LT01},
  [\href{https://arxiv.org/abs/1803.02759}{{\ttfamily 1803.02759}}].

\bibitem{Dittrich:2017hnl}
B.~Dittrich, C.~Goeller, E.~Livine and A.~Riello, \emph{{Quasi-local
  holographic dualities in non-perturbative 3d quantum gravity I: Convergence
  of multiple approaches and examples of Ponzano-Regge statistical duals}},
  \href{http://dx.doi.org/10.1016/j.nuclphysb.2018.06.007}{\emph{Nucl. Phys.}
  {\bfseries B938} (2019) 807--877},
  [\href{https://arxiv.org/abs/1710.04202}{{\ttfamily 1710.04202}}].

\bibitem{Dittrich:2017rvb}
B.~Dittrich, C.~Goeller, E.~R. Livine and A.~Riello, \emph{{Quasi-local
  holographic dualities in non-perturbative 3d quantum gravity II: From
  coherent quantum boundaries to BMS3 characters}},
  \href{http://dx.doi.org/10.1016/j.nuclphysb.2018.06.010}{\emph{Nucl. Phys.}
  {\bfseries B938} (2019) 878--934},
  [\href{https://arxiv.org/abs/1710.04237}{{\ttfamily 1710.04237}}].

\bibitem{Goeller:2019apd}
C.~Goeller, \emph{{Quasi-Local 3D Quantum Gravity : Exact Amplitude and
  Holography}}.
\newblock PhD thesis, Lyon, Ecole Normale Superieure, Perimeter Inst. Theor.
  Phys., 2019.
\newblock \href{https://arxiv.org/abs/2005.09985}{{\ttfamily 2005.09985}}.

\bibitem{Asante:2019ndj}
S.~K. Asante, B.~Dittrich and F.~Hopfmueller, \emph{{Holographic formulation of
  3D metric gravity with finite boundaries}},
  \href{http://dx.doi.org/10.3390/universe5080181}{\emph{Universe} {\bfseries
  5} (2019) 181}, [\href{https://arxiv.org/abs/1905.10931}{{\ttfamily
  1905.10931}}].

\bibitem{Witten:1988hf}
E.~Witten, \emph{{Quantum Field Theory and the Jones Polynomial}},
  \href{http://dx.doi.org/10.1007/BF01217730}{\emph{Commun.Math.Phys.}
  {\bfseries 121} (1989) 351}.

\bibitem{Witten:1989rw}
E.~Witten, \emph{{Gauge Theories, Vertex Models and Quantum Groups}},
  \href{http://dx.doi.org/10.1016/0550-3213(90)90115-T}{\emph{Nucl. Phys. B}
  {\bfseries 330} (1990) 285--346}.

\bibitem{Bais:2002ye}
F.~A. Bais, N.~M. Muller and B.~J. Schroers, \emph{{Quantum group symmetry and
  particle scattering in (2+1)-dimensional quantum gravity}},
  \href{http://dx.doi.org/10.1016/S0550-3213(02)00572-2}{\emph{Nucl. Phys.}
  {\bfseries B640} (2002) 3--45},
  [\href{https://arxiv.org/abs/hep-th/0205021}{{\ttfamily hep-th/0205021}}].

\bibitem{Noui:2006ku}
K.~Noui, \emph{{Three Dimensional Loop Quantum Gravity: Particles and the
  Quantum Double}},
  \href{http://dx.doi.org/10.1063/1.2352860}{\emph{J.Math.Phys.} {\bfseries 47}
  (2006) 102501}, [\href{https://arxiv.org/abs/gr-qc/0612144}{{\ttfamily
  gr-qc/0612144}}].

\bibitem{Meusburger:2008bs}
C.~Meusburger and K.~Noui, \emph{{The Hilbert space of 3d gravity: quantum
  group symmetries and observables}},
  \href{http://dx.doi.org/10.4310/ATMP.2010.v14.n6.a3}{\emph{Adv. Theor. Math.
  Phys.} {\bfseries 14} (2010) 1651--1715},
  [\href{https://arxiv.org/abs/0809.2875}{{\ttfamily 0809.2875}}].

\bibitem{Ballesteros:2010zq}
A.~Ballesteros, F.~J. Herranz and C.~Meusburger, \emph{{Three-dimensional
  gravity and Drinfel'd doubles: spacetimes and symmetries from quantum
  deformations}},
  \href{http://dx.doi.org/10.1016/j.physletb.2010.03.043}{\emph{Phys. Lett. B}
  {\bfseries 687} (2010) 375--381},
  [\href{https://arxiv.org/abs/1001.4228}{{\ttfamily 1001.4228}}].

\bibitem{Dupuis:2020ndx}
M.~Dupuis, L.~Freidel, F.~Girelli, A.~Osumanu and J.~Rennert, \emph{{On the
  origin of the quantum group symmetry in 3d quantum gravity}},
  [\href{https://arxiv.org/abs/2006.10105}{{\ttfamily 2006.10105}}].

\bibitem{Deser:1981wh}
S.~Deser, R.~Jackiw and S.~Templeton, \emph{{Topologically Massive Gauge
  Theories}}, \href{http://dx.doi.org/10.1006/aphy.2000.6013,
  10.1016/0003-4916(82)90164-6}{\emph{Annals Phys.} {\bfseries 140} (1982)
  372--411}.

\bibitem{Deser:1982vy}
S.~Deser, R.~Jackiw and S.~Templeton, \emph{{Three-Dimensional Massive Gauge
  Theories}}, \href{http://dx.doi.org/10.1103/PhysRevLett.48.975}{\emph{Phys.
  Rev. Lett.} {\bfseries 48} (1982) 975--978}.

\bibitem{Deser:1984kw}
S.~Deser and R.~Jackiw, \emph{{'Selfduality' of Topologically Massive Gauge
  Theories}}, \href{http://dx.doi.org/10.1016/0370-2693(84)91833-1}{\emph{Phys.
  Lett.} {\bfseries 139B} (1984) 371--373}.

\bibitem{Bergshoeff:2009aq}
E.~A. Bergshoeff, O.~Hohm and P.~K. Townsend, \emph{{More on Massive 3D
  Gravity}}, \href{http://dx.doi.org/10.1103/PhysRevD.79.124042}{\emph{Phys.
  Rev. D} {\bfseries 79} (2009) 124042},
  [\href{https://arxiv.org/abs/0905.1259}{{\ttfamily 0905.1259}}].

\bibitem{Bergshoeff:2009hq}
E.~A. Bergshoeff, O.~Hohm and P.~K. Townsend, \emph{{Massive Gravity in Three
  Dimensions}},
  \href{http://dx.doi.org/10.1103/PhysRevLett.102.201301}{\emph{Phys. Rev.
  Lett.} {\bfseries 102} (2009) 201301},
  [\href{https://arxiv.org/abs/0901.1766}{{\ttfamily 0901.1766}}].

\bibitem{Bergshoeff:2013xma}
E.~A. Bergshoeff, S.~de~Haan, O.~Hohm, W.~Merbis and P.~K. Townsend,
  \emph{{Zwei-Dreibein Gravity: A Two-Frame-Field Model of 3D Massive
  Gravity}}, \href{http://dx.doi.org/10.1103/PhysRevLett.111.111102,
  10.1103/PhysRevLett.111.259902}{\emph{Phys. Rev. Lett.} {\bfseries 111}
  (2013) 111102}, [\href{https://arxiv.org/abs/1307.2774}{{\ttfamily
  1307.2774}}].

\bibitem{Alexandrov:2014oda}
S.~Alexandrov and C.~Deffayet, \emph{{On Partially Massless Theory in 3
  Dimensions}},
  \href{http://dx.doi.org/10.1088/1475-7516/2015/03/043}{\emph{JCAP} {\bfseries
  1503} (2015) 043}, [\href{https://arxiv.org/abs/1410.2897}{{\ttfamily
  1410.2897}}].

\bibitem{Bergshoeff:2015zga}
E.~Bergshoeff, W.~Merbis, A.~J. Routh and P.~K. Townsend, \emph{{The Third Way
  to 3D Gravity}},
  \href{http://dx.doi.org/10.1142/S0218271815440150}{\emph{Int. J. Mod. Phys.}
  {\bfseries D24} (2015) 1544015},
  [\href{https://arxiv.org/abs/1506.05949}{{\ttfamily 1506.05949}}].

\bibitem{Afshar:2014ffa}
H.~R. Afshar, E.~A. Bergshoeff and W.~Merbis, \emph{{Extended massive gravity
  in three dimensions}},
  \href{http://dx.doi.org/10.1007/JHEP08(2014)115}{\emph{JHEP} {\bfseries 08}
  (2014) 115}, [\href{https://arxiv.org/abs/1405.6213}{{\ttfamily 1405.6213}}].

\bibitem{Bergshoeff:2014pca}
E.~Bergshoeff, O.~Hohm, W.~Merbis, A.~J. Routh and P.~K. Townsend,
  \emph{{Minimal Massive 3D Gravity}},
  \href{http://dx.doi.org/10.1088/0264-9381/31/14/145008}{\emph{Class. Quant.
  Grav.} {\bfseries 31} (2014) 145008},
  [\href{https://arxiv.org/abs/1404.2867}{{\ttfamily 1404.2867}}].

\bibitem{Bergshoeff:2014bia}
E.~A. Bergshoeff, O.~Hohm, W.~Merbis, A.~J. Routh and P.~K. Townsend,
  \emph{{Chern-Simons-like Gravity Theories}},
  \href{http://dx.doi.org/10.1007/978-3-319-10070-8_7}{\emph{Lect. Notes Phys.}
  {\bfseries 892} (2015) 181--201},
  [\href{https://arxiv.org/abs/1402.1688}{{\ttfamily 1402.1688}}].

\bibitem{Merbis:2014vja}
W.~Merbis, \emph{{Chern-Simons-like Theories of Gravity}},  other thesis, 11,
  2014.

\bibitem{Geiller:2018ain}
M.~Geiller and K.~Noui, \emph{{A remarkably simple theory of 3d massive
  gravity}}, \href{http://dx.doi.org/10.1007/JHEP04(2019)091}{\emph{JHEP}
  {\bfseries 04} (2019) 091},
  [\href{https://arxiv.org/abs/1812.01018}{{\ttfamily 1812.01018}}].

\bibitem{Geiller:2019dpc}
M.~Geiller and K.~Noui, \emph{{Metric formulation of the simple theory of 3d
  massive gravity}},
  \href{http://dx.doi.org/10.1103/PhysRevD.100.064066}{\emph{Phys. Rev. D}
  {\bfseries 100} (2019) 064066},
  [\href{https://arxiv.org/abs/1905.04390}{{\ttfamily 1905.04390}}].

\bibitem{Ashtekar:1996cd}
A.~Ashtekar, J.~Bicak and B.~G. Schmidt, \emph{{Asymptotic structure of
  symmetry reduced general relativity}},
  \href{http://dx.doi.org/10.1103/PhysRevD.55.669}{\emph{Phys. Rev. D}
  {\bfseries 55} (1997) 669--686},
  [\href{https://arxiv.org/abs/gr-qc/9608042}{{\ttfamily gr-qc/9608042}}].

\bibitem{Barnich:2006av}
G.~Barnich and G.~Compere, \emph{{Classical central extension for asymptotic
  symmetries at null infinity in three spacetime dimensions}},
  \href{http://dx.doi.org/10.1088/0264-9381/24/5/F01}{\emph{Class. Quant.
  Grav.} {\bfseries 24} (2007) F15--F23},
  [\href{https://arxiv.org/abs/gr-qc/0610130}{{\ttfamily gr-qc/0610130}}].

\bibitem{Barnich:2014kra}
G.~Barnich and B.~Oblak, \emph{{Notes on the BMS group in three dimensions: I.
  Induced representations}},
  \href{http://dx.doi.org/10.1007/JHEP06(2014)129}{\emph{JHEP} {\bfseries 06}
  (2014) 129}, [\href{https://arxiv.org/abs/1403.5803}{{\ttfamily 1403.5803}}].

\bibitem{Barnich:2015uva}
G.~Barnich and B.~Oblak, \emph{{Notes on the BMS group in three dimensions: II.
  Coadjoint representation}},
  \href{http://dx.doi.org/10.1007/JHEP03(2015)033}{\emph{JHEP} {\bfseries 03}
  (2015) 033}, [\href{https://arxiv.org/abs/1502.00010}{{\ttfamily
  1502.00010}}].

\bibitem{Barnich:2015sca}
G.~Barnich, L.~Donnay, J.~Matulich and R.~Troncoso, \emph{{Super-BMS$_{3}$
  invariant boundary theory from three-dimensional flat supergravity}},
  \href{http://dx.doi.org/10.1007/JHEP01(2017)029}{\emph{JHEP} {\bfseries 01}
  (2017) 029}, [\href{https://arxiv.org/abs/1510.08824}{{\ttfamily
  1510.08824}}].

\bibitem{Barnich:2015mui}
G.~Barnich, H.~A. Gonzalez, A.~Maloney and B.~Oblak, \emph{{One-loop partition
  function of three-dimensional flat gravity}},
  \href{http://dx.doi.org/10.1007/JHEP04(2015)178}{\emph{JHEP} {\bfseries 04}
  (2015) 178}, [\href{https://arxiv.org/abs/1502.06185}{{\ttfamily
  1502.06185}}].

\bibitem{Oblak:2016eij}
B.~Oblak, \emph{{BMS Particles in Three Dimensions}}.
\newblock PhD thesis, Brussels U., 2016.
\newblock \href{https://arxiv.org/abs/1610.08526}{{\ttfamily 1610.08526}}.
\newblock 10.1007/978-3-319-61878-4.

\bibitem{Garbarz:2015lua}
A.~Garbarz and M.~Leston, \emph{{Quantization of BMS$_3$ orbits: a perturbative
  approach}},
  \href{http://dx.doi.org/10.1016/j.nuclphysb.2016.02.038}{\emph{Nucl. Phys. B}
  {\bfseries 906} (2016) 133--146},
  [\href{https://arxiv.org/abs/1507.00339}{{\ttfamily 1507.00339}}].

\bibitem{Strominger:2017zoo}
A.~Strominger, \emph{{Lectures on the Infrared Structure of Gravity and Gauge
  Theory}},  [\href{https://arxiv.org/abs/1703.05448}{{\ttfamily 1703.05448}}].

\bibitem{Bagchi:2019unf}
A.~Bagchi, A.~Saha and Zodinmawia, \emph{{BMS Characters and Modular
  Invariance}}, \href{http://dx.doi.org/10.1007/JHEP07(2019)138}{\emph{JHEP}
  {\bfseries 07} (2019) 138},
  [\href{https://arxiv.org/abs/1902.07066}{{\ttfamily 1902.07066}}].

\bibitem{Grumiller:2016pqb}
D.~Grumiller and M.~Riegler, \emph{{Most general AdS$_{3}$ boundary
  conditions}}, \href{http://dx.doi.org/10.1007/JHEP10(2016)023}{\emph{JHEP}
  {\bfseries 10} (2016) 023},
  [\href{https://arxiv.org/abs/1608.01308}{{\ttfamily 1608.01308}}].

\bibitem{Grumiller:2017sjh}
D.~Grumiller, W.~Merbis and M.~Riegler, \emph{{Most general flat space boundary
  conditions in three-dimensional Einstein gravity}},
  \href{http://dx.doi.org/10.1088/1361-6382/aa8004}{\emph{Class. Quant. Grav.}
  {\bfseries 34} (2017) 184001},
  [\href{https://arxiv.org/abs/1704.07419}{{\ttfamily 1704.07419}}].

\bibitem{Mielke:1991nn}
E.~W. Mielke and P.~Baekler, \emph{{Topological gauge model of gravity with
  torsion}}, \href{http://dx.doi.org/10.1016/0375-9601(91)90715-K}{\emph{Phys.
  Lett. A} {\bfseries 156} (1991) 399--403}.

\bibitem{Baekler:1992ab}
P.~Baekler, E.~Mielke and F.~Hehl, \emph{{Dynamical symmetries in topological
  3-D gravity with torsion}},
  \href{http://dx.doi.org/10.1007/BF02726888}{\emph{Nuovo Cim. B} {\bfseries
  107} (1992) 91--110}.

\bibitem{Cacciatori:2005wz}
S.~L. Cacciatori, M.~M. Caldarelli, A.~Giacomini, D.~Klemm and D.~S. Mansi,
  \emph{{Chern-Simons formulation of three-dimensional gravity with torsion and
  nonmetricity}},
  \href{http://dx.doi.org/10.1016/j.geomphys.2006.01.006}{\emph{J. Geom. Phys.}
  {\bfseries 56} (2006) 2523--2543},
  [\href{https://arxiv.org/abs/hep-th/0507200}{{\ttfamily hep-th/0507200}}].

\bibitem{Giacomini:2006dr}
A.~Giacomini, R.~Troncoso and S.~Willison, \emph{{Three-dimensional
  supergravity reloaded}},
  \href{http://dx.doi.org/10.1088/0264-9381/24/11/005}{\emph{Class. Quant.
  Grav.} {\bfseries 24} (2007) 2845--2860},
  [\href{https://arxiv.org/abs/hep-th/0610077}{{\ttfamily hep-th/0610077}}].

\bibitem{Blagojevic:2004hj}
M.~Blagojevic and B.~Cvetkovic, \emph{{Canonical structure of 3-D gravity with
  torsion}},  [\href{https://arxiv.org/abs/gr-qc/0412134}{{\ttfamily
  gr-qc/0412134}}].

\bibitem{Blagojevic:2006jk}
M.~Blagojevic and B.~Cvetkovic, \emph{{Black hole entropy in 3-D gravity with
  torsion}}, \href{http://dx.doi.org/10.1088/0264-9381/23/14/013}{\emph{Class.
  Quant. Grav.} {\bfseries 23} (2006) 4781},
  [\href{https://arxiv.org/abs/gr-qc/0601006}{{\ttfamily gr-qc/0601006}}].

\bibitem{Blagojevic:2006hh}
M.~Blagojevic and B.~Cvetkovic, \emph{{Black hole entropy from the boundary
  conformal structure in 3D gravity with torsion}},
  \href{http://dx.doi.org/10.1088/1126-6708/2006/10/005}{\emph{JHEP} {\bfseries
  10} (2006) 005}, [\href{https://arxiv.org/abs/gr-qc/0606086}{{\ttfamily
  gr-qc/0606086}}].

\bibitem{Blagojevic:2005pd}
M.~Blagojevic and B.~Cvetkovic, \emph{{Asymptotic charges in 3-D gravity with
  torsion}}, \href{http://dx.doi.org/10.1088/1742-6596/33/1/026}{\emph{J. Phys.
  Conf. Ser.} {\bfseries 33} (2006) 248--253},
  [\href{https://arxiv.org/abs/gr-qc/0511162}{{\ttfamily gr-qc/0511162}}].

\bibitem{Blagojevic:2013bu}
M.~Blagojevic, B.~Cvetkovic, O.~Miskovic and R.~Olea, \emph{{Holography in 3D
  AdS gravity with torsion}},
  \href{http://dx.doi.org/10.1007/JHEP05(2013)103}{\emph{JHEP} {\bfseries 05}
  (2013) 103}, [\href{https://arxiv.org/abs/1301.1237}{{\ttfamily 1301.1237}}].

\bibitem{Cvetkovic:2018ati}
B.~Cvetkovi\'c and D.~Simi\'c, \emph{{Near-horizon geometry with torsion}},
  \href{http://dx.doi.org/10.1103/PhysRevD.99.024032}{\emph{Phys. Rev. D}
  {\bfseries 99} (2019) 024032},
  [\href{https://arxiv.org/abs/1809.00555}{{\ttfamily 1809.00555}}].

\bibitem{Klemm:2007yu}
D.~Klemm and G.~Tagliabue, \emph{{The CFT dual of AdS gravity with torsion}},
  \href{http://dx.doi.org/10.1088/0264-9381/25/3/035011}{\emph{Class. Quant.
  Grav.} {\bfseries 25} (2008) 035011},
  [\href{https://arxiv.org/abs/0705.3320}{{\ttfamily 0705.3320}}].

\bibitem{Ning:2018gfm}
C.-H. Wei and B.~Ning, \emph{{Quasi-local Energy in 3D Gravity with Torsion}},
  [\href{https://arxiv.org/abs/1807.08736}{{\ttfamily 1807.08736}}].

\bibitem{Peleteiro:2020ubv}
J.~Peleteiro and C.~Valc\'arcel, \emph{{Spin-3 fields in Mielke-Baekler
  gravity}}, \href{http://dx.doi.org/10.1088/1361-6382/ab9882}{\emph{Class.
  Quant. Grav.} {\bfseries 37} (2020) 185010},
  [\href{https://arxiv.org/abs/2003.02627}{{\ttfamily 2003.02627}}].

\bibitem{artamonov2016introduction}
D.~V. Artamonov, \emph{{Introduction to finite $W$-algebras}},
  [\href{https://arxiv.org/abs/1607.01697}{{\ttfamily 1607.01697}}].

\bibitem{Freidel:2020xyx}
L.~Freidel, M.~Geiller and D.~Pranzetti, \emph{{Edge modes of gravity - I:
  Corner potentials and charges}},
  [\href{https://arxiv.org/abs/2006.12527}{{\ttfamily 2006.12527}}].

\bibitem{Freidel:2020svx}
L.~Freidel, M.~Geiller and D.~Pranzetti, \emph{{Edge modes of gravity - II:
  Corner metric and Lorentz charges}},
  [\href{https://arxiv.org/abs/2007.03563}{{\ttfamily 2007.03563}}].

\bibitem{Freidel:2020ayo}
L.~Freidel, M.~Geiller and D.~Pranzetti, \emph{{Edge modes of gravity - III:
  Corner simplicity constraints}},
  [\href{https://arxiv.org/abs/2007.12635}{{\ttfamily 2007.12635}}].

\bibitem{Chen_2016}
X.~Chen, A.~Tiwari and S.~Ryu, \emph{Bulk-boundary correspondence in
  (3+1)-dimensional topological phases},
  \href{http://dx.doi.org/10.1103/physrevb.94.045113}{\emph{Physical Review B}
  {\bfseries 94} (Jul, 2016) }.

\bibitem{Chen_2017}
X.~Chen, A.~Tiwari, C.~Nayak and S.~Ryu, \emph{Gauging (3+1)-dimensional
  topological phases: An approach from surface theories},
  \href{http://dx.doi.org/10.1103/physrevb.96.165112}{\emph{Physical Review B}
  {\bfseries 96} (Oct, 2017) }.

\bibitem{Wen:2016snr}
X.~Wen, S.~Matsuura and S.~Ryu, \emph{{Edge theory approach to topological
  entanglement entropy, mutual information and entanglement negativity in
  Chern-Simons theories}},
  \href{http://dx.doi.org/10.1103/PhysRevB.93.245140}{\emph{Phys. Rev.}
  {\bfseries B93} (2016) 245140},
  [\href{https://arxiv.org/abs/1603.08534}{{\ttfamily 1603.08534}}].

\bibitem{Delcamp:2016eya}
C.~Delcamp, B.~Dittrich and A.~Riello, \emph{{On entanglement entropy in
  non-Abelian lattice gauge theory and 3D quantum gravity}},
  \href{http://dx.doi.org/10.1007/JHEP11(2016)102}{\emph{JHEP} {\bfseries 11}
  (2016) 102}, [\href{https://arxiv.org/abs/1609.04806}{{\ttfamily
  1609.04806}}].

\bibitem{Aldrovandi:2013wha}
R.~Aldrovandi and J.~G. Pereira, \emph{{Teleparallel Gravity}: {An
  Introduction}}, vol.~173.
\newblock Springer, 2013,
  \href{http://dx.doi.org/10.1007/978-94-007-5143-9}{10.1007/978-94-007-5143-9}.

\bibitem{Delcamp:2018sef}
C.~Delcamp, L.~Freidel and F.~Girelli, \emph{{Dual loop quantizations of 3d
  gravity}},  [\href{https://arxiv.org/abs/1803.03246}{{\ttfamily
  1803.03246}}].

\bibitem{Dupuis:2019unm}
M.~Dupuis, F.~Girelli, A.~Osumanu and W.~Wieland, \emph{{First-order
  formulation of teleparallel gravity and dual loop gravity}},
  \href{http://dx.doi.org/10.1088/1361-6382/ab7bb7}{\emph{Class. Quant. Grav.}
  {\bfseries 37} (2020) 085023},
  [\href{https://arxiv.org/abs/1906.02801}{{\ttfamily 1906.02801}}].

\bibitem{Banerjee:2012jn}
R.~Banerjee and D.~Roy, \emph{{Trivial symmetries in a 3D topological torsion
  model of gravity}},
  \href{http://dx.doi.org/10.1088/1742-6596/405/1/012028}{\emph{J. Phys. Conf.
  Ser.} {\bfseries 405} (2012) 012028},
  [\href{https://arxiv.org/abs/1212.4238}{{\ttfamily 1212.4238}}].

\bibitem{Banerjee:2009vf}
R.~Banerjee, S.~Gangopadhyay, P.~Mukherjee and D.~Roy, \emph{{Symmetries of
  topological gravity with torsion in the Hamiltonian and Lagrangian
  formalisms}}, \href{http://dx.doi.org/10.1007/JHEP02(2010)075}{\emph{JHEP}
  {\bfseries 02} (2010) 075},
  [\href{https://arxiv.org/abs/0912.1472}{{\ttfamily 0912.1472}}].

\bibitem{Banerjee:2011cu}
R.~Banerjee and D.~Roy, \emph{{Poincare gauge symmetries, Hamiltonian
  symmetries and trivial gauge transformations}},
  \href{http://dx.doi.org/10.1103/PhysRevD.84.124034}{\emph{Phys. Rev. D}
  {\bfseries 84} (2011) 124034},
  [\href{https://arxiv.org/abs/1110.1720}{{\ttfamily 1110.1720}}].

\bibitem{Barnich:2012aw}
G.~Barnich, A.~Gomberoff and H.~A. Gonzalez, \emph{{The Flat limit of three
  dimensional asymptotically anti-de Sitter spacetimes}},
  \href{http://dx.doi.org/10.1103/PhysRevD.86.024020}{\emph{Phys. Rev. D}
  {\bfseries 86} (2012) 024020},
  [\href{https://arxiv.org/abs/1204.3288}{{\ttfamily 1204.3288}}].

\bibitem{Barnich:2012rz}
G.~Barnich, A.~Gomberoff and H.~A. Gonz\'alez, \emph{{Three-dimensional
  Bondi-Metzner-Sachs invariant two-dimensional field theories as the flat
  limit of Liouville theory}},
  \href{http://dx.doi.org/10.1103/PhysRevD.87.124032}{\emph{Phys. Rev. D}
  {\bfseries 87} (2013) 124032},
  [\href{https://arxiv.org/abs/1210.0731}{{\ttfamily 1210.0731}}].

\bibitem{Compere:2014cna}
G.~Comp\`ere, L.~Donnay, P.-H. Lambert and W.~Schulgin, \emph{{Liouville theory
  beyond the cosmological horizon}},
  \href{http://dx.doi.org/10.1007/JHEP03(2015)158}{\emph{JHEP} {\bfseries 03}
  (2015) 158}, [\href{https://arxiv.org/abs/1411.7873}{{\ttfamily 1411.7873}}].

\bibitem{Compere:2015knw}
G.~Comp\`ere, P.~Mao, A.~Seraj and M.~Sheikh-Jabbari, \emph{{Symplectic and
  Killing symmetries of AdS$_{3}$ gravity: holographic vs boundary gravitons}},
  \href{http://dx.doi.org/10.1007/JHEP01(2016)080}{\emph{JHEP} {\bfseries 01}
  (2016) 080}, [\href{https://arxiv.org/abs/1511.06079}{{\ttfamily
  1511.06079}}].

\bibitem{Adami:2020ugu}
H.~Adami, M.~Sheikh-Jabbari, V.~Taghiloo, H.~Yavartanoo and C.~Zwikel,
  \emph{{Symmetries at null boundaries: two and three dimensional gravity
  cases}}, \href{http://dx.doi.org/10.1007/JHEP10(2020)107}{\emph{JHEP}
  {\bfseries 10} (2020) 107},
  [\href{https://arxiv.org/abs/2007.12759}{{\ttfamily 2007.12759}}].

\bibitem{Speranza:2017gxd}
A.~J. Speranza, \emph{{Local phase space and edge modes for
  diffeomorphism-invariant theories}},
  \href{http://dx.doi.org/10.1007/JHEP02(2018)021}{\emph{JHEP} {\bfseries 02}
  (2018) 021}, [\href{https://arxiv.org/abs/1706.05061}{{\ttfamily
  1706.05061}}].

\bibitem{Harlow:2019yfa}
D.~Harlow and J.-Q. Wu, \emph{{Covariant phase space with boundaries}},
  [\href{https://arxiv.org/abs/1906.08616}{{\ttfamily 1906.08616}}].

\bibitem{Geiller:2019bti}
M.~Geiller and P.~Jai-akson, \emph{{Extended actions, dynamics of edge modes,
  and entanglement entropy}},
  \href{http://dx.doi.org/10.1007/JHEP09(2020)134}{\emph{JHEP} {\bfseries 20}
  (2020) 134}, [\href{https://arxiv.org/abs/1912.06025}{{\ttfamily
  1912.06025}}].

\bibitem{Chandrasekaran:2020wwn}
V.~Chandrasekaran and A.~J. Speranza, \emph{{Anomalies in gravitational charge
  algebras of null boundaries and black hole entropy}},
  [\href{https://arxiv.org/abs/2009.10739}{{\ttfamily 2009.10739}}].

\bibitem{Grumiller:2019ygj}
D.~Grumiller, M.~Sheikh-Jabbari, C.~Troessaert and R.~Wutte,
  \emph{{Interpolating Between Asymptotic and Near Horizon Symmetries}},
  \href{http://dx.doi.org/10.1007/JHEP03(2020)035}{\emph{JHEP} {\bfseries 03}
  (2020) 035}, [\href{https://arxiv.org/abs/1911.04503}{{\ttfamily
  1911.04503}}].

\bibitem{Jacobson:2015uqa}
T.~Jacobson and A.~Mohd, \emph{{Black hole entropy and Lorentz-diffeomorphism
  Noether charge}},
  \href{http://dx.doi.org/10.1103/PhysRevD.92.124010}{\emph{Phys. Rev.}
  {\bfseries D92} (2015) 124010},
  [\href{https://arxiv.org/abs/1507.01054}{{\ttfamily 1507.01054}}].

\bibitem{Prabhu:2015vua}
K.~Prabhu, \emph{{The First Law of Black Hole Mechanics for Fields with
  Internal Gauge Freedom}},
  \href{http://dx.doi.org/10.1088/1361-6382/aa536b}{\emph{Class. Quant. Grav.}
  {\bfseries 34} (2017) 035011},
  [\href{https://arxiv.org/abs/1511.00388}{{\ttfamily 1511.00388}}].

\bibitem{DePaoli:2018erh}
E.~De~Paoli and S.~Speziale, \emph{{A gauge-invariant symplectic potential for
  tetrad general relativity}},
  \href{http://dx.doi.org/10.1007/JHEP07(2018)040}{\emph{JHEP} {\bfseries 07}
  (2018) 040}, [\href{https://arxiv.org/abs/1804.09685}{{\ttfamily
  1804.09685}}].

\bibitem{Oliveri:2019gvm}
R.~Oliveri and S.~Speziale, \emph{{Boundary effects in General Relativity with
  tetrad variables}},  [\href{https://arxiv.org/abs/1912.01016}{{\ttfamily
  1912.01016}}].

\bibitem{Oliveri:2020xls}
R.~Oliveri and S.~Speziale, \emph{{A note on dual gravitational charges}},
  [\href{https://arxiv.org/abs/2010.01111}{{\ttfamily 2010.01111}}].

\bibitem{Godazgar:2018qpq}
H.~Godazgar, M.~Godazgar and C.~Pope, \emph{{New dual gravitational charges}},
  \href{http://dx.doi.org/10.1103/PhysRevD.99.024013}{\emph{Phys. Rev. D}
  {\bfseries 99} (2019) 024013},
  [\href{https://arxiv.org/abs/1812.01641}{{\ttfamily 1812.01641}}].

\bibitem{Godazgar:2018dvh}
H.~Godazgar, M.~Godazgar and C.~Pope, \emph{{Tower of subleading dual BMS
  charges}}, \href{http://dx.doi.org/10.1007/JHEP03(2019)057}{\emph{JHEP}
  {\bfseries 03} (2019) 057},
  [\href{https://arxiv.org/abs/1812.06935}{{\ttfamily 1812.06935}}].

\bibitem{Godazgar:2019dkh}
H.~Godazgar, M.~Godazgar and C.~Pope, \emph{{Dual gravitational charges and
  soft theorems}}, \href{http://dx.doi.org/10.1007/JHEP10(2019)123}{\emph{JHEP}
  {\bfseries 10} (2019) 123},
  [\href{https://arxiv.org/abs/1908.01164}{{\ttfamily 1908.01164}}].

\bibitem{Godazgar:2020kqd}
H.~Godazgar, M.~Godazgar and M.~J. Perry, \emph{{Hamiltonian derivation of dual
  gravitational charges}},
  \href{http://dx.doi.org/10.1007/JHEP09(2020)084}{\emph{JHEP} {\bfseries 20}
  (2020) 084}, [\href{https://arxiv.org/abs/2007.07144}{{\ttfamily
  2007.07144}}].

\bibitem{Godazgar:2020gqd}
H.~Godazgar, M.~Godazgar and M.~J. Perry, \emph{{Asymptotic gravitational
  charges}},
  \href{http://dx.doi.org/10.1103/PhysRevLett.125.101301}{\emph{Phys. Rev.
  Lett.} {\bfseries 125} (2020) 101301},
  [\href{https://arxiv.org/abs/2007.01257}{{\ttfamily 2007.01257}}].

\bibitem{Kol:2019nkc}
U.~Kol and M.~Porrati, \emph{{Properties of Dual Supertranslation Charges in
  Asymptotically Flat Spacetimes}},
  \href{http://dx.doi.org/10.1103/PhysRevD.100.046019}{\emph{Phys. Rev. D}
  {\bfseries 100} (2019) 046019},
  [\href{https://arxiv.org/abs/1907.00990}{{\ttfamily 1907.00990}}].

\bibitem{Kol:2020vet}
U.~Kol, \emph{{Subleading BMS Charges and The Lorentz Group}},
  [\href{https://arxiv.org/abs/2011.06008}{{\ttfamily 2011.06008}}].

\bibitem{Balasubramanian:1999re}
V.~Balasubramanian and P.~Kraus, \emph{{A Stress tensor for Anti-de Sitter
  gravity}}, \href{http://dx.doi.org/10.1007/s002200050764}{\emph{Commun. Math.
  Phys.} {\bfseries 208} (1999) 413--428},
  [\href{https://arxiv.org/abs/hep-th/9902121}{{\ttfamily hep-th/9902121}}].

\bibitem{Fareghbal:2013ifa}
R.~Fareghbal and A.~Naseh, \emph{{Flat-Space Energy-Momentum Tensor from
  BMS/GCA Correspondence}},
  \href{http://dx.doi.org/10.1007/JHEP03(2014)005}{\emph{JHEP} {\bfseries 03}
  (2014) 005}, [\href{https://arxiv.org/abs/1312.2109}{{\ttfamily 1312.2109}}].

\bibitem{Barnich:2013yka}
G.~Barnich and H.~A. Gonzalez, \emph{{Dual dynamics of three dimensional
  asymptotically flat Einstein gravity at null infinity}},
  \href{http://dx.doi.org/10.1007/JHEP05(2013)016}{\emph{JHEP} {\bfseries 05}
  (2013) 016}, [\href{https://arxiv.org/abs/1303.1075}{{\ttfamily 1303.1075}}].

\bibitem{Carlip:2016lnw}
S.~Carlip, \emph{{The Dynamics of Supertranslations and Superrotations in 2+1
  Dimensions}},  [\href{https://arxiv.org/abs/1608.05088}{{\ttfamily
  1608.05088}}].

\bibitem{Carlip:1989nz}
S.~Carlip, \emph{{Exact Quantum Scattering in (2+1)-Dimensional Gravity}},
  \href{http://dx.doi.org/10.1016/0550-3213(89)90183-1}{\emph{Nucl. Phys. B}
  {\bfseries 324} (1989) 106--122}.

\bibitem{MR1080700}
P.~de~Sousa~Gerbert, \emph{On spin and (quantum) gravity in {$2+1$}
  dimensions},
  \href{http://dx.doi.org/10.1016/0550-3213(90)90288-O}{\emph{Nuclear Phys. B}
  {\bfseries 346} (1990) 440--472}.

\bibitem{Hooft:1993gz}
G.~'t~Hooft, \emph{{The Evolution of gravitating point particles in (2+1)-
  dimensions}}, {\emph{Class. Quant. Grav.} {\bfseries 10} (1993) 1023--1038}.

\bibitem{Hooft:1993nj}
G.~'t~Hooft, \emph{{Canonical quantization of gravitating point particles in
  (2+1)-dimensions}}, {\emph{Class. Quant. Grav.} {\bfseries 10} (1993)
  1653--1664}, [\href{https://arxiv.org/abs/gr-qc/9305008}{{\ttfamily
  gr-qc/9305008}}].

\bibitem{Matschull:1997du}
H.-J. Matschull and M.~Welling, \emph{{Quantum mechanics of a point particle in
  2+1 dimensional gravity}},
  \href{http://dx.doi.org/10.1088/0264-9381/15/10/008}{\emph{Class. Quant.
  Grav.} {\bfseries 15} (1998) 2981--3030},
  [\href{https://arxiv.org/abs/gr-qc/9708054}{{\ttfamily gr-qc/9708054}}].

\bibitem{Matschull:2001ec}
H.-J. Matschull, \emph{{The Phase space structure of multi particle models in
  2+1 gravity}},
  \href{http://dx.doi.org/10.1088/0264-9381/18/17/309}{\emph{Class.Quant.Grav.}
  {\bfseries 18} (2001) 3497--3560},
  [\href{https://arxiv.org/abs/gr-qc/0103084}{{\ttfamily gr-qc/0103084}}].

\bibitem{Buffenoir:2003zu}
E.~Buffenoir and K.~Noui, \emph{{Unfashionable observations about
  three-dimensional gravity}},
  [\href{https://arxiv.org/abs/gr-qc/0305079}{{\ttfamily gr-qc/0305079}}].

\bibitem{Dittrich:2016typ}
B.~Dittrich and M.~Geiller, \emph{{Quantum gravity kinematics from extended
  TQFTs}}, \href{http://dx.doi.org/10.1088/1367-2630/aa54e2}{\emph{New J.
  Phys.} {\bfseries 19} (2017) 013003},
  [\href{https://arxiv.org/abs/1604.05195}{{\ttfamily 1604.05195}}].

\bibitem{Noui:2004iz}
K.~Noui and A.~Perez, \emph{{Three dimensional loop quantum gravity: Coupling
  to point particles}}, {\emph{Class. Quant. Grav.} {\bfseries 22} (2005)
  4489--4514}, [\href{https://arxiv.org/abs/gr-qc/0402111}{{\ttfamily
  gr-qc/0402111}}].

\bibitem{Schroers:2007ey}
B.~Schroers, \emph{{Lessons from (2+1)-dimensional quantum gravity}},
  \href{http://dx.doi.org/10.22323/1.043.0035}{\emph{PoS} {\bfseries QG-PH}
  (2007) 035}, [\href{https://arxiv.org/abs/0710.5844}{{\ttfamily 0710.5844}}].

\bibitem{FGL-toappear}
L.~Freidel, C.~Goeller and E.~Livine, \emph{The quantum gravity disk: Discrete
  current algebra}, {\emph{, to appear} (2020) }.

\bibitem{Noui:2004ja}
K.~Noui and A.~Perez, \emph{{Dynamics of loop quantum gravity and spin foam
  models in three dimensions}},
  [\href{https://arxiv.org/abs/gr-qc/0402112}{{\ttfamily gr-qc/0402112}}].

\bibitem{Freidel:2000uq}
L.~Freidel, \emph{{A Ponzano-Regge model of Lorentzian 3-dimensional gravity}},
  \href{http://dx.doi.org/10.1016/S0920-5632(00)00775-1}{\emph{Nucl. Phys. B
  Proc. Suppl.} {\bfseries 88} (2000) 237--240},
  [\href{https://arxiv.org/abs/gr-qc/0102098}{{\ttfamily gr-qc/0102098}}].

\bibitem{Freidel:2004vi}
L.~Freidel and D.~Louapre, \emph{{Ponzano-Regge model revisited. I: Gauge
  fixing, observables and interacting spinning particles}},
  \href{http://dx.doi.org/10.1088/0264-9381/21/24/002}{\emph{Class. Quant.
  Grav.} {\bfseries 21} (2004) 5685--5726},
  [\href{https://arxiv.org/abs/hep-th/0401076}{{\ttfamily hep-th/0401076}}].

\bibitem{Barrett:2008wh}
J.~W. Barrett and I.~Naish-Guzman, \emph{{The Ponzano-Regge model}},
  \href{http://dx.doi.org/10.1088/0264-9381/26/15/155014}{\emph{Class. Quant.
  Grav.} {\bfseries 26} (2009) 155014},
  [\href{https://arxiv.org/abs/0803.3319}{{\ttfamily 0803.3319}}].

\bibitem{1992PhRvL..68.1795M}
S.~{Mizoguchi} and T.~{Tada}, \emph{{Three-dimensional gravity from the
  Turaev-Viro invariant}},
  \href{http://dx.doi.org/10.1103/PhysRevLett.68.1795}{\emph{Phys. Rev. Lett.}
  {\bfseries 68} (1992) 1795--1798},
  [\href{https://arxiv.org/abs/arXiv:hep-th/9110057}{{\ttfamily
  arXiv:hep-th/9110057}}].

\bibitem{Turaev:1992hq}
V.~Turaev and O.~Viro, \emph{{State sum invariants of 3 manifolds and quantum
  6j symbols}},
  \href{http://dx.doi.org/10.1016/0040-9383(92)90015-A}{\emph{Topology}
  {\bfseries 31} (1992) 865--902}.

\bibitem{Grumiller:2002nm}
D.~Grumiller, W.~Kummer and D.~Vassilevich, \emph{{Dilaton gravity in
  two-dimensions}},
  \href{http://dx.doi.org/10.1016/S0370-1573(02)00267-3}{\emph{Phys. Rept.}
  {\bfseries 369} (2002) 327--430},
  [\href{https://arxiv.org/abs/hep-th/0204253}{{\ttfamily hep-th/0204253}}].

\bibitem{Mertens:2018fds}
T.~G. Mertens, \emph{{The Schwarzian Theory - Origins}},
  [\href{https://arxiv.org/abs/1801.09605}{{\ttfamily 1801.09605}}].

\bibitem{Gaikwad:2018dfc}
A.~Gaikwad, L.~K. Joshi, G.~Mandal and S.~R. Wadia, \emph{{Holographic dual to
  charged SYK from 3D Gravity and Chern-Simons}},
  \href{http://dx.doi.org/10.1007/JHEP02(2020)033}{\emph{JHEP} {\bfseries 02}
  (2020) 033}, [\href{https://arxiv.org/abs/1802.07746}{{\ttfamily
  1802.07746}}].

\bibitem{Skenderis:2009nt}
K.~Skenderis, M.~Taylor and B.~C. van Rees, \emph{{Topologically Massive
  Gravity and the AdS/CFT Correspondence}},
  \href{http://dx.doi.org/10.1088/1126-6708/2009/09/045}{\emph{JHEP} {\bfseries
  09} (2009) 045}, [\href{https://arxiv.org/abs/0906.4926}{{\ttfamily
  0906.4926}}].

\bibitem{Concha:2018zeb}
P.~Concha, N.~Merino, O.~Miskovic, E.~Rodr\'\i{}guez, P.~Salgado-Rebolledo and
  O.~Valdivia, \emph{{Asymptotic symmetries of three-dimensional Chern-Simons
  gravity for the Maxwell algebra}},
  \href{http://dx.doi.org/10.1007/JHEP10(2018)079}{\emph{JHEP} {\bfseries 10}
  (2018) 079}, [\href{https://arxiv.org/abs/1805.08834}{{\ttfamily
  1805.08834}}].

\bibitem{Adami:2020xkm}
H.~Adami, P.~Concha, E.~Rodriguez and H.~Safari, \emph{{Asymptotic symmetries
  of Maxwell Chern\textendash{}Simons gravity with torsion}},
  \href{http://dx.doi.org/10.1140/epjc/s10052-020-08537-z}{\emph{Eur. Phys. J.
  C} {\bfseries 80} (2020) 967},
  [\href{https://arxiv.org/abs/2005.07690}{{\ttfamily 2005.07690}}].

\end{thebibliography}\endgroup
\bibliographystyle{Biblio}

\end{document}